\documentclass[longauth]{aa}  
\makeatletter
\renewcommand*\aa@pageof{, page \thepage{} of \pageref*{LastPage}}
\makeatother

\usepackage{hyperref}
\hypersetup{breaklinks=true,colorlinks=true,urlcolor=blue, linkcolor=blue,  citecolor=blue, bookmarksopen=true}
\usepackage{graphicx}
\usepackage{txfonts}
\usepackage{placeins}
\usepackage{float}
\usepackage{xcolor}
\graphicspath{{plots/}{./}}
\newcommand{\Lsun}{\mathrm{L}_\odot}
\newcommand{\Msun}{\mathrm{M}_\odot}
\begin{document}

   \title{Mid-infrared evidence for iron-rich dust in the multi-ringed inner disk of HD 144432
   \thanks{Based on observations collected at the European Southern Observatory under ESO programmes 190.C-0963(D), 190.C-0963(E), 190.C-0963(F), 0100.C-0278(E), 0103.D-0153(C), 0103.C-0347(C), 0103.D-0153(G), 108.225V.003, 108.225V.011, and 108.225V.006.}}
    \titlerunning{Mid-infrared evidence for iron-rich dust in the multi-ringed inner disk of HD 144432}

   \author{J. Varga\inst{1,2,3}
          \and
          L. B. F. M. Waters\inst{4,5}
          \and
          M. Hogerheijde\inst{3,6}
          \and
          R. van Boekel\inst{7}
          \and
          A. Matter\inst{8}
          \and
          B. Lopez \inst{8}
          \and
          K. Perraut\inst{9}
          \and
          L. Chen\inst{1,2}
          \and          
          D. Nadella\inst{3}
          \and
          S. Wolf\inst{10}
          \and
          C. Dominik\inst{6}
          \and
          Á. K\'osp\'al\inst{1,2,7,11}
          \and
          P. \'Abrah\'am\inst{1,2,11}
          \and
          J.-C. Augereau\inst{9}
          \and
          P. Boley\inst{12}
          \and
          {G. Bourdarot}\inst{13}
          \and
          {A. Caratti o Garatti}\inst{7,14}
          \and
          F. Cruz-S\'aenz de Miera\inst{1,2,{15}}
          \and
          W. C. Danchi\inst{16} 
          \and
          {V. Gámez Rosas}\inst{3}
          \and
          Th. Henning\inst{7}
          \and
          K.-H. Hofmann\inst{17}
          \and
          M. Houll\'e\inst{8}
          \and
          {J. W. Isbell}\inst{7}
          \and
          W. Jaffe\inst{3}
          \and
          T. Juh\'asz\inst{1,2,11}
          \and
          V. Kecskem\'ethy\inst{3}
          \and
          J. Kobus\inst{10}
          \and
          E. Kokoulina\inst{18,8}
          \and
          {L. Labadie}\inst{19}
          \and
          F. Lykou\inst{1,2}
          \and
          F. Millour\inst{8}
          \and
          A. Mo\'or\inst{1,2}
          \and
          {N. Morujão}\inst{20}
          \and
          E. Pantin\inst{21}
          \and
          D. Schertl\inst{17}
          \and
          M. Scheuck\inst{7}
          \and
          L. van Haastere\inst{3}
          \and
          G. Weigelt\inst{17}
          \and
          J. Woillez\inst{22}
          \and
          P. Woitke\inst{23}
          \and
          {MATISSE \&} GRAVITY Collaborations
          }
          
   \institute{HUN-REN Research Centre for Astronomy and Earth Sciences, Konkoly Observatory, Konkoly-Thege Miklós út 15-17, H-1121 Budapest, Hungary
    \\
              \email{varga.jozsef@csfk.org}
        \and
        CSFK, MTA Centre of Excellence, Konkoly-Thege Miklós út 15-17, H-1121 Budapest, Hungary
         \and Leiden Observatory, Leiden University, P.O. Box 9513, 2300 RA Leiden, The Netherlands
\and
             Institute for Mathematics, Astrophysics and Particle Physics, Radboud University, P.O. Box 9010, MC 62 NL-6500 GL Nijmegen, the Netherlands
            \and SRON Netherlands Institute for Space Research, Niels Bohrweg 4, 2333 CA Leiden, The Netherlands
            \and Anton Pannekoek Institute for Astronomy, University of Amsterdam, Science Park 904, 1090 GE Amsterdam, The Netherlands \and
            Max-Planck-Institut f\"ur Astronomie, K\"onigstuhl 17, D-69117 Heidelberg, Germany \and
            Universit\'e C\^ote d'Azur, Observatoire de la C\^ote d'Azur, CNRS, Laboratoire Lagrange, France \and
            Univ. Grenoble Alpes, CNRS, IPAG, 38000 Grenoble, France \and
            Institute of Theoretical Physics and Astrophysics, University of Kiel, Leibnizstr. 15, 24118 Kiel, Germany \and
            ELTE Eötvös Loránd University, Institute of Physics, Pázmány Péter sétány 1/A, 1117 Budapest, Hungary \and
            Visiting astronomer, Laboratoire Lagrange, Universit\'e C\^ote d'Azur, Observatoire de la C\^ote d'Azur, CNRS, Boulevard de l'Observatoire, CS 34229, 06304 Nice Cedex 4, France \and 
            Max Planck Institute for Extraterrestrial Physics, Giessenbachstrasse, 85741 Garching bei München, Germany \and
            INAF-Osservatorio Astronomico di Capodimonte, via Moiariello 16, 80131 Napoli, Italy \and
            Institut de Recherche en Astrophysique et Plan\'etologie, Universit\'e de Toulouse, UT3-PS, OMP, CNRS, 9 av. du Colonel Roche, 31028 Toulouse Cedex 4, France \and
            NASA Goddard Space Flight Center, Astrophysics Division, Greenbelt, MD, 20771, USA \and
            Max-Planck-Institut für Radioastronomie, Auf dem Hügel 69, 53121, Bonn, Germany \and
            STAR Institute, University of Li\`ege, Li\`ege, Belgium \and
            I. Physikalisches Institut, Universität zu Köln, Zülpicher Str. 77, 50937, Köln, Germany \and
            CENTRA, Centro de Astrofísica e Gravitação, IST, Universidade de Lisboa, 1049-001 Lisboa, Portugal and Faculdade de Engenharia, Universidade do Porto, Rua Dr. Roberto Frias, P-4200-465 Porto, Portugal \and 
            AIM, CEA, CNRS, Université Paris-Saclay, Université Paris Diderot, Sorbonne Paris Cité, 91191, Gif-sur-Yvette, France \and
            European Southern Observatory, Karl-Schwarzschild-Stra{\ss}e 2, 85748 Garching, Germany \and
            Space Research Institute, Austrian Academy of Sciences, Schmiedlstr. 6, 8042, Graz, Austria
             }

   \date{Received September 15, 1996; accepted March 16, 1997}


  \abstract
   {Rocky planets form by the concentration of solid particles in the inner few au regions of planet-forming disks. Their chemical composition reflects the materials in the disk available in the solid phase at the time the planets were forming. Studying the dust before it gets incorporated in planets provides a valuable diagnostic for the material composition.
   }
   {We aim to constrain the structure and dust composition of the inner disk of the young {Herbig Ae} star HD~144432, using an extensive set of infrared interferometric data taken by the Very Large Telescope Interferometer (VLTI), combining PIONIER, GRAVITY, and MATISSE observations.}
   {We introduced a new physical disk model, \texttt{TGMdust}, to image the interferometric data, and to fit the disk structure and dust composition. 
   We also performed equilibrium condensation calculations with \textsc{GGchem} to assess the hidden diversity of minerals occurring in a planet-forming disk such as HD~144432.}
   {Our best-fit model has three disk zones with ring-like structures at $0.15$, $1.3$, and $4.1$~au. Assuming that the dark regions in the disk at $\sim$$0.9$~au and at $\sim$$3$~au are gaps opened by planets, we estimate the masses of the putative gap-opening planets to be around a Jupiter mass. We find evidence for an optically thin emission ($\tau<0.4$) from the inner two disk zones ($r < 4$~au) at $\lambda>3\ \mu$m. Our silicate compositional fits confirm radial mineralogy gradients, as for the mass fraction of crystalline silicates we get around $61\%$ in the innermost zone ($r<1.3$~au), mostly from enstatite, while only $\sim$$20\%$ in the outer two zones. 
   To identify the dust component responsible for the infrared continuum emission, we explore two cases for the dust composition, one with a silicate+iron mixture and the other with a silicate+carbon one.    
   We find that the iron-rich model provides a better fit to the spectral energy distribution. 
   Our \textsc{GGchem} calculations also support an iron-rich and carbon-poor dust composition in the warm disk regions ($r<5$~au, $T > 300$~K).}
   {We propose that in the warm inner regions ($r<5$~au) of typical planet-forming disks, most if not all carbon is in the gas phase, while iron and iron sulfide grains are major constituents of the solid mixture along with forsterite and enstatite. Our analysis demonstrates the need for detailed studies of the dust in inner disks with new mid-infrared instruments such as MATISSE and JWST/MIRI.}


   \maketitle
%

\section{Introduction}

Young low- and intermediate-mass pre-main-sequence stars are often surrounded by a gas and dust disk. This disk forms early in the star formation process as a result of angular momentum conservation, as matter from the collapsing molecular cloud core accretes onto the forming protostar. Recent observations with high-angular resolution facilities -- for example, the Atacama Large Millimeter/Submillimeter Array (ALMA) and the Spectro-Polarimetric High-contrast Exoplanet REsearch at the Very Large Telescope (VLT/SPHERE) -- have revealed {a rich variety of substructures in those disks} \citep[e.g.,][]{ALMA2015_HLTau,Andrews2018,Huang2018,Avenhaus2018,Boccaletti2020}. These observations show rings, gaps, spiral arms, crescents, and clumps. {This is accompanied with strong observational evidence that planets form and evolve in those disks} \citep{Marois2008,Keppler2018,Muller2018,Teague2018,Pinte2019}. Planet formation may start very soon after the formation of the disk, as evidenced by structures detected in very young disks {($<2$~Myr)} using high-resolution images at millimeter wavelengths \citep{ALMA2015_HLTau,Sheehan2018,Segura-Cox2020,Nakatani2020,Sheehan2020,Sheehan2022}. 

These observations are put into context using thermochemical and hydrodynamical disk models that describe the changes expected in the disk structure and chemical composition of the gas and the solids as a result of planet formation \citep[e.g.,][]{Kley2012,Flock2015,Woitke2018}. Key questions to address are how to link the observed diversity of planetary systems to the properties of the disks in which they form; what are the processes that determine the composition of exoplanets and their atmospheres; and how can we constrain the chemical composition of the rocky planets that form from the refractory dust in the inner regions of disks.

As the disk evolves, the small (sub-$\mu$m sized) and mostly amorphous silicate dust grains that are accreted from the molecular cloud experience grain growth through coagulation and begin to settle toward the midplane, and, once reaching a critical size that depends on the local gas pressure, drift inward toward the star \citep{Testi2014}. While in the outer disk grain growth dominates, in the dense inner disk grain-grain collisions result in a collisional equilibrium, which may include planetesimal-sized bodies. As the temperature increases inward, the chemical composition and lattice structure of the grains are modified, eventually leading to annealed, crystalline grains \citep{Bouwman2001,Henning2010}. While irradiation from the central star heats up the disk surface, viscous heating due to accretion acts in the midplane. Both processes cause part of the dust to constantly evaporate and
reform \citep{Min2011}. Gas phase condensation of solid material in the inner disk regions provides a source of freshly condensed crystalline material that are mixed in with the grains that are accreted from the molecular cloud \citep{Sargent2009,Matter2020}. 

These processes tend to produce a radial distribution of silicate grains with a gradient in size and crystallinity \citep{vanBoekel2004nature,Varga2018}. The opening of disk gaps and associated gas pressure bumps \citep{Ruge2016,Dullemond2018} can lead to size sorting and {thus} substantially different spatial distributions of large and small grains in the disk \citep{Pinilla2012}. In addition, parent body processing followed by a collisional release of small grains may also modify the chemical composition of the grains in the inner disk \citep{Thebault2007}.  
Finally, episodic eruptions \citep{Abraham2009}, {photoevaporative winds \citep{Owen2011},} and local heating processes (for instance, related to the formation of gas giant planets, or electrical discharges in the disk) also affect the nature and spatial distribution of the small grains \citep{Pilipp1998,Desch2000,Harker2002}. 

Spectral analysis can uncover the mineral buildup of the dust in planet-forming disks \citep{vanBoekel2005survey,Juhasz2010,Olofsson2010}, {thus providing clues about the composition of future planets that formed from that material to be compared with the Solar System.}
Silicate dust grains in the size range up to an $\sim$$5\ \mu$m radius are easily detected in the surface layers of passively heated disks, because of the strong vibrational resonances at infrared (IR) wavelengths that cause emission bands and allow a measure of their chemical composition and size \citep[e.g.,][]{Molster2003}. Observations have indeed shown the presence of $\mu$m-sized grains in the disk surface that are substantially larger than those found in the interstellar medium, and of crystalline silicates such as olivines and pyroxenes with very low Fe/Mg ratios, consistent with forsterite (Mg$_2$SiO$_4$) and enstatite (MgSiO$_3$) \citep{Henning2010}. These minerals are also abundant in the inner Solar System bodies, as an example, Mg, Si, and O constitute 59\% of the mass of Earth \citep{Morgan1980}. However, the single most abundant element present in the inner Solar System planets is Fe (e.g., it makes up 32 \% of the mass of Earth, \citealp{Morgan1980}), which is mostly concentrated in the cores of the planets. This implies that during the formation of planets, a large reservoir of iron-containing dust was available. Yet, the detection of iron in the interstellar and circumstellar dust proves to be elusive\footnote{{There is some observational evidence for metallic iron and iron-enriched silicate dust grains in the circumstellar envelopes of evolved stars \citep[e.g., ][]{Habing1996,Kemper2002,Marini2019}.}}. Metallic iron does not show any conspicuous spectral features at optical and IR wavelengths, in contrast to silicates. The same applies for amorphous carbon dust, which is also presumed to be present in the cosmic dust \citep[e.g.,][]{Henning1998,Tielens2022}. 

In this paper one of our main aims is to attempt to constrain the solid iron content of a planet-forming disk, using IR photometric and spectro-interferometric observations. {For that we analyzed} the spectral fingerprints of the different dust species, aiming to constrain their mass ratios. The novel aspect in this work is that we performed a consistent modeling of high angular resolution IR interferometric data that enables us to uncover the disk structure at sub-au resolution, and at the same time determine the composition and radial distribution of the dust. Spatially resolving the warm inner parts of the disk ($<10$~au) where the relevant dust emission features originate is a key aspect, because it enables a more direct inference of the optical properties of the solids, that can be compared to optical data of minerals from laboratory experiments \citep{Dorschner1995,Jager1998}.

Here we study the planet-forming disk around HD~144432 A, a nearby Herbig Ae star {belonging to the Upper Sco subgroup of the Sco-Cen OB association \citep{Galli2018,Luhman2020}}, at a distance of $154.1$~pc \citep{Bailer-Jones2021}, with a spectral type of A9/F0V \citep{Houk1982}. \citet{Muller2011} showed that HD~144432 is a {hierarchical} triple system, consisting of the A-type primary, and a close binary system located 1\farcs47 ($227$~au) away. The close binary pair, HD~144432~B and C has a separation of 0\farcs1 ($15$~au). Components B and C are T Tauri type stars of spectral type K7V and M1V, respectively, and the triple system seems to be coeval with an age of $6\pm 3$~Myr \citep{Muller2011}. HD~144432 has a prominent silicate spectral emission feature in the $N$ band ($8-13\ \mu$m), with one of the highest peak-to-continuum ratios ($\sim$$4$) observed so far \citep{vanBoekel2005survey}. {From a comparison of ISOPHOT-S and Spitzer InfraRed Spectrograph (IRS) spectra of the object, \citet{kospal_atlas} pointed out that the $N$ band silicate feature of HD 144432 is variable.} \citet{vanBoekel2004nature} presented $N$ band ($8-13\ \mu$m) spectro-interferometric observations of HD~144432 from the VLTI/MIDI instrument. The authors found a radial mineralogy gradient in the disk, where the region within $r\sim2$~au is richer in crystalline silicate grains, compared to the rest of the disk. 

\begin{table*}
\caption{Overview of VLTI observations of HD~144432 used in our work. The $\tau_0 $ is the atmospheric coherence time. LDD {(limb darkened diameter)} is the estimated angular diameter of the calibrator. {GRA4MAT is the new instrument mode of MATISSE, which uses the GRAVITY fringe tracker to stabilize the MATISSE fringes.}  }
\begin{center}
    \label{tab:obs}
    \small
    \begin{tabular}{c c c c c c c c c c}
        \hline
        \hline
        Instrument & \multicolumn{5}{c}{Target} & \multicolumn{3}{c}{Calibrator} & {Band}\\
        \hline
        & Date and time & Seeing & $\tau_0$ & Stations & Array & Name & LDD & Time \\
        & (UTC) & (\arcsec) & (ms) &  &  & & (mas)& (UTC) \\
        \hline
PIONIER & 2013-06-07T02:52 & 0.7 & 8.0 & K0-A1-G1-J3 & large &  &  &  & $H$\\
PIONIER & 2013-06-17T02:46 & 0.6 & 5.2 & D0-G1-H0-I1 & medium  &  &  &  & $H$\\
PIONIER & 2013-07-03T04:03 & 1.3 & 2.7 & D0-A1-C1-B2 & small &  &  &  & $H$\\
GRAVITY & 2018-03-05T08:29 & 0.5 & 11.0 & A0-G1-J2-J3 & large &  &  &  & $K$\\
MATISSE & 2019-05-06T07:33 & 0.5 & 6.2 & K0-G1-D0-J3 & large & 56 Hya & 1.2 & 07:17 & $L$\\
GRAVITY & 2019-05-24T04:38 & 0.9 & 2.8 & A0-G1-J2-J3 & large &  &  &  & $K$\\
MATISSE & 2019-09-02T01:19 & 0.6 & 6.2 & A0-B2-D0-C1 & small & $H$ Sco & 4.7 & 01:57 & $L$ (chopped)\\
GRA4MAT & 2022-03-09T09:10 & 0.4 & 16.8 & K0-G2-D0-J3 & medium & HD 138816 & 2.1 & 08:41 & $L$ (chopped)\tablefootmark{(a)}\\
GRA4MAT & 2022-03-09T09:10 & 0.4 & 16.8 & K0-G2-D0-J3 & medium & HD 138816 & 2.1 & 08:41 & $M$ (chopped)\tablefootmark{(a)}\\
GRA4MAT & 2022-03-22T09:11 & 0.9 & 3.8 & U1-U2-U3-U4 & UTs & HD 151051 & 2.3 & 08:39 & $L$\\
MATISSE & 2022-03-22T09:11 & 0.8 & 5.2 & U1-U2-U3-U4 & UTs & HD 151051 & 2.3 & 08:39 & $M$ (chopped)\\
GRA4MAT & 2022-03-22T09:11 & 0.9 & 3.8 & U1-U2-U3-U4 & UTs & HD 151051 & 2.3 & 08:39 & $N$\\
GRA4MAT & 2022-04-04T08:23 & 0.8 & 3.8 & A0-G1-J2-J3 & large & 27 Sco & 2.6 & 08:51 & $L$ (chopped)\\
GRA4MAT & 2022-04-04T08:23 & 0.8 & 3.8 & A0-G1-J2-J3 & large & 27 Sco & 2.6 & 08:51 & $M$ (chopped)\\
        \hline
    \end{tabular}
\end{center}
\tablefoot{\tablefoottext{a}{Only the BCD IN-IN data were used.}
}
\end{table*}

\citet{Chen2012} studied near-IR ($H$ and $K$ bands) VLTI/AMBER interferometric observations of HD~144432. Using a geometric ring model they found a nearly face-on disk (inclination $< 28\degr$) with a $K$ band ring radius of $0.17 \pm 0.01$~au, which may be consistent with the location of the dust sublimation radius. They also needed a spatially extended halo component to fit their data. In their subsequent paper \citep{Chen2016}, they presented radiative transfer simulations with RADMC-3D using $H$ (IOTA, AMBER), $K$ (Keck Interferometer, AMBER), and $N$ band (MIDI) data. Their best-fit model features an optically thin inner disk component between $0.2$ and $0.3$~au, and an optically thick outer disk between $1.4$ and $10$~au. They conclude that {the disk of HD~144432 has a} gap-like discontinuity at around $1$~au radius, characteristic of pre-transitional disks. Such disks are thought to represent an evolutionary step between an earlier stage where optically thick material fills the entire warm region ($\lesssim10$~au), and the later transitional disk stage where the warm disk region is mostly cleared of dust. \citet{Monnier2017} were not able to detect the disk of HD~144432 in $H$ band scattered light (using GPI), {concluding} that the nearby companion {may} truncate the outer disk. {HD 144432 was part of a VLT/SPHERE survey of self-shadowed disks \citep{Garufi2022}. The authors suggest that the disk is detected out to $0.25\arcsec$ ($\sim$$40$~au), and that the strong asymmetry in the polarimetric image may be because of shadowing (see Fig.~2, and page 5 therein).}

In this paper we model new $L$, $M$, and $N$ band VLTI/MATISSE and $K$ band VLTI/GRAVITY interferometric observations of HD~144432, complemented with archival data from VLTI/PIONIER ($H$ band). To our knowledge, this is the most complete IR interferometric data-set {on} a young stellar object {analyzed} so far, in terms of wavelength coverage, ranging between $1.6-13\ \mu$m. We apply a new flat disk temperature gradient model with simple radiative transfer to fit the data, that is able to reveal disk substructures, and constrain the dust composition.
The paper is structured as follows: in Section~\ref{sec:obsdata} we describe the observations and data processing, in Section~\ref{sec:model} we present our modeling approach, in Section~\ref{sec:results} we show our results, followed by a discussion and summary in Sections~\ref{sec:discussion} and \ref{sec:summary}, respectively.

\section{Observations and data processing}
\label{sec:obsdata}

European Southern Observatory's (ESO) Very Large Telescope Interferometer (VLTI) is located at Paranal Observatory, Chile. Currently VLTI has three instruments: PIONIER, working in $H$ band ($1.6\ \mu$m); GRAVITY, working in $K$ band ($2.2\ \mu$m); and MATISSE, working in $L$ ($2.8-4\ \mu$m), $M$ ($4.6-5\ \mu$m), and $N$ ($8-13\ \mu$m) bands.
All of these instruments combine the light of four telescopes, either of the $8.2$~m {diameter} Unit Telescopes (UTs), or of the $1.8$~m {diameter} Auxiliary Telescopes (ATs). {Using} interferometric beam combination, VLTI instruments sample the visibility function, a complex quantity, which is the Fourier-transform of the object's brightness distribution on the sky. We collected a large set of VLTI data on HD~144432 partly from our observations, partly from {the Optical interferometry DataBase (OiDB) at the Jean-Marie Mariotti Center}. Table~\ref{tab:obs} shows the overview of the VLTI observations used. 
Additionally, we collected archival IR photometric and spectroscopic data {using the VizieR catalog access tool}. 

VLTI instruments provide spectrally resolved measurements. As the probed IR bands are quite broad, the angular resolution ($\vartheta$) varies considerably over a band, modulating the spectral response. Additionally, spectral modulations can arise due to well-resolved structures in the disk. In the interferometric data, those spatial signatures are mixed with spectral emission or absorption features of the dust grains. A key aspect in our analysis is to disentangle the spatial signatures from true spectral information. 

The angular resolution of VLTI instruments depends on the wavelength and the actual baseline lengths of the observations. The fixed UT array has a baseline range of $46-130$~m, while, as of 2023, the movable ATs can be {positioned} to three configurations: small ($11-34$~m), medium ($40-104$~m), and large ($58-132$~m). The approximate {angular resolution}\footnote{{Following \citet{Matter2020},} the {angular resolution} is estimated as $\vartheta = {0.77} \lambda / \left(B_\mathrm{p}\right)$, where $\lambda$ is the wavelength, and $B_\mathrm{p}$ is the projected baseline length.} at long baselines ($100-130$~m) is {2}~mas for PIONIER, {3}~mas for GRAVITY, {4.5}~mas for MATISSE $L$ band {(at $3.5\ \mu$m)}, and {14}~mas for MATISSE $N$ band {(at $10\ \mu$m)}. These numbers translate to physical scales in the range of {$0.3-2$}~au at the distance of our object.

\subsection{MATISSE observations and data processing}
\label{sec:MATISSE_obsdata}

We observed HD~144432 with MATISSE \citep{Lopez2021} between 2019 and 2022 in the frame of our Guaranteed Time Observing (GTO) program for surveying young stellar objects (YSOs). {Data from 2019 were taken using the standard instrument mode (MATISSE standalone), while the 2022 data were taken using GRA4MAT. GRA4MAT \citep[GRAVITY for MATISSE][]{Woillez+2023} is a new instrument mode, which uses the fringe tracker of the GRAVITY instrument \citep{Gravity2017} to stabilize the MATISSE fringes. This allows us to have longer integration time (on the order of seconds), in order to reach higher signal-to-noise ratio (S/N). In contrast, the integration time of MATISSE standalone is limited by the coherence time of the atmosphere, typically $\sim$$0.1$~s in $L$ band.} We reduced and calibrated the data, and selected a set of data products to be used in this work. The selection was necessary to ensure that only good quality data (taken in good atmospheric conditions, and without technical issues) ended up in our final data set for modeling. Table~\ref{tab:obs} lists those data sets. The selected data sets were observed at seeing values ranging between $0.4\arcsec$ and $1.3\arcsec$, and atmospheric coherence time ($\tau_0$) values between $2.8$ and $16.8$~ms.

We used versions 1.5.0 (for the 2019 data) and 1.5.8 (for the 2022 data) of the standard MATISSE data reduction pipeline (DRS) \citep{Millour2016} to process the data\footnote{{The pipeline versions 1.5.0 and 1.5.8 give essentially the same results for $LM$ band data.}}. Additional processing steps, like Beam Commuting Device (BCD) calibration and flux calibration was performed using the MATISSE tools python software package\footnote{\url{https://gitlab.oca.eu/MATISSE/tools}}. Final processing (averaging and merging different data types to a final OIFITS file) was done using our own python tool. We applied the same data processing workflow as in \citet{Varga2021}. For more details, we refer to Sect.~3.1 in that work. Each calibrated data set contains the following: spectrally resolved visibilities on six baselines, closure phases on 4 baseline triplets, {flux-calibrated} correlated spectra\footnote{The correlated flux is the fringes amplitude in flux units. If it is normalized to the single-dish flux of the object, it is known as {fringes contrast or} (absolute) visibility.  By correlated spectrum we mean a series of correlated flux values along the wavelength direction. For more details we refer to Appendix~\ref{sec:app_viscalc}.} on six baselines, and a {flux-calibrated} total spectrum (i.e., single-dish spectrum). {The final calibrated MATISSE data products are shown in Appendix~\ref{sec:app_dataplots}, in Figs~\ref{fig:data_L}, \ref{fig:data_M}, and \ref{fig:data_N}\footnote{{Our data products are publicly available in OIFITS version 2 format \citep{Duvert2017} at the Optical interferometry Database (OiDB, \url{http://oidb.jmmc.fr}).}}.}
The spectral resolution in $LM$ bands is 34, while it is 30 in the $N$ band. More details on the data selection and processing can be found in Appendix~\ref{sec:app_dataproc}.

 \paragraph{{First look at the data.}} The $LM$ band data show a moderately resolved structure, with the visibility dropping to $0.3$ at the longest baseline (Fig.~\ref{fig:data_L}), with some contribution from the unresolved central star which accounts for about $10\%$ of the total $L$ band emission (for more details on how the stellar contribution is calculated, we refer to Sect.~\ref{sec:model}). The $L$ band closure phases do not deviate significantly from zero ($<6\degr$), indicating a brightness distribution which is nearly centrally symmetric. 
 The $L$ band spectra are featureless. In the $N$ band, the visibilities on the UT baselines are between $0.03$ and $0.35$ {(Fig.~\ref{fig:data_N})}, indicating a much more resolved brightness distribution, compared to that in $L$ band. {The contribution of the central star in $N$ band is only $0.5\%-3\%$, practically negligible. The N-band closure phases reach $\sim$$30\degr$, indicating some asymmetry of the N-band brightness distribution.} The $N$ band single-dish spectrum and correlated spectra are dominated by the silicate spectral feature in emission (Fig.~\ref{fig:silfit}). The ratio of the peak flux with respect to the continuum is $3.3$ in case of the MATISSE single-dish spectrum. This remarkable feature strength has already been known from earlier single-dish \citep{vanBoekel2005survey,Juhasz2010} and interferometric observations \citep{vanBoekel2004nature,Varga2018}. As a 0th order approximation, the correlated spectra may be interpreted as inner disk spectra, corresponding to the radial region of the object which remains unresolved at a given baseline. 
Thus, increasing the baseline length allows us to hone in to the inner regions of the object (for more on the applicability of this notion, we refer to Sect.~\ref{sec:res_sil}).  
In the case of HD~144432, the shape of the silicate feature shows a gradual change with baseline: at the largest spatial scales (single-dish and short baseline spectra), we can see a triangular shape peaking at around $10~\mu$m, while with increasing baseline the shoulder at $11.3~\mu$m becomes more and more prominent. At relatively long baselines ($90-100$~m) the $11.3~\mu$m peak overtakes the $10~\mu$m peak. 
This behavior indicates a radial gradient in the silicate mineralogy, in agreement with \cite{vanBoekel2004nature}. 

\subsection{GRAVITY and PIONIER data}

Our target was observed in the $K$ band with the GRAVITY instrument \citep{Gravity2017} in the large AT configuration (A0-G1-J2-J3) on 2018-03-05 and on 2019-05-24{, as part of the GRAVITY GTO program for YSOs}. We used the single-field mode that feeds both the fringe tracker (FT) and the science instrument (SC) with the same object. The FT operates at low spectral resolution (5 spectral channels over the whole K-band) and high speed ($\sim$900~Hz; \citealp{Lacour2019}) to freeze the atmospheric effects and lock the fringes for the SC instrument allowing long integrations of a few tens of seconds and high spectral resolution observations to be performed. Each observation file corresponds to five minutes on the object and contains the interferometric observables for all spectral channels of both instruments (FT and SC). We calibrated the instrument transfer function by observing interferometric calibrators (HD~103125, HD~132763, and HD~163495)\footnote{All these calibrators have been selected, with the help of the SearchCal tool of the Jean-Marie Mariotti Center (JMMC) (https://www.jmmc.fr/search-cal/), to be single (according to the Washington Visual Double Star Catalog, \citealp{Mason2001}), non-variable, and bright stars ($K < 5.6$ mag), close enough ($<60\degr$ angular separation) to the target.} before and after our science target. We reduced the data with the GRAVITY data reduction
pipeline \citep{Lapeyrere2014} and following the approach of \citet{Perraut2019,Perraut2021}, we only used the FT data for our fits and discarded the blueish spectral channel that is polluted by the fluorescence induced by the metrology laser operating at 1.908~$\mu$m.\\


Calibrated PIONIER data were downloaded from the OiDB. We selected one data set for each AT configuration, {the best ones to avoid errors introduced by lower-quality data}. The 2013-06-07 and 2013-07-03 data sets were provided by Jean-Philippe Berger, and the 2013-06-17 data by Bernard Lazareff. All of these PIONIER data products are graded as L2, that is, calibrated but unpublished data.

{\paragraph{First look at the data.} GRAVITY and PIONIER data are shown in Appendix~\ref{sec:app_dataplots}, in Figs.~\ref{fig:data_K} and \ref{fig:data_H}. Both instruments resolve the disk emission, with lowest visibilities of $0.25$ (GRAVITY) and $0.5$ (PIONIER) at the longest baselines. Considering the flux contribution of the unresolved central star ($\sim$$45\%$ in $H$ band, and $\sim$$30\%$ in $K$ band), we can see that the disk emission gets fully resolved both in the GRAVITY and PIONIER data. The closure phases remain small ($<6\degr$), indicating that the near-IR disk emission is nearly centrally symmetric.} 


\subsection{Photometric data}

We collected IR photometric and spectroscopic data in order to construct the IR spectral energy distribution (SED) of HD~144432. The $JHK$ photometry comes from the Two Micron All Sky Survey (2MASS) catalog \citep{Cutri2003}. The $LMN$ band spectroscopy comes from our MATISSE observations. Finally we use a low spectral resolution Spitzer IRS spectrum from 2004-08-08 (AOR: 3587072) that covers the $5.3-36.9\ \mu$m wavelength range {(Fig.~\ref{fig:modelfit_Fe} right panel). We use the MATISSE $N$ band single-dish spectrum, which we average with the Spitzer spectrum in the overlapping wavelength range.} We deredden the photometric data using the interstellar dust model of \citet{Jones2013}, and assuming an optical extinction of $A_V = 0.4$ \citep{Chen2016photometric}. 
 
\section{Model-based imaging}
\label{sec:model}

Our aim is to describe the IR interferometric and photometric data of HD~144432 with a relatively simple flat-disk semi-physical model, in which the physical quantities have radial dependence only. The model {outputs} images of the object at the wavelengths of the data, as well as constrain several physical parameters of the disk. For this purpose we have developed a new temperature gradient model, which we call \texttt{TGMdust}, with the following main characteristics:
\begin{itemize}
    \item The model disk starts at a specific inner radius, $R_\mathrm{in}$, representing the inner edge of the dusty disk.
    \item The disk can be divided into several radial zones, each having its own surface density profile, allowing us to represent radial disk substructure. The surface density profiles are power laws. 
    \item The disk temperature profile is described by a {single} power law {over the whole radial extent of the disk.}
    \item The emission of several dust components is taken into account. The dust composition between radial zones can differ, accounting for mineralogy gradients.
\end{itemize}

The following equation gives the radial temperature profile:
\begin{equation}
\label{eq:T_profile}
T\left( r \right) = T_\text{in} \left( \frac{r}{R_\text{in}} \right)^{q},
\end{equation}
where $T_\text{in}$ is the temperature at the inner edge of the dusty disk, and $q$ is the power law exponent of the temperature profile. In each radial zone, the surface density profile is also a power law:
\begin{equation}
\label{eq:surfdens_profile}
\Sigma_{i}\left( r \right) = \Sigma_{0,i} \left( \frac{r}{R_{i}} \right)^{p_{i}}.
\end{equation}
Here $i$ denotes the i-th radial zone (between $R_{i}$ and $R_{i+1}$), where the surface density at the zone inner edge ($R_{i}$) is $\Sigma_{0,i}$, and the power law index is $p_{i}$. In the first, innermost zone $R_{1} \equiv R_\text{in}$. 

The various dust components are taken into account by their individual opacity curves. We calculate the vertical optical depth of the disk in the following way:
\begin{equation}
\label{eq:tau}
\tau_{\nu,i}\left(r\right) = \sum_{j=1}^{N}{c_{i,j} \Sigma_{i}\left( r \right)  \kappa_{\nu,j}},
\end{equation}
where $\kappa_{\nu,j}$ is the wavelength-dependent opacity and $c_{i,j}$ is the mass fraction of the j-th dust component, in the i-th zone. 

Assuming a geometrically thin and homogeneous emitting layer, we use the following radiative transfer equation to calculate the surface brightness profile of the disk:
\begin{equation}
\label{eq:I_nu_profile}
I_{\nu,i} \left( r \right) = \left(1-\text{e}^{-\tau_{\nu,i}\left(r\right)} \right) B_{\nu}\left( T\left( r \right)\right), 
\end{equation}
where the source function is the blackbody radiation ($B_{\nu}$), and $\tau_{\nu,i}\left(r\right)$ is the 
radially and wavelength-dependent 
{vertical} optical depth {profile} in the i-th zone.

To get observable quantities, we calculate the flux density ($F_{\nu}$) from the surface brightness profiles of each zone, then add them together along with the flux density of the central star ($F_{\nu,\text{star}}$):
\begin{eqnarray}
\label{eq:F_nu}
F_{\nu,i} = \int_{R_{i}}^{R_{i+1}} 2\pi r I_{\nu,i} \left( r \right)  \text{d}r, \\
F_{\nu} = \sum_{i=1}^{M} F_{\nu,i} + F_{\nu,\text{star}}.
\end{eqnarray}
We note that $F_{\nu}$ can be directly compared to single-dish photometric and spectroscopic observations. For the outermost radius of the integration ($R_\text{out}$), we choose $30$~au which, at the distance of our object, roughly corresponds to the field of view of MATISSE in the $N$ band on the UTs. We expect that this radius is large enough that it encloses practically all the flux emitted in the wavelength range of our data\footnote{{We also tested a larger outer radius, $60$~au, with which we got practically the same result as with the smaller value.}}.
For the calculation of the interferometric visibility and correlated flux, we refer to Appendix \ref{sec:app_viscalc}. Although our model has only radial dependence, we can account for the inclination of the disk ($i$) by the deprojection of the baselines (detailed in Appendix~\ref{sec:app_viscalc}). Several authors analyzing near-IR VLTI data found that the disk is close to face-on \citep{Chen2012,Lazareff2017,Perraut2019,Kluska2020}. We selected a subset of the MATISSE data with the highest quality to fit the disk orientation with a simple geometric model. {We applied a Markov Chain Monte Carlo (MCMC) algorithm for the parameter search, and the disk size and orientation were fit.} We found a $\cos i$ of $0.79 \pm 0.1$, and the position angle of the major axis $\theta = 108\degr \pm 15\degr$ (measured from north to east). In our subsequent modeling runs we use these values, keeping them fixed.

The flux contribution of the central star is also kept fixed in the modeling. We estimate the stellar spectrum by fitting PHOENIX stellar atmosphere models\footnote{\url{ftp://phoenix.astro.physik.uni-goettingen.de/HiResFITS}} \citep{Husser2013} to optical photometric data of HD~144432. The $B_T$ and $V_T$ magnitudes were obtained from the Tycho-2 catalog \citep{Hog2000}, $G_\text{BP}$, $G$, and $G_\text{RP}$ magnitudes were obtained from the Gaia Early Data Release 3 \citep[EDR3,][]{Gaia2016,GaiaEDR3}, and $U$ and $I$ magnitudes were obtained from \citet{Mendigutia2012}. {While Gaia was able to spatially resolve the A and the BC components, as the EDR3 catalog provides separate photometry for the objects, in the $B_T$, $V_T$, $U$, and $I$ band observations the stars are likely measured together. According to the Gaia EDR3, the BC binary in the $G$ band is $3.68$ mag fainter than the primary. Using the stellar parameters by \citet{Muller2011}, we estimate the contamination by the BC components to be negligible ($<3\%$) in all our photometric bands except the $I$ band where it is $\sim$$10\%$. 
The stellar atmosphere fit was done with a modified version of SED Fitter \citep{Robitaille2007}. We permitted a relatively small fit range around the typical literature values of $T_\text{eff}$ and $\log g$. We got $T_\text{eff} = 8000$~K and $\log g = 5.5$ as best fits. For comparison, the typical range for the effective temperature of HD 144432(A) found in the literature is $7100-7500$~K. A spectroscopic determination of $T_\text{eff}$ by \citet{Fairlamb2015} using XSHOOTER yielded $7500 \pm 250$~K. Although our fit is several hundreds of K higher, it is still reasonable.}


\subsection{Dust opacities}

\begin{table}
 \caption[]{\label{tab:dustopac}Overview of the dust species used in this work.}
 \small
\begin{tabular}{p{3.1cm}lcc}
 \hline \hline
  Species &
  Chemical &
  Lattice &
  References \\
 &
  formula &
  structure &
 \\ \hline
{Mg-}silicate of olivine stoichiometry & Mg$_2$SiO$_4$ & amorphous & 1\\
Forsterite  & Mg$_2$SiO$_4$ & crystalline  & 2\\
{Mg-}silicate of pyroxene stoichiometry  & MgSiO$_3$ & amorphous & 3\\
Enstatite  & MgSiO$_3$ & crystalline & 4\\
Iron & Fe & metallic & 5 \\
Carbon & C & amorphous & 6 \\
\hline
\end{tabular}
\tablebib{(1)~\citet{Jager2003};
(2) \citet{Sogawa2006}; (3) \citet{Dorschner1995}; (4) \citet{Jager1998};
(5) \citet{Henning1996}; (6) \citet{Zubko1996}.
}
\end{table}

The $N$ band spectral window of MATISSE gives a handle on constraining the dust composition of the disk. Silicate minerals have distinct spectral emission features in this wavelength region. The shape of the feature mostly depends on the following factors: (1) the chemical composition of the mineral, (2) the crystallinity, and (3) the grain size and shape. Models aiming to explain the SEDs of planet-forming disks have to include a continuum opacity source to add more dust opacity at wavelengths where silicates are too transparent. For that purpose, carbon is usually added to the dust mix \citep[e.g.,][]{Woitke2016}. However, metallic iron may also be considered \footnote{{Dust spectral features tend to be weaker with increasing grain size. Thus, very large ($10-100\ \mu$m-sized) silicate grains may be also considered as the source of the IR continuum emission. However, even at that grain sizes there are distinct trends in the silicate opacity curves, such as a very broad dip in the $\sim$$2-7\ \mu$m wavelength range, that could be distinguished from the truly featureless opacity curves of, for instance, carbon or iron. Another argument is that while $\mu$m-sized carbon grains have IR opacity values in the range of $10^4-10^5$, the same range for $10-100\ \mu$m-sized silicate grains is only $10^1-10^2$. This means that in order to produce the same amount of radiation, we would need $\sim$$1000$ times as much big silicate grains as $\mu$m-sized carbon grains, in terms of mass.
}}.
While the variety of silicates found in the Solar System is enormous, in the interstellar space a relatively small set of silicate minerals has been identified so far. In our modeling we consider the following key dust components, thought to be main ingredients of the cosmic dust, from earlier results of, for example, \citet{Bouwman2001}, \citet{vanBoekel2005survey}, \citet{Juhasz2010}, {\citet{Jones2017}, and \citet{Tielens2022}}: amorphous {magnesium-}silicate of olivine stoichiometry (referred as olivine in the following), amorphous {magnesium-}silicate with pyroxene stoichiometry (referred as pyroxene in the following), crystalline forsterite, crystalline enstatite, metallic iron, and amorphous carbon. In our modeling we assume that the dust species are isolated, that is, they are not in thermal contact with each other. Carbonaceous material can cover a wide range of properties, for instance, graphitic (conducting) or diamond-like (not conducting) types have very different optical constants \citep{Jager1998_carbon}. We use the optical data for amorphous carbon by \citet{Zubko1996} who noted that their carbon samples were semiconductors rather than metals. 
For each {dust} component we consider two grain radii, $0.1\ \mu$m ("small") and $2\ \mu$m ("large"), except for enstatite for which we have a third size of $5\ \mu$m ("very large") in addition. The motivation for the latter choice came from our initial modeling tests where we noticed that better fit can be achieved if we include $5\ \mu$m-sized enstatite grains in the dust mix. In contrast, including $5\ \mu$m-sized grains of any other dust species did not improve the fit noticeably.

Metallic iron might become oxidized. Iron oxide solids have several resonances in their mid-IR spectra, mostly between $10$ and $50\ \mu$m. In Fig.~\ref{fig:app_oxides} we compare the observed Spitzer spectrum of HD 144432 in that spectral region with the opacity curves of FeO (w\"ustite), Fe$_2$O$_3$ (hematite), and Fe$_3$O$_4$ (magnetite). As we see no signs of the iron oxide spectral features in the Spitzer data, we conclude that iron oxides are not major
constituents of the dust in the inner disk of HD 144432.

In Table~\ref{tab:dustopac} we present an overview on the dust species used in our modeling, with references to the refractive index data. The opacities were calculated using the distribution of hollow spheres (DHS) scattering theory \citep{Min2005}. The upper boundary for the hollow sphere distribution ($f_\text{max}$) is $0.7$ for the amorphous species and iron, and $1.0$ for the crystalline silicates. We used optool\footnote{\url{https://github.com/cdominik/optool}} \citep{Dominik2021optool} to calculate the opacity curves of carbon and iron grains. {They} are shown in Fig.~\ref{fig:app_opac} (N-band only) and in Fig.~\ref{fig:app_opac2} (covering the full wavelength range of the data). An important aim of this work is to check whether our data are consistent with an iron-rich or carbon-rich dust composition. Therefore, we run models where we include either iron or carbon, to represent the extremities in the abundance ratios of those two species. 

\subsection{Optimization}

We differentiate between model parameters describing the disk structure and parameters of the dust composition. In case of a single zone, the model has 5 structural parameters ($T_\text{in}$, $q$, $R_\text{in}$, $\Sigma_{0,i}$, $p_{i}$). Each further zone adds 3 parameters ($R_i$, $\Sigma_{0,i}$, and $p_i$). The number of dust components in our initial model is 11, consisting of four silicate species and one species for the featureless continuum emission (either iron or carbon), in two or, in the case of enstatite, three grain sizes. As we want to model radial variations in the mineralogy, {each radial zone may have a specific dust composition}. We run fits with 1, 2, and 3 zones. In the 3-zone model, there are 11 structural parameters, and 33 dust mass fractions (11 per each zone). Given the large number of free parameters, we employ a two-step optimization procedure, in which we first constrain the silicate composition using a spectral decomposition tool called \texttt{specfit}, then in the second step, we fit the overall disk structure with \texttt{TGMdust}. \texttt{specfit} employs a genetic fitting algorithm, while \texttt{TGMdust} performs MCMC sampling. For more details, we refer to Appendix \ref{sec:app_optimization}.

An important note is that the dust mass fractions derived from our modeling only make sense if the disk emission at the relevant wavelengths ($N$ band) is optically thin. In the following section, we show whether this assumption holds.

\section{Results}
\label{sec:results}

\begin{figure*}
 \includegraphics[width=0.503\textwidth,trim=30 20 40 42,clip]{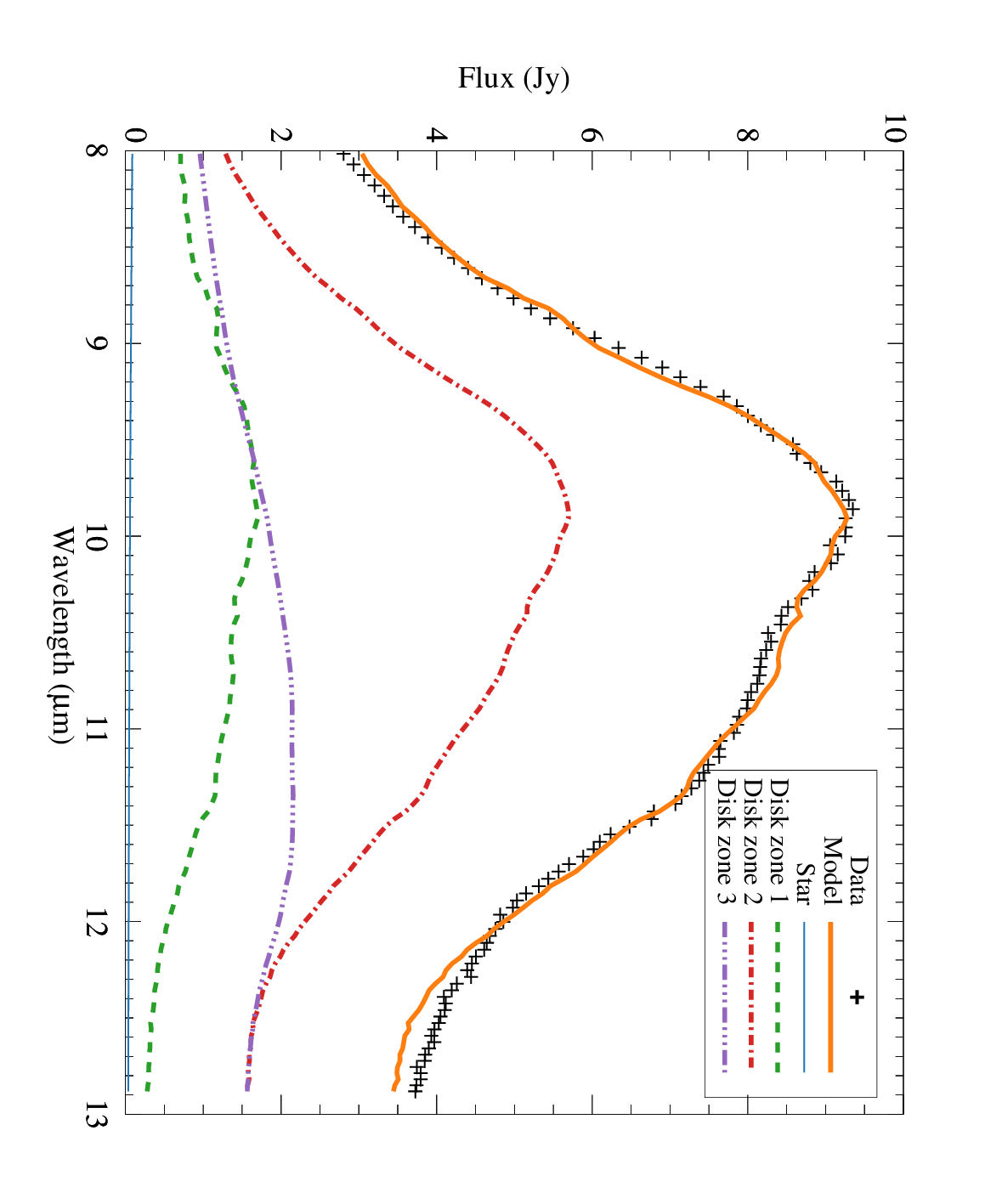}
 \includegraphics[width=0.487\textwidth,trim=18  8 24 20,clip]{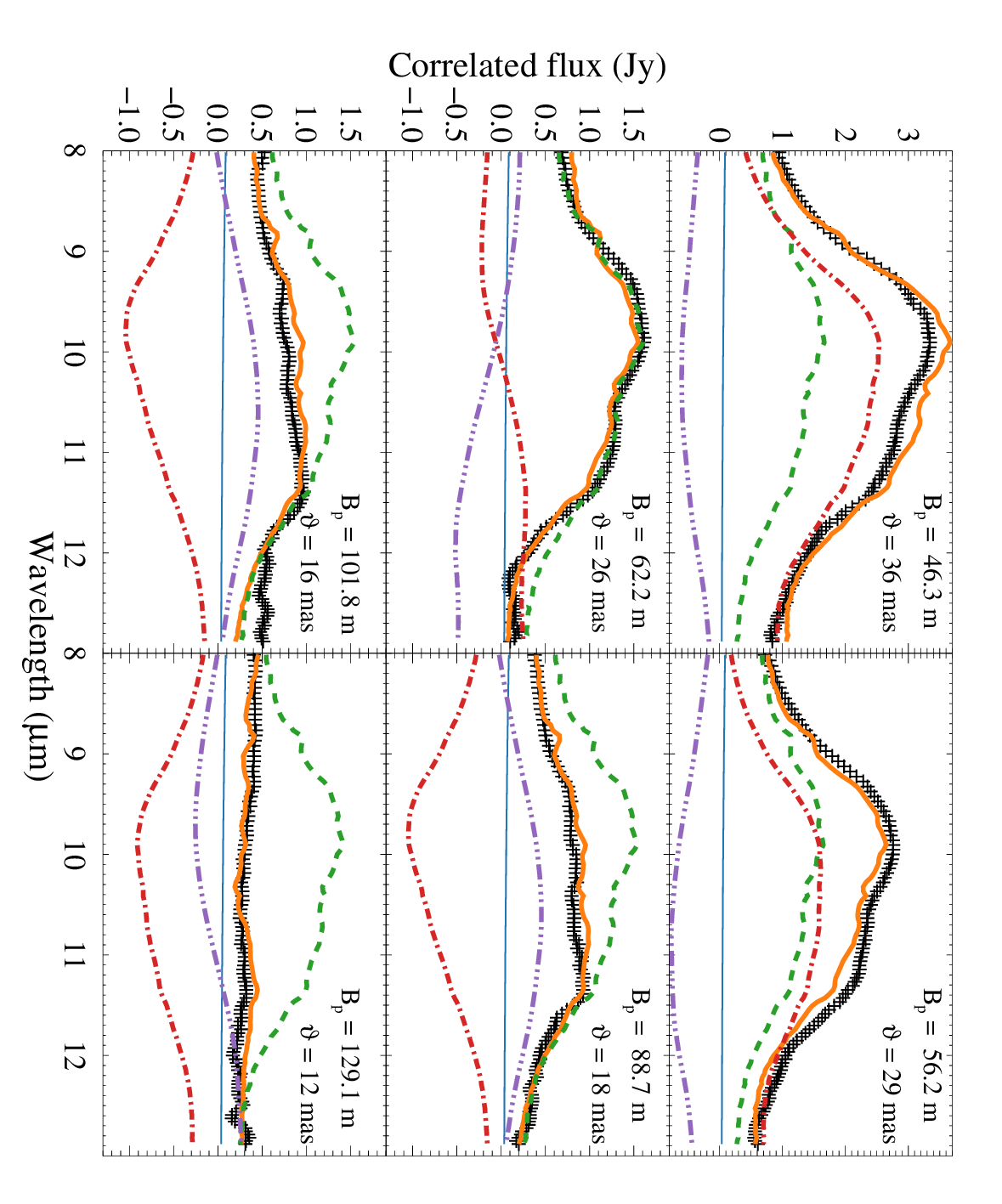}
      \caption{Dust compositional fits to the N-band MATISSE data. Left panel: fit to the single-dish spectrum. Right panel: fits to the six correlated spectra. The solid yellow curve is the total model spectrum, while the green, red, and purple lines show the contributions from the individual disk zones. The thin blue line is the model stellar spectrum.} 
         \label{fig:silfit}
\end{figure*}

\begin{figure}
 \includegraphics[width=0.98\columnwidth,trim=20 25 35 22,clip]{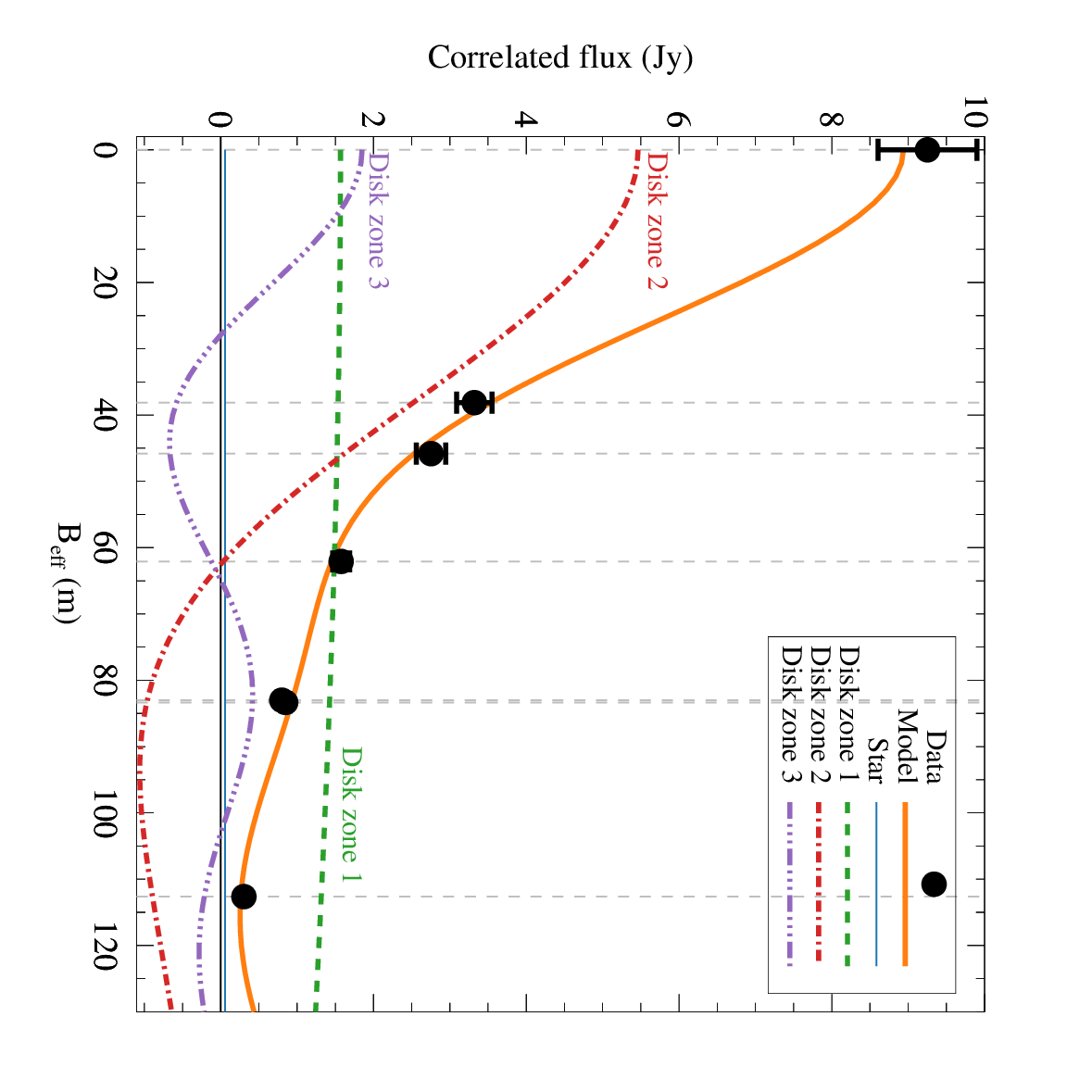}
      \caption{
      Correlated flux density as function of the baseline length at a selected wavelength, $10\  \mu$m. The contributions of the three disk zones are the green, red, and purple lines, while the correlated flux curve of the whole disk is the orange solid line. Filled circles indicate the MATISSE N-band data points. The zero baseline data point is the single-dish flux. The thin blue line is the contribution of the central star. 
      } 
         \label{fig:silfit_bl}
\end{figure}

An important finding is that we need at least three zones to fit the data well. In the following, we first present our results on the dust composition, then on the disk structure.

\subsection{{Runs with \texttt{specfit}}}
\label{sec:res_sil}

\begin{table}
 \caption[]{\label{tab:specfit_res}Results of the dust compositional fit. The abundances are 
 given as fractions of the total dust mass in percent, excluding the {hidden} dust component responsible for the continuum emission.} 
    \begin{tabular}{p{2.4cm}ccc} 
    \hline \hline 
    Species & 
    Zone 1 & 
    Zone 2 & 
    Zone 3 \\ 
    & $0.2-1.3$~au & $1.3-4.1$~au & $>4.1$~au \\
    \hline
$\text{Olivine}\ 2.0\ \mu\text{m}$ &  $25.2^{+3.6}_{-12.8}$  & $44.2^{+5.8}_{-15.0}$  & $81.2^{+0.1}_{-65.2}$  \\[1mm]
$\text{Pyroxene}\ 0.1\ \mu\text{m}$ &  $5.1^{+3.0}_{-2.7}$  & $35.1^{+12.3}_{-7.0}$  & $0.0^{+53.0}_{-0.0}$  \\[1mm]
$\text{Pyroxene}\ 2.0\ \mu\text{m}$ &  $8.3^{+9.3}_{-2.0}$  & $0.0^{+0.0}_{-0.0}$  & $0.0^{+0.0}_{-0.0}$  \\[1mm]
$\text{Forsterite}\ 0.1\ \mu\text{m}$ &  $2.0^{+0.7}_{-0.2}$  & $0.8^{+0.7}_{-0.7}$  & $0.0^{+2.5}_{-0.0}$  \\[1mm]
$\text{Enstatite}\ 5.0\ \mu\text{m}$ &  $59.3^{+3.7}_{-4.7}$  & $20.0^{+3.3}_{-10.5}$  & $18.8^{+11.6}_{-15.9}$  \\[1mm] 
    \hline 
{Crystallinity fraction} & {$61.3^{+3.8}_{-4.7}$} & {$20.8^{+3.4}_{-10.5}$} & {$18.8^{+11.9}_{-15.9}$} \\[1mm]
\hline 
    \end{tabular} \\
\end{table}

\begin{figure*}
 \includegraphics[width=\hsize]{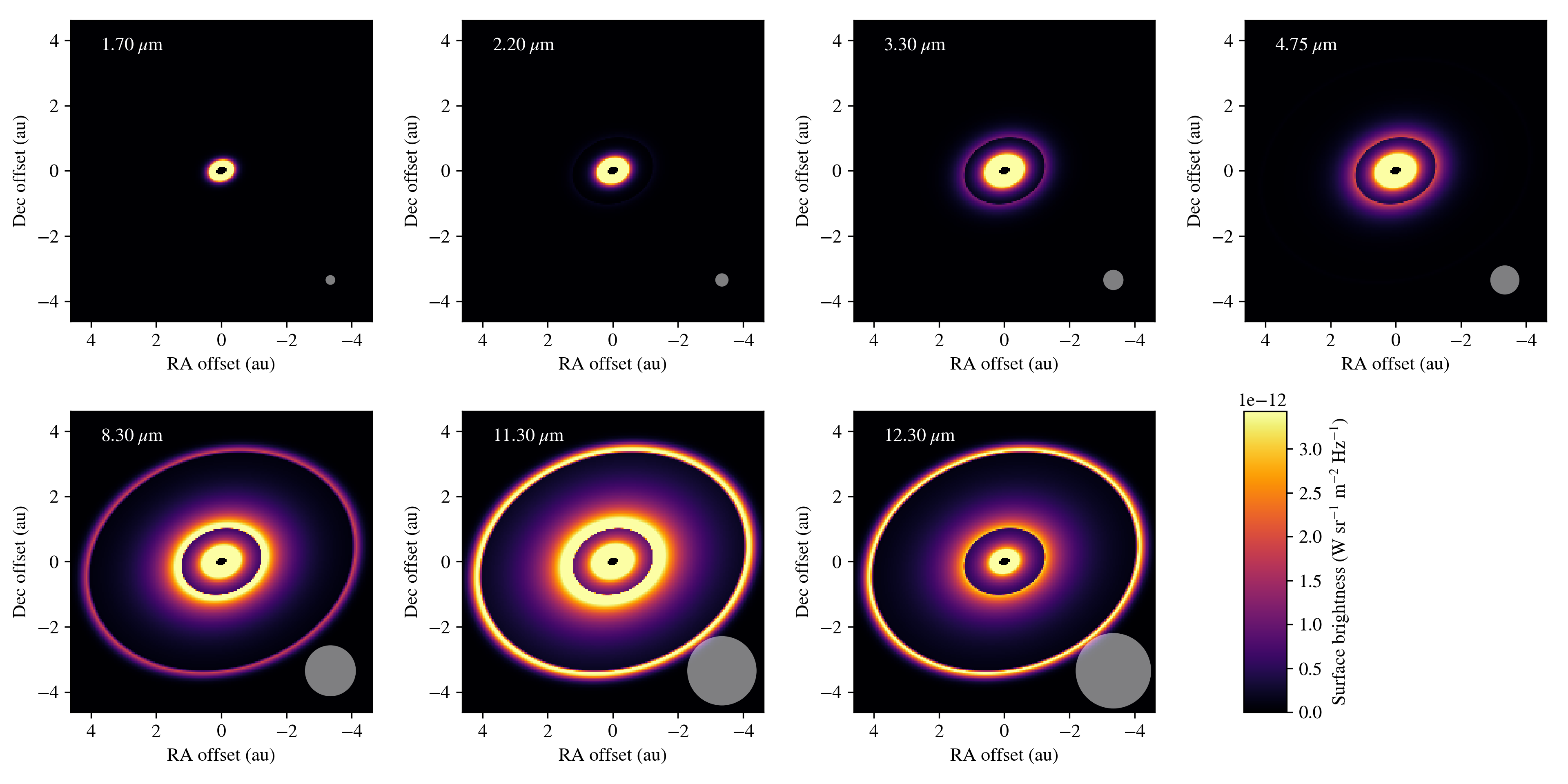}
 \caption{Best-fit model images at various data wavelengths with the \texttt{TGMdust} run with iron grains. The brightness scaling is linear and homogeneous across all wavelengths. {For a better perception} of the fainter structural features, the images become saturated at $3.4\times10^{-12}$ W~m$^{-2}$~Hz$^{-1}$~sr$^{-1}$. The gray circles in the bottom left corner indicate the approximate beam size {(estimated as $0.77 \lambda / (B_\text{p,max})$, where $B_\text{p,max}$ is the maximum baseline length)}. }
         \label{fig:modelimg_Fe}
\end{figure*}

\begin{figure}
 \includegraphics[width=\hsize]{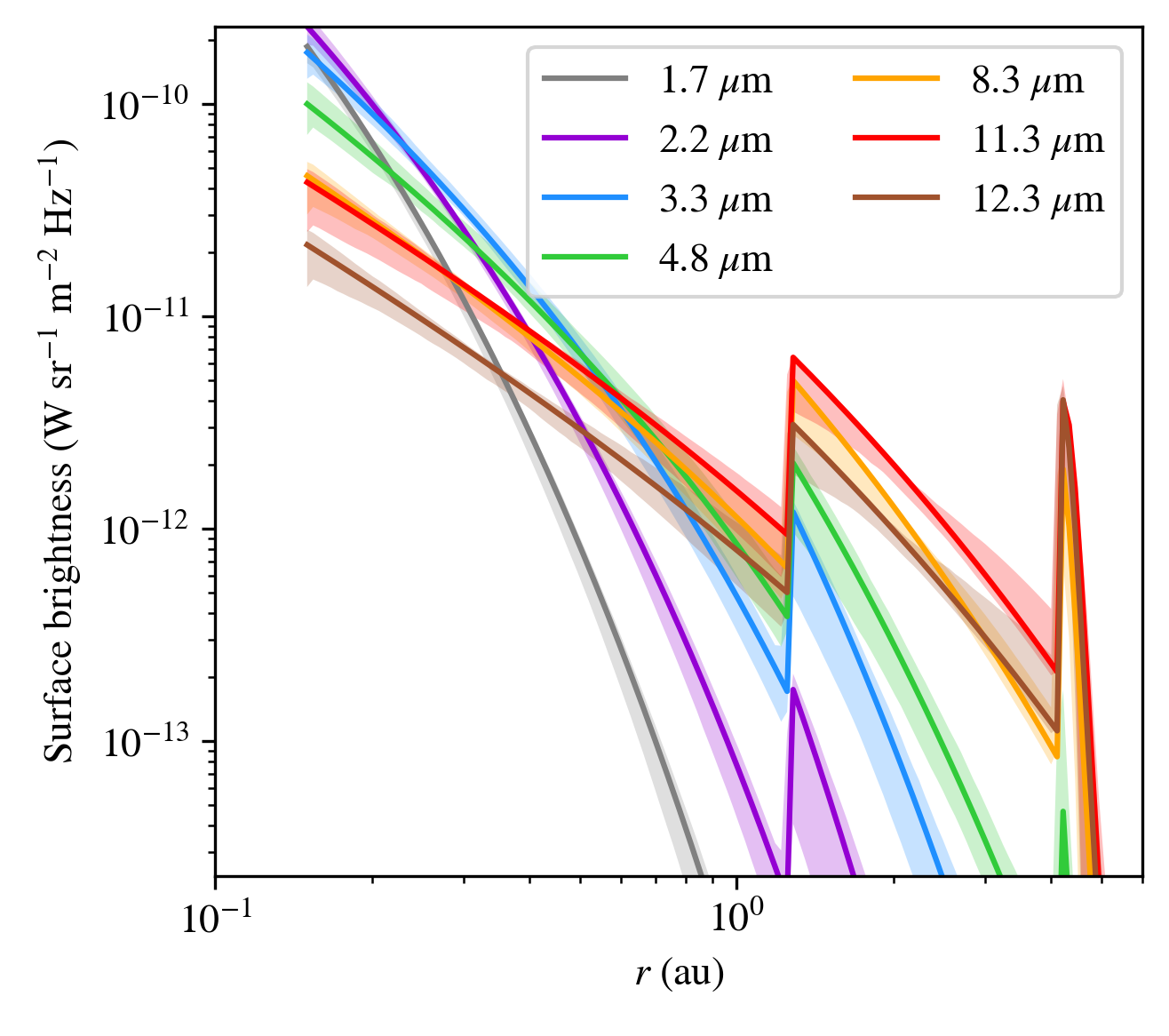}
      \caption{Radial surface brightness profiles of the disk corresponding to the best-fit model with iron grains. The shaded area around each curve is the $16-84$ percentile range from the last $5000$ samples of the chain. 
      } 
         \label{fig:I_nu_prof_Fe}
\end{figure}

\paragraph{{Fitting approach.}} 
Initial runs with \verb|specfit| showed that the fit prefers to have no or very small amounts ($<0.05\%$) of small olivine, large forsterite, small enstatite, and large enstatite. Thus, in order to decrease complexity of the model, we removed those dust components from our further fits, leaving only six dust components to fit (five silicates and the hidden dust component). A good fit to the data is reached after a few dozen generations with a population of 100~models. In order to get a denser sampling of the parameter space, our final modeling was run for 600 generations. The fits to the N-band MATISSE data are shown in Fig.~\ref{fig:silfit}, and the best-fit silicate mass fractions are presented in Table~\ref{tab:specfit_res}. {The reduced $\chi^2$ corresponding to the best fit is $10.2$. In contrast, with a two-zone model we got a much worse value of $40.6$. The two-zone model could fit the single-dish spectrum, and the correlated spectra on the three shortest baselines well, but the longer-baseline correlated spectra are entirely missed. The three-zone model provides equally good fit on all baselines. This demonstrates that the long-baseline ($>88$~m) MATISSE $N$ band data contain crucial information on the inner-disk substructure of HD 144432.}

Because the convergence of the genetic algorithm is different from that of an MCMC sampler, we cannot simply use the sample distributions to estimate the uncertainties. Thus, we filter the samples by imposing a very conservative upper limit of 10 times the best-fit $\chi^2$, and the confidence intervals are taken as the $16$th and $84$th percentiles of the filtered marginalized parameter distributions. 

\paragraph{Disk structure and radial dust composition.}
The most important disk structural parameters which are constrained by the fit are the zone radii: $R_1 = 0.22^{+0.01}_{-0.03}$~au, $R_2 = 1.26^{+0.05}_{-0.08}$~au, and $R_3 = 4.13^{+0.03}_{-0.24}$~au. {As shown in} Table~\ref{tab:specfit_res}, the most abundant dust components are large olivine, small pyroxene, and very large enstatite. The dust composition of the individual zones differs significantly. The fit results provide strong evidence for radial mineralogy gradients. Specifically, the crystalline species (forsterite, enstatite) are concentrated in the innermost zone ($<1.3$~au) where {around $61\%$} of the silicate dust mass is in crystalline form, {predominantly as enstatite}. In contrast, the crystalline fraction in the outer zones is only $\sim20\%$ {(Table~\ref{tab:specfit_res})}. Most of the dust abundances are relatively well constrained, except for large olivine and small pyroxene in the third zone. We cannot exclude that the third zone contains significant amounts of small pyroxene. We note that silicate compositional fits are subject to some degeneracy, due to similarly shaped dust opacity curves. We investigate this issue further in Appendix \ref{sec:app_alt_specfit}.

Another notable finding is that the fraction of small grains ($0.1\ \mu$m size) is rather small, only $\sim7\%$ in the inner zone, and $\sim36\%$ in the second zone. The high crystallinity fraction and the high ratio of larger, $\mu$m-sized grains {can} indicate that the dust in the inner au of the disk is significantly processed. {Alternatively, evaporation and recondensation may be important in the innermost disk region, that can also produce crystalline grains. The latter idea is further explored in Sect.~\ref{sec:disc:ggchem}.} {For the hidden dust component, we got mass fractions of $\sim$$30\%$ in the first, $\sim$$50\%$ in the second, and $\sim$$95\%$ in the third zone.}
As the opacity level of that component was chosen arbitrarily, its mass fraction values have limited physical meaning. However, it is worth to note that the mass fraction of the hidden dust increases toward the outer zones. 

Comparing the abundances of Mg$_2$SiO$_4$ {(crystalline forsterite and amorphous olivine together)} vs. MgSiO$_3$ {(crystalline enstatite and amorphous pyroxene together)}, we see that MgSiO$_3$ is $2.7$ times more abundant in the innermost zone, and $1.2$ times in the second zone. In the outer zone the same ratio is $0.2$, so there Mg$_2$SiO$_4$ appears to be more abundant, although this is a highly uncertain estimate. The elemental abundances of the disk zones are similar; each zone contains roughly $30\%$ Mg, $24\%$ Si, and $46\%$ O by mass. These numbers relate only to the silicates, as the elemental composition of the hidden dust component is not considered at this stage. The abundance ratios of Mg/Si and Si/O are $\sim0.08$~dex and $\sim-0.3$~dex, respectively. 

The left panel of Fig.~\ref{fig:silfit} shows the model fit to the single-dish $N$ band MATISSE spectrum. Green, red, and purple lines indicate the individual contributions from each zone. It turns out that the silicate spectral feature mostly originates from the second zone ($1.3$~au $- 4.1$~au) where the temperatures are in the range of $300-600$~K. There is a large difference in the shape of the silicate spectral feature in the different zones. The inner zone has a spectrum characteristic of crystalline silicates (especially enstatite), the spectrum of the second zone is dominated by small and large amorphous grains, while the third zone shows a very weak silicate feature with an emission mainly from large amorphous grains. 

\begin{table}
 \caption[]{\label{tab:fit_res}Final results of the fits. 
 } 
 \centering
    \begin{tabular}{lcc} 
\hline \hline
 & with Fe & with C \\
\hline \multicolumn{3}{c}{Best-fit parameter values with \texttt{TGMdust}} \\
$R_1$ (au)  &  $0.15^{+0.01}_{-0.00}$  &  $0.15^{+0.01}_{-0.00}$  \\[1mm]
$R_2$ (au)\tablefootmark{(a)}  & \multicolumn{2}{c}{ $1.26^{+0.05}_{-0.08} $} \\[1mm]
$R_3$ (au)\tablefootmark{(a)}  & \multicolumn{2}{c}{ $4.13^{+0.03}_{-0.24}$ } \\[1mm]
$T_\mathrm{in}$ (K)  &  $1780^{+69}_{-193}$  &  $1606^{+86}_{-111}$  \\[1mm]
$q$ &  $-0.53^{+0.04}_{-0.05}$  &  $-0.50^{+0.04}_{-0.06}$  \\[1mm]
$\Sigma_{0,1}$ ($10^{-3} \text{g\ cm}^{-2}$)  & $1.21^{+0.35}_{-0.56}$ & $0.10^{+0.04}_{-0.07}$ \\[1mm]
$\Sigma_{0,2}$ ($10^{-3} \text{g\ cm}^{-2}$)  & $0.37^{+0.06}_{-0.11}$ & $0.17^{+0.05}_{-0.06}$ \\[1mm]
$\Sigma_{0,3}$ ($10^{-3} \text{g\ cm}^{-2}$)  & $10.5^{+632.5}_{-4.1}$ & $78.3^{+196.5}_{-76.2}$ \\[1mm]
$p_1$ &  $-0.86^{+0.38}_{-0.29}$  &  $-0.74^{+0.45}_{-0.28}$  \\[1mm]
$p_2$ &  $-1.25^{+0.79}_{-0.17}$  &  $-1.24^{+0.68}_{-0.21}$  \\[1mm]
$p_3$ &  $-40.8^{+4.4}_{-49.1}$  &  $-67.5^{+37.1}_{-23.1}$  \\[1mm]
$c_{1,\text{small}}$ ($\%$) & $0.04^{+1.20}_{-0.03}$ & $21.79^{+12.68}_{-7.05}$ \\[1mm]
$c_{1,\text{large}}$ ($\%$) & $91.98^{+2.63}_{-1.65}$ & $3.45^{+13.24}_{-1.75}$ \\[1mm]
$c_{2,\text{small}}$ ($\%$) & $45.48^{+7.30}_{-13.83}$ & $9.86^{+9.31}_{-7.89}$ \\[1mm]
$c_{2,\text{large}}$ ($\%$) & $10.56^{+12.85}_{-10.54}$ & $0.41^{+7.03}_{-0.39}$ \\[1mm]
$c_{3,\text{small}}$ ($\%$) & $2.92^{+44.54}_{-2.82}$ & $0.23^{+50.30}_{-0.14}$ \\[1mm]
$c_{3,\text{large}}$ ($\%$) & $0.01^{+14.46}_{-0.00}$ & $0.02^{+11.61}_{-0.00}$ \\[1mm]
\hline \multicolumn{3}{c}{Fixed parameters} \\
$\theta$ ($\degr$)  & \multicolumn{2}{c}{ $108.0$ } \\
$\cos\,i$  & \multicolumn{2}{c}{ $0.79$ } \\
$R_\mathrm{out}$ (au)  & \multicolumn{2}{c}{ $30.00$ } \\
$d$ (pc) & \multicolumn{2}{c}{ $154.1$ } \\
\hline \multicolumn{3}{c}{Resulting half-flux radii (au)} \\
$2.2\ \mu$m  & 0.23 & 0.23\\
$3.3\ \mu$m  & 0.31 & 0.30\\
$4.75\ \mu$m  & 0.48 & 0.46\\
$11.3\ \mu$m  & 2.25 & 2.31\\
\hline
$\chi^2_V/N_V$  & $4.9$ & $5.3$ \\[1mm]
$\chi^2_\text{SED}/N_\text{SED}$  & $5.5$ & $6.8$ \\[1mm]
$\chi^2_\text{total}$ &  $20.2$ & $22.8$ \\[1mm]
\hline \hline
    \end{tabular} 
\tablefoot{\tablefoottext{a}{These parameters were fit in the \texttt{specfit} run, but were fixed in the \texttt{TGMdust} runs.}
}
\end{table}

\paragraph{{Spatial and spectral signatures in the correlated spectra.}} In earlier IR interferometric studies (e.g., \citealp{vanBoekel2004nature} or \citealp{Varga2018}) the correlated spectra were usually treated as they were proper spectra. However, in the general sense that notion is not correct. If the object has a significant substructure (e.g., rings, a stellar disk, or a binary) which is well resolved by the interferometer, the spatial signature emerges as a modulation in the correlated spectra that becomes mixed with the real spectral signature of the object \citep[e.g.,][]{Jamialahmadi2018_HD100456}.
This effect can be seen in the right panel of Fig.~\ref{fig:silfit}, showing the fits to the correlated spectra, and in Fig.~\ref{fig:silfit_bl} where we plot the correlated flux as a function of baseline length at a single wavelength of $10\ \mu$m. 
The contributions of the various disk zones show the effects of resolution: while the correlated flux of the inner zone barely decreases in the sampled baseline range, the other two zones show a sharper decrease, and {an oscillation pattern associated with the resolved ring-like emission}. The frequency of this pattern is related to the location of the zones: the further out a ring is located, the higher is the frequency\footnote{The exact mathematical formula for the correlated flux of an infinitesimally thin ring is $F_{\nu,\text{corr}} = F_{\nu,\text{total}} J_0 \left(2\pi r B_\text{p}/\lambda  \right)$ where $F_{\nu,\text{total}}$ is the single-dish flux, $J_0$ is the Bessel function of the first kind of order zero, and $r$ is the ring radius.}. The modulation effect gets more prominent with increasing baseline. 

The contributions of the disk zones are added up in the correlated flux of the full disk (orange line). We note that the correlated fluxes of the different disk zones may have negative values. At the longest baseline ($B_\mathrm{p} = 129.1$~m, Fig.~\ref{fig:silfit} right panel), we can even see the silicate feature to completely disappear, due to the positive and negative correlated flux contributions from the different rings canceling each other. This is because in the general case the visibility function of a ring is the $J_0$ Bessel function which oscillates around zero. The correlated flux which we measure with our interferometer is the absolute value of the sum of the correlated fluxes of all zones, thus it is always positive\footnote{There is an inherent assumption in our analysis that the real value of the complex visibility function of the whole object equals to its absolute value. This is only true if the phase of the visibility function is zero. We are not able to measure the visibility phase directly with VLTI, instead we measure the closure phase which is the sum of the three phases along a closed triangle of baselines (for more information, we refer to \citealp{Monnier2003_closurephase}). As the closure phases of our object are relatively small (typically $<10\degr$), we conclude that our above assumption holds well.}. 

The bottom line is that sharp edges in the object image cause strong modulations in the visibility function, both in the spectral and spatial directions. In the HD~144432 data this effect is prominent in the N-band correlated spectra. As a general note, one should be aware of spatial modulations when working with correlated spectra. Interpreting those data as proper spectra should be generally avoided, instead, interferometric modeling or image reconstruction should be applied. Still, the old notion that the decreasing silicate feature amplitude with baseline indicates radial mineralogy gradient or radiative transfer effects, can hold in systems where the disk has no significant substructure at the sampled spatial scales.

\subsection{\texttt{TGMdust} runs}
\label{sec:res_TGMdust}

\begin{figure*}
  \includegraphics[width=0.373\hsize]{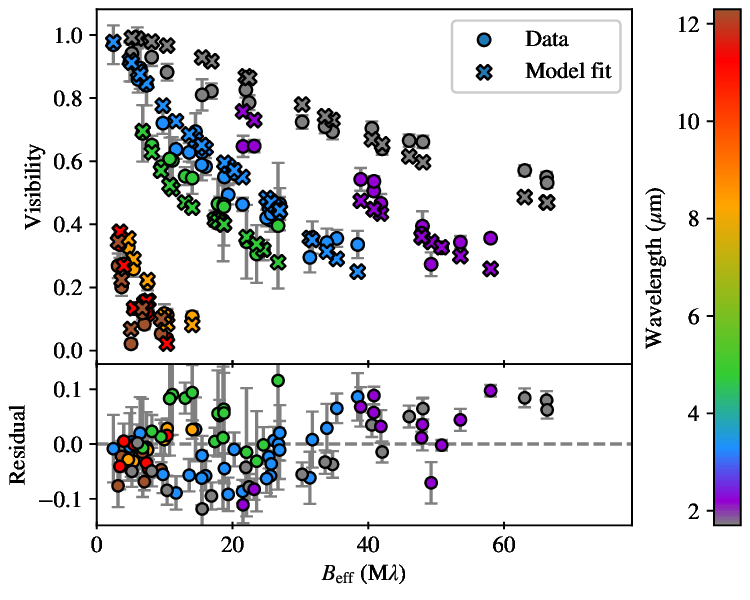}
  \includegraphics[width=0.623\hsize]{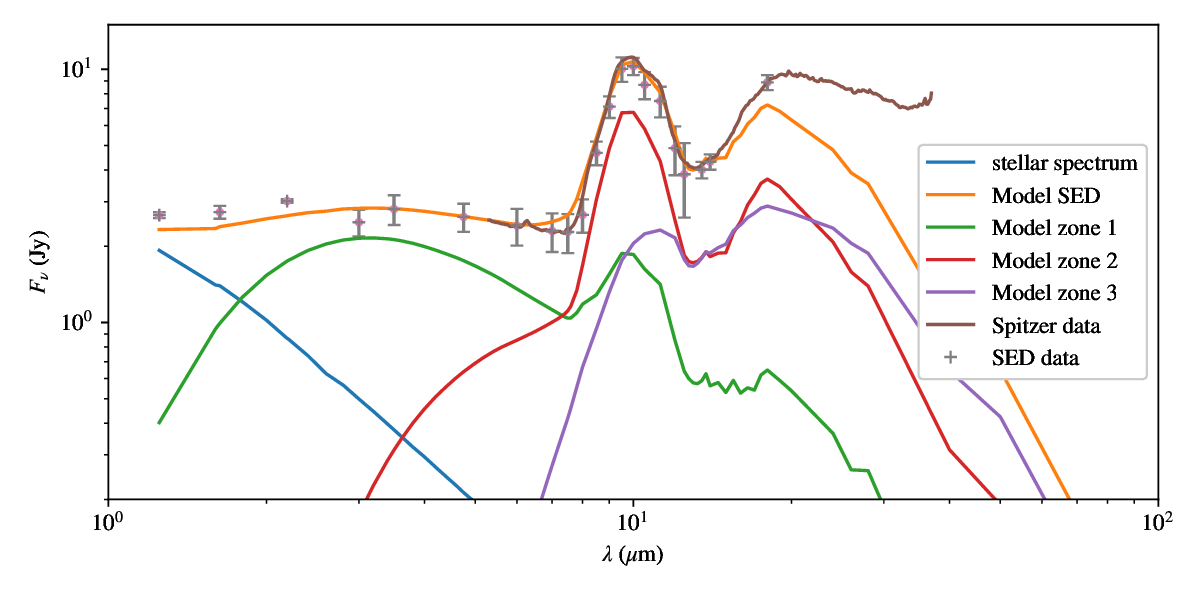}
      \caption{Results of the \texttt{TGMdust} model run with iron grains. Left panel: fit to the visibilities with respect to the effective baseline length. The points are color coded for wavelength. {The following wavelengths are plotted: $1.7\ \mu$m (PIONIER, gray), $2.2\ \mu$m (GRAVITY, purple), $3.3\ \mu$m (MATISSE  $L$, blue), $4.75\ \mu$m (MATISSE $M$, green), $8.3\ \mu$m (orange), $11.3\ \mu$m (red), and $12.3\ \mu$m (brown, last three are MATISSE $N$ data).} Right panel: fit to the SED. {The first three data points are 2MASS $JHK$ measurements, the three points between $3$ and $5\ \mu$m are extracted from our MATISSE $LM$-band data, and the points between $6$ and $18\ \mu$m are taken from a Spitzer observation, except the $N$ band where the points are the average of the Spitzer and our MATISSE single-dish spectra.}} 
         \label{fig:modelfit_Fe}
\end{figure*}
\begin{figure*}
  \includegraphics[width=0.373\hsize]{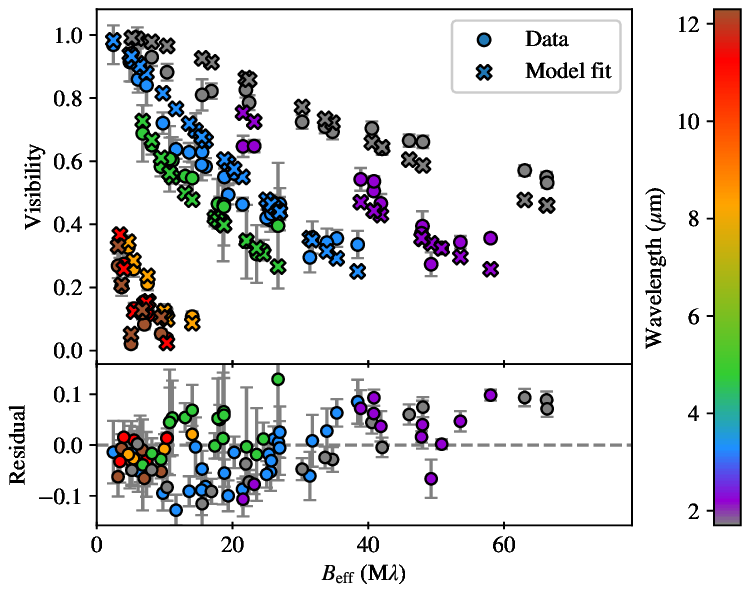}
  \includegraphics[width=0.623\hsize]{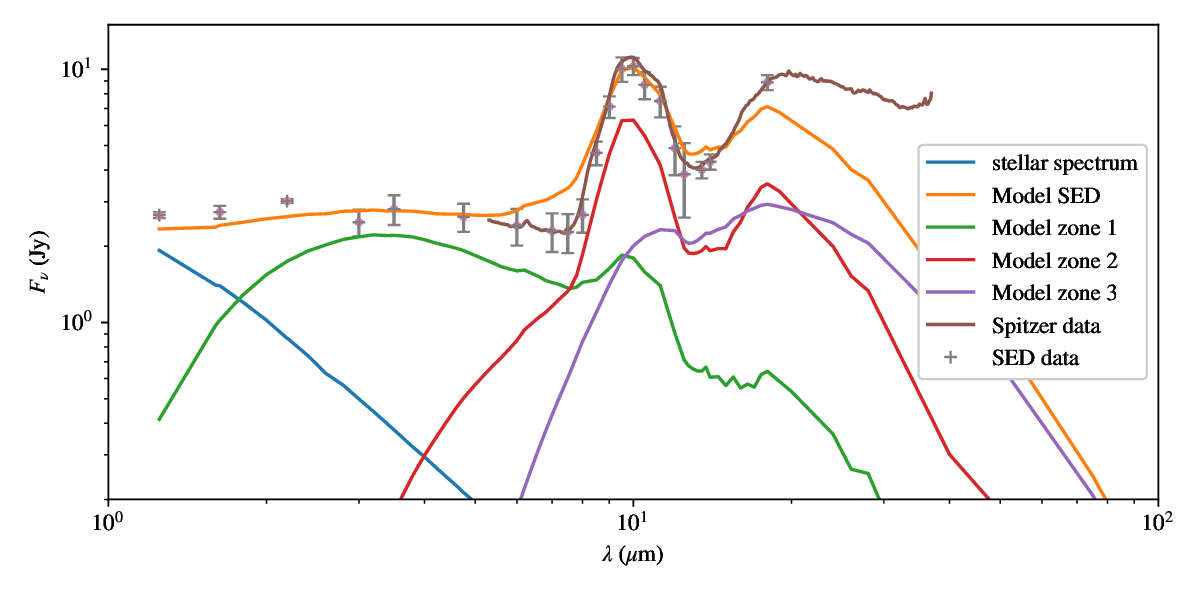}
      \caption{{Same as Fig.~\ref{fig:modelfit_Fe}, but for the \texttt{TGMdust} model run with carbon grains.}
      } 
         \label{fig:modelfit_C}
\end{figure*}

After the retrieval of the silicate composition with \verb|specfit|, we run MCMC fits with our \verb|TGMdust| code, using the full data-set covering the $HKLMN$ bands. We fit the visibility data in the $1.6-12.3\ \mu$m wavelength range, and the SED between $1.25$ and $18\ \mu$m. In these runs we {fixed} the {relative} mass fractions {between the different silicate components}, and the inner radii of the outer two zones. This is because those parameters are well constrained by the N-band only modeling with \verb|specfit|. We choose to fit the inner radius of the disk, because that is better constrained with the shorter wavelength VLTI data ($HKL$ bands). In addition {to} silicates, here we consider dust components responsible for the continuum opacity of the disk, namely iron and amorphous carbon. We attempt to determine their mass fractions, expecting that the featureless opacity curves of those species (Fig.~\ref{fig:app_opac2}) might be better constrained using the wide wavelength span of the full data set. We perform two separate runs, one with silicates and iron, the other with silicates and carbon grains. We choose a population of two grain sizes, $0.1\ \mu$m and $2\ \mu$m, like in the case of the silicate grains. We have $15$ fit parameters in total: the inner radius of the disk ($R_1$), the temperature at the inner radius ($T_\text{in}$), the power law exponent of the temperature profile ($q$), the three surface densities at the zone inner edges ($\Sigma_{0,i}$), the three surface density gradients ($p_i$), and the mass fractions of either iron or carbon per grain size per zone ({$c_{i,\text{small/large}}$}, six mass fractions).

Each MCMC run consists of $50000$ steps with $32$ walkers. When estimating the best-fit values, the first 20000 steps are discarded. For the $\chi^2$, we first calculate separately the $\chi^2$ values for the SED ($\chi^2_\text{SED}$) and for the visibility data ($\chi^2_V$). In case of the SED the $\chi^2$ is calculated using the logarithms of the values, which helps to better fit the overall shape and flux levels of the SED. To equalize the weight of the visibilities in the fitting, we multiply $\chi^2_V$ by a factor of $w = 3$. This ensures a similarly good fit for both data types. {With other values for $w$ we found that either the SED fit or the visibility fit is much worse than the other.} The total $\chi^2$, used in the optimization, is then $\chi^2_\text{total} = \chi^2_\text{SED}/N_\text{SED} + w \chi^2_V/N_V$, where $N_\text{SED}$ and $N_V$ are the number of fit data points in the SED and in the visibility data, respectively. The best-fit parameter values are chosen to correspond to the sample with the lowest $\chi^2_\text{total}$ in the chain, and the uncertainties of the parameters are taken as the $16-84$ percentile ranges of the respective posterior distributions. 

\begin{figure}
 \includegraphics[width=0.495\hsize]{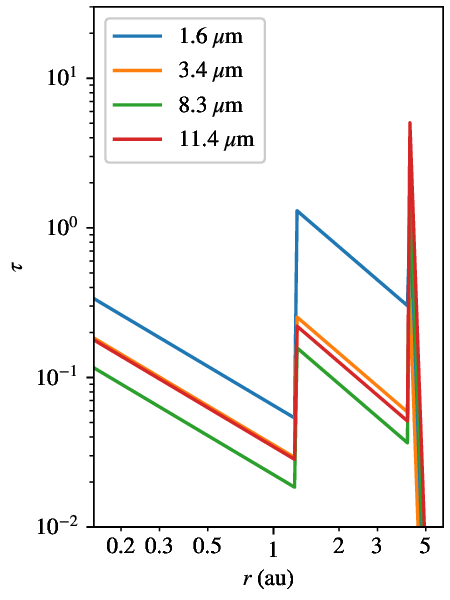}
  \includegraphics[width=0.495\hsize]{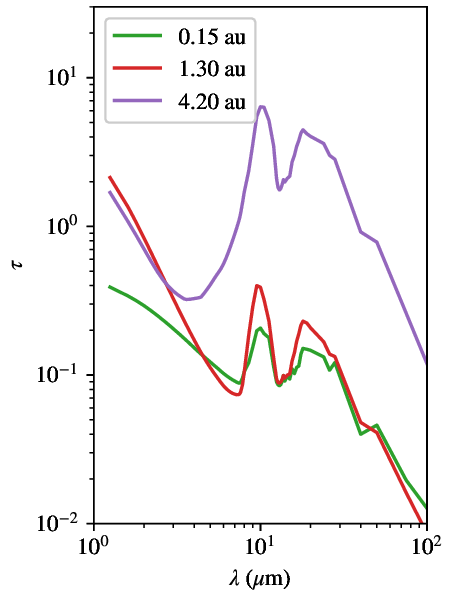}
      \caption{Vertical optical depth of the best-fit disk model with iron grains. Left panel: Radial optical depth profiles at selected wavelengths. Right panel: Optical depth spectra at the inner edges {of each zone}.  } 
         \label{fig:tau_Fe}
\end{figure}

In Table~\ref{tab:fit_res} we list the best-fit parameters from both modeling runs, along with the fixed parameters and $\chi^2$ values. The results of the modeling with iron grains are presented in the following figures: Fig.~\ref{fig:modelimg_Fe} shows the model images at several wavelengths, Fig.~\ref{fig:I_nu_prof_Fe} shows the radial surface brightness profiles, Fig.~\ref{fig:modelfit_Fe} shows the fit to the data, and Fig.~\ref{fig:tau_Fe} shows the vertical optical depth profiles. The corresponding figures for the model with carbon grains are 
Fig.~\ref{fig:modelimg_C} (model images), Fig.~\ref{fig:I_nu_prof_C} (radial surface brightness profiles), Fig.~\ref{fig:modelfit_C} (fit to the data), and Fig.~\ref{fig:tau_C} (vertical optical depth). The corner plots of the MCMC runs are presented in Fig.~\ref{fig:cornerplot_Fe} (run with iron) and Fig.~\ref{fig:cornerplot_C} (run with carbon). The model images were produced by interpolating Eq.~\ref{eq:I_nu_profile} on a rectangular grid, also applying the inclination effect.

The iron and carbon runs provide similar fits to the data, with slightly better $\chi^2$ values for the run with iron (Table~\ref{tab:fit_res}). Looking at the left panels in Figs.~\ref{fig:modelfit_Fe} and \ref{fig:modelfit_C}, the fits to the visibilities are nearly identical. The fit quality is generally good, especially in the $N$ band, while at shorter wavelengths there are small to moderate systematic differences between the model and the data. For instance, the $HKL$ data points at $<30$~M$\lambda$ spatial frequencies\footnote{{The spatial frequency is calculated by dividing the baseline length with the wavelength of the data.}} are systematically lower than the model points, indicating that at those wavelengths the disk is actually more extended than the model prediction. In contrast, in the $M$ band, the data points exceed the model points, indicating that the disk emission in $M$ band is slightly more compact than the model. These differences suggest that the disk has more substructure than {what} our 3-zone model can capture. Although the $\chi^2_\text{SED}$ values are close to each other, the model with iron can reproduce the SED much better than the model with carbon (right panels in Figs.~\ref{fig:modelfit_Fe} and \ref{fig:modelfit_C}). The difference is most pronounced in the spectral region around the silicate feature, as the model with iron correctly fits both the silicate peak and the continuum level, while the model with carbon overestimates the continuum between $6$ and $15\ \mu$m. 

\subsection{Disk structure}

Both models yield very similar model images (Figs.~\ref{fig:modelimg_Fe} and \ref{fig:modelimg_C}). There is a general trend that the emission gets more extended with increasing wavelength. In the $H$ and $K$ band, the flux comes almost exclusively from the innermost zone. {From} around $3\ \mu$m the second zone starts to appear, and in the $N$ band emission all three zones contribute significantly. This is also reflected in the half-flux radius of the disk emission, listed in Table~\ref{tab:fit_res} for several wavelengths, that 
increases from $0.2$~au at $2.2\ \mu$m to $2.3$~au at $11.3\ \mu$m.
The zone inner edges are sharply defined. The inner two zones are much brighter at $11.3\ \mu$m than at $8.3$ or $12.5\ \mu$m; this is due to the increased emission of the silicate grains at that wavelength. The same can also be seen in the radial surface brightness profiles (Figs.~\ref{fig:I_nu_prof_Fe} and \ref{fig:I_nu_prof_C}). The shaded areas in the surface brightness profiles, that reflect the uncertainty of the model fit, suggest that the profiles are quite well constrained by the data. We note that in the \verb|TGMdust| run we do not fit the inner radii of the second and third zones, so the uncertainty of those parameters are not represented in the surface brightness profile plots. For the inner radius of the disk ($R_1 \equiv R_\mathrm{in}$), the \verb|TGMdust| fit prefers a very small value $<0.16$~au, which differs significantly the value found in the \verb|specfit| run ($0.22^{+0.01}_{-0.03}$~au). Moreover, $0.16$~au should be regarded as an upper limit, as the fit range had a lower boundary of $0.15$~au, and the corner plots indicate that the fit would prefer even lower values. However, permitting a smaller inner radius would be unphysical, as the dust species in our model cannot exist at those regions because of the high temperatures there ($>1500$~K). Consequently, the inner rim may have a distinct radial brightness profile or a special dust composition, or both, that our model cannot sufficiently reproduce. 

Our models prefer a rather large brightness contrast at the inner edge of the third zone ($\sim$$4$~au), and a very steep decrease of the surface brightness outward. The edge at $4$~au is well constrained by the $N$ band data, however, the physical parameters of the third zone, such as the surface density profile and dust composition, are poorly constrained. This is because most of that radial zone is likely too cold to emit significantly in the $N$ band. The emission of that zone mostly contributes at wavelengths longer than $20\ \mu$m, for that spectral region the model underestimates the SED. To conclude, our modeling can constrain the disk structure between $\sim$$0.15$ and $\sim$$5$~au.

\subsection{Optical depth, temperature, and density profiles}
\label{sec:res_tau}

The surface brightness profile  (Eq. \ref{eq:I_nu_profile}) is constrained both by the SED (which sets the flux density) and by the VLTI data (which constrain the area from which we receive the flux). We assume that the source function is a blackbody, which is generally accepted for the radiation of circumstellar dust at IR wavelengths. The temperature profile is also well constrained by the simultaneous modeling of the SED and VLTI data, thanks to the wide spectral coverage of both data-sets. Thus, we argue that $\tau$, the only remaining variable in Eq.~\ref{eq:I_nu_profile}, is robustly measured. Both Fig.~\ref{fig:tau_Fe} and Fig.~\ref{fig:tau_C} indicate that the disk emission is not optically thick in the inner two zones ($r<4$~au), as at $\lambda > 3\ \mu$m $\tau$ remains below $0.4$. Consequently, our assumption that the disk emission is optically thin, holds relatively well. The \verb|TGMdust| model runs suggest an optically thick inner edge of the third zone, even in $N$ band, while in the silicate compositional fits we generally assumed optically thin radiation. Thus, the \verb|specfit| results for that zone are probably biased. At the same time, we have already noticed in Sect.~\ref{sec:res_sil} that the dust composition in the third zone is poorly constrained by \verb|specfit|.

The temperature profile follows a relatively shallow decrease ($q \approx -0.5$). The temperature at the inner edge is considerably higher than the generally assumed $1500$~K (dust sublimation temperature), as it is $\approx$$1800$~K in the run with iron, and $\approx$$1600$~K in the run with carbon. There are several minerals which can endure such high temperatures. We discuss this subject further in Sect.~\ref{sec:disc_inner_edge}. 

The dust surface densities at the inner edges of the first two zones are in the range of $10^{-4}-10^{-3}$ g~cm$^{-2}$. The iron-rich model gives a total dust mass of $5.7 \times 10^{-5}$~M$_\text{Earth}$ in zone 1, and $2.6 \times 10^{-4}$~M$_\text{Earth}$ in zone 2, while with the carbon-rich model we find $5.9 \times 10^{-6}$~M$_\text{Earth}$ in zone 1 and $1.2 \times 10^{-4}$~M$_\text{Earth}$ in zone 2. The model with carbon has a lower surface density and dust mass, compared to the model with iron. This is because the IR opacity of small carbon grains is roughly an order of magnitude above that of iron, so in order to produce the same amount of radiation less carbon is needed than iron, in terms of mass. 
Both \verb|TGMdust| runs imply a very steep falloff of the surface density after 4 au. However, as was mentioned previously, our data cannot really constrain the disk parameters in the outer zone.

{We note that in our modeling we consider only three, relatively small grain sizes ($0.1$, $2.0$, and $5.0\ \mu$m), while it is very likely that the disk has considerable amount of larger ($>10\ \mu$m) grains which can hide a significant amount of mass because of their very low IR opacities. Thus, our derived masses may only reflect the mass present in small grains, and underestimate the total dust mass. A more general approach for this problem would be to apply a continuous grain size distribution \citep[e.g.,][]{Mathis1977_MRN}, but that is out of scope of the current study.
}

The models indicate that we need significant amounts of either iron or carbon in the dust mix in order to be able to reproduce the data. However, their radial distribution and grain size ratios cannot be well ascertained. This is primarily due to the featureless opacity curves of these dust types, that makes it difficult to identify the specific dust components in our spectro-interferometric data. Still, based on the fits to the SED (Sect.~\ref{sec:res_TGMdust}), we have a preference for the model with iron grains. Looking at the iron dust mass fractions ($c_{i,\text{small}}$, $c_{i,\text{large}}$) one by one, the large grains in the inner zone ($92^{+3}_{-2}\%$), and the small grains in the middle zone ($45^{+7}_{-14}\%$) are relatively well constrained, while for the other components either we have upper limits, or no constraints within the fit ranges at all. In case of the model run with carbon we find that only the small grains in the inner zone ($22^{+13}_{-7}\%$) are constrained well. 



\section{Discussion}
\label{sec:discussion}

\subsection{Disk structure}

Our results indicate a structured inner disk of HD~144432, with not less that three rings in the inner $5$~au. This is a robust finding, as our multiband data set has a sufficient baseline sampling and resolution ($0.2-1.4$~au, depending on the wavelength) to recover those structures. As the closure phases of the VLTI data remain typically within $\pm 10\degr$, deviations from central symmetry are small, hence the observed structural features indeed should be concentric rings viewed slightly inclined. In our modeling the ring-like features arise as a result of a sudden increase of the radial surface brightness at the zone edges. In reality, the rings may have a rounded radial profile, but it is out of the scope of this study to constrain that aspect. \citet{Chen2016} performed radiative transfer simulations to fit a set of $H$, $K$, and $N$ band interferometric data on HD~144432. Their best model (coined DA0) features an inner disk between $0.2$ and $0.3$~au, which is optically thin at $\lambda = 2\ \mu$m in the midplane in the radial direction; and an optically thick outer disk between $1.4$ and $10$~au. Their inner disk may be the same structure as our zone 1 ($0.15-1.3$~au), while their outer disk coincides with our disk zones 2 and 3 ($r>1.3$~au). As they only had a single-baseline $N$ band interferometric observation with MIDI, that was not enough to detect the ring at $4$~au.

\paragraph{{Nature of the dust inner rim.}} We now investigate whether the inner radius of our model disk could be the dust sublimation radius. The following equation from \citet{Dullemond2010} describes the temperature -- radius relation of the dust, assuming thermal equilibrium with the stellar radiation:
\begin{equation}
\label{eq:R_dust}
    R_\text{dust} = \left(\frac{L_*}{16 \pi \epsilon \sigma T_\text{dust}^4} \right)^{1/2},
\end{equation}
where $T_\text{dust}$ is the dust temperature, $L_*$ is the stellar luminosity, taken as {$11.6\ \Lsun$ from our stellar atmosphere fitting (Sect.~\ref{sec:model})}, $\epsilon$ is the cooling factor, and $\sigma$ is the Stefan-Boltzmann constant. For regular grains we can assume sublimation at $T_\text{dust} \approx 1500$~K, {the usual value found in the literature \citep[e.g., ][]{Dullemond2010}.} Following Eq.~10 in \citet{Dullemond2010} we calculate the values of $\epsilon$ for the dust mixes in the inner zone. For the dust mix with carbon, we get $\epsilon = 0.23$, and for the mix with iron $\epsilon = 0.47$. With these values we estimate a sublimation radius of {$0.24$}~au for the carbon-rich dust, and {$0.17$}~au for the iron-rich dust. Thus, the preference for a relatively small inner radius ($< 0.16$~au) by the \verb|TGMdust| fit can be still consistent with the dust sublimation radius. However, the inner temperatures in our best fits are $100-200$~K above $1500$~K. We can use Eq.~\ref{eq:R_dust} to calculate the radius ($R_\text{dust}$) corresponding to our fit inner temperatures ($T_\text{in}$). For the model with carbon-rich dust we get $0.17$~au, and for the model with iron-rich dust we get $0.15$~au. These values are remarkably consistent with our values for the inner rim radius ($R_\text{in}$). Still, regular (sub-)$\mu$m-sized dust grains are not expected to survive at temperatures significantly above $1500$~K. We note that the uniform sublimation temperature which we applied in our modeling oversimplifies the complex nature of various
grain materials with significantly different evaporation temperatures \citep[e.g.,][]{Gail2004}. We explore this issue further in Sect.~\ref{sec:disc_inner_edge}. 
There is also a possibility that the disk structure and dust composition at the inner rim {is} more complicated than that our model can represent. For more accurate results on the physical conditions near the dust inner rim, one has to run proper radiative transfer models with scattering included, also accounting for the height and shape of the inner rim.

To put our results into context, \citet{Chen2012} calculated the radius of the $K$ band emission from interferometric model fitting.
The $K$ band radiation is expected to come almost exclusively from the dust inner rim, so the $K$ band disk size is a good proxy for the dust sublimation radius. {Their} value for the {$K$ band ring radius ($0.17$~au) is in agreement with our finding for the inner rim}. 




\begin{figure}
 \includegraphics[width=\hsize]{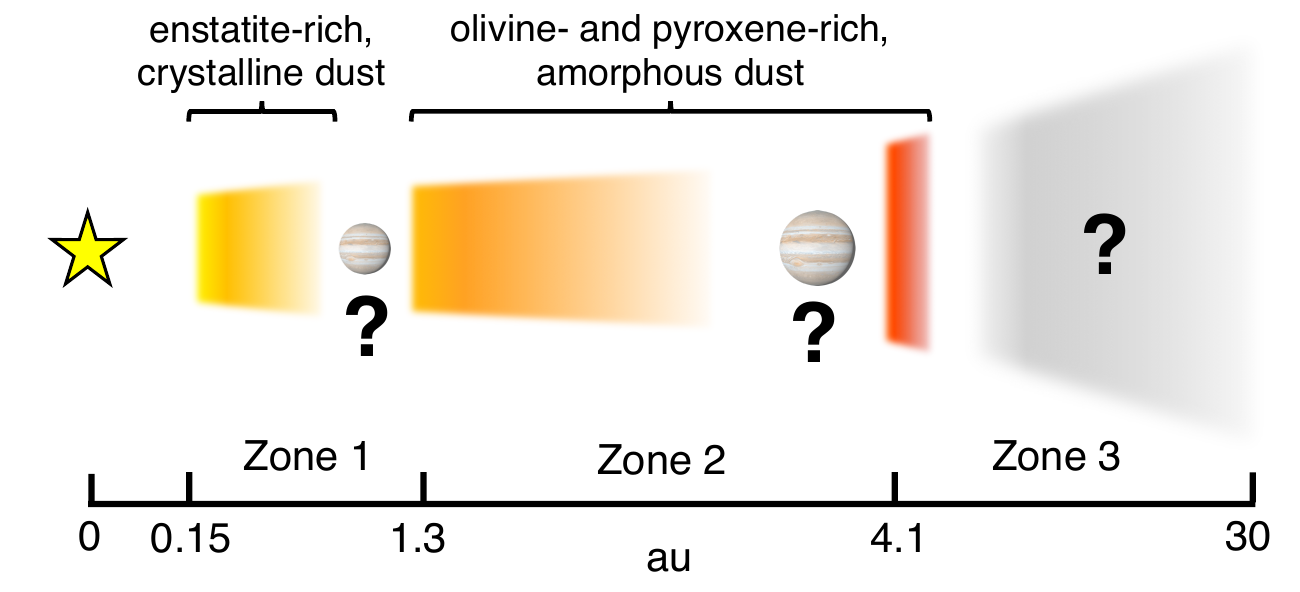}
      \caption{{Sketch of the disk of HD 144432, as seen by VLTI. We show the three ring-like structures, and the possible gaps between them. In the gaps we indicate two putative Jupiter mass planets. The cold outer disk, shown as a gray patch, cannot be constrained by our data.}  } 
         \label{fig:cartoon}
\end{figure}

\paragraph{{Two planet-induced gaps?}} One might ask whether the dark regions between the ring-like features are real disk gaps, or just shadowed regions. As in our modeling we do not represent the vertical extent of the disk, we cannot address this question. \citet{Chen2016} explored this aspect in their RADMC-3D simulations with a puffed-up inner rim as the source of shadowing. They found that the self-shadowed model does not fit the data well, as it provides too low flux in the near-IR, and too much flux at $10-100\ \mu$m. 

If the dark regions are indeed gaps, they might have been opened by massive planets residing in the disk \citep{Asensio-Torres2021}. It has been shown that the width and depth of disk gaps in the gas surface density profile are closely related to the planetary mass \citep{Kanagawa2015}. {We assume that the $\mu$m-sized dust which we detect in our IR data is colocated with the gas, as expected for such small grains \citep[e.g., ][]{Fouchet2010,Pohl2016}.} Here we use the Eq.~5 in \citet{Kanagawa2016} to estimate the masses of the putative gap-opening planets residing in the inner disk of HD~144432. The equation requires the gap width which is not straightforward to derive from our modeling, as we represent the surface brightness profile in each disk zone by a power law, and the dark gap-like regions in our images (Fig.~\ref{fig:modelimg_Fe}) have just one well defined edge (the outer one). Here we assume that the surface density at the gap inner edge in zone 1 is equal to the surface density at the gap outer edge which coincides with the inner radius of zone 2. Thus, we get a gap inner edge at $r\approx0.6$~au, and an outer edge at $1.26$~au. We place our planet in the middle position at $R_{\text{p}1}=0.93$~au. From the temperature profile in our best-fit model with iron grains we calculate the aspect ratio ($H_g/R_{\text{p}1}=0.038$, {where $H_g$ is the gas pressure scale height}) of the disk at the location of the planet assuming hydrostatic equilibrium. For the stellar mass, we take $1.8\ \Msun$ \citep{Guzman-Diaz2021}. The gap width is also influenced by the $\alpha$ viscosity parameter of the disk which is the source of considerable uncertainty. For a relatively low $\alpha$ value ($10^{-4}$) we get a planet mass of $0.4$~M$_\text{Jup}$, while for a higher value ($\alpha=10^{-3}$) we get $1.3$~M$_\text{Jup}$. 

For the gap-like region in zone 2, we cannot define the gap inner edge as we did it in the first case, because the surface density at the outer edge of the second gap (i.e., the inner edge of zone 3) is much higher than the surface density anywhere in zone 2. Thus, for simplicity, we assume that the gap edges coincide with the inner ($1.3$~au) and outer ($4.1$~au) limits of zone 2. We place the planet in the middle of the zone at $R_{\text{p}2}=2.7$~au, where we estimate the aspect ratio to be $H_g/R_{\text{p}2}=0.048$. We get a planet mass of $1.3$~M$_\text{Jup}$ and $4.1$~M$_\text{Jup}$ for the low and high viscosity case ($\alpha=10^{-4}$ and $10^{-3}$), respectively. 
Thus, we {argue} that both gap-like regions (one at $\sim$$0.9$~au, the other at $\sim$$3$~au) in the inner disk of HD~144432 might host gas giant planets, {assuming that the structures we detect are tracing the overall density profile. In Fig.~\ref{fig:cartoon} we present a sketch of the disk of HD 144432, as we conceive it from our VLTI data and modeling, with the locations of the putative planets indicated.}

\subsection{{An optically thin inner disk}}

In physical models of optically thick planet-forming disks the surface layer, which is directly illuminated by the central star, gets superheated, while the deeper regions remain cold \citep[e.g.,][]{Chiang1997}. In this case the warm to hot surface layer dominates the near-IR continuum and also gives rise to the dust spectral emission features, while the layer immediately below the superheated disk photosphere is expected to be significantly colder and to emit only at longer, {mid-IR} wavelengths. 
This requires a strong vertical temperature gradient in the disk, otherwise the thermal IR emission becomes partially optically thick, and the silicate spectral feature gets weaker. It is not straightforward to distinguish between this scenario and the case for a really optically thin disk. It might happen that the radiation we observe in HD~144432 comes from a superheated layer, mimicking an optically thin component. {Still, the fact that we were able to reproduce both the level of the IR continuum emission and the silicate feature, using only one temperature gradient component per radial zone, is a strong indication for an optically thin emission. }

{Furthermore}, \citet{Chen2016} have already demonstrated in their RADMC-3D simulations that an optically thin inner component (between $0.2$ and $0.3$~au) is indeed compatible with the SED and IR interferometric data of HD~144432. For comparison, in our modeling we found evidence for optically thin IR emission coming from the inner two disk zones ($<4$~au, Sect.~\ref{sec:res_tau}). We note that \citet{Chen2016} reported the optical depth measured in the midplane in the radial direction ($\tau_1\left(2\ \mu\text{m}\right)=0.14$), that is not directly comparable to our $\tau$, which is measured in the vertical direction.
Thus, instead of the optical depths we opt to compare dust masses. They reported a dust mass for their inner disk component  of $1.8 \times 10^{-11} \Msun$. Our dust mass estimations for the inner zone of our model (between $0.15$ and $1.3$~au) are $1.7 \times 10^{-10} \Msun$ with the iron-rich dust mix, and $1.8 \times 10^{-11} \Msun$ with the carbon-rich dust. A word of caution here is that the dust mixes in these models may have significantly different opacities, so having similar dust masses does not guarantee that the optical depths are also similar. Still, as \citet{Chen2016} used carbon grains in their modeling, it seems plausible that their opacities are similar to our carbon-rich model opacities. In that case their model is in remarkably good agreement with ours, in terms of total dust mass and optical depth. To further investigate the optical depth aspect, we would need proper radiative transfer modeling of our data set which is out of scope of this study.



As a final note, the reduced optical depth in the inner disk of HD~144432 might indicate that the disk is being depleted, emphasizing the transitory nature of this object. Massive planets may halt the inward radial drift of dust grains, trapping them in outer rings or vortices \citep{Fedele2017}. Alternatively, photoevaporative and magnetohydrodynamic disk winds can be also be responsible for the clearing of the inner disk \citep{Pascucci2009,Lesur2022}. 

\subsection{Dust composition}
\label{sec:discussion_dust}

\paragraph{{Radial mineralogy gradients.}} An important result from our modeling is the confirmation of radial mineralogy gradients which were first observed by \citet{vanBoekel2004nature}. From an analysis of VLTI/MIDI N-band spectro-interferometric data they measured a crystallinity fraction of $55^{+30}_{-20}\%$ in the $1-2$~au disk region, and $10^{+10}_{-5}\%$ in the $2-20$~au disk region. The radii of their disk regions were estimated from the baseline lengths of the observations. This method only gives a rough approximation, assuming that the interferometer collects the light from an aperture corresponding to the spatial resolution. In our modeling we were able to constrain the spatial extents of the disk zones with different mineralogical compositions much more precisely. In the \verb|specfit| run, we found that the inner zone ($0.2-1.3$~au) has a crystallinity fraction of $61^{+4}_{-5}\%$, while in the outer zones the same number is $20\pm10\%$. These numbers match well with the values found by \citet{vanBoekel2004nature} within the error bars, and also agree with our previous findings presented in \cite{Varga2018}.

Alternatively, optically thick emission from the hot inner rim can also cause the inner disk spectra to show weaker silicate peaks, as shown by \citet{Meijer2007}. However, in our \verb|TGMdust| fits we found that the IR radiation from the disk inner edge is optically thin ($\tau < 0.4$). Thus, we do not see much support for a bright, optically thick inner rim in the disk of HD~144432 from our data.

\paragraph{Comparison with disk-integrated mid-IR spectroscopy.} \citet{vanBoekel2005survey} and later \citet{Juhasz2010} performed dust compositional fits to N-band single-dish (TIMMI2 and Spitzer) spectra of HD~144432. \citet{vanBoekel2005survey} found a fit with $53\%$ small ($0.1\ \mu$m) amorphous olivine, $42\%$ large ($1.5\ \mu$m) amorphous pyroxene, and a small fraction of crystalline dust, mainly small forsterite ($1.9\%$). \citet{Juhasz2010} in their fits to Spitzer spectra between wavelengths $5$ and $17\ \mu$m found that the main amorphous constituents are large ($2\ \mu$m) amorphous olivine ($42\%$), small ($0.1\ \mu$m) amorphous pyroxene ($19\%$), and large amorphous pyroxene ($15\%$). Among the crystalline species, their fit preferred more enstatite ($6.1\%$) than forsterite ($2.3\%$). The results of these studies differ significantly, indicating that there is an inherent ambiguity in dust compositional fits to mid-IR silicate spectra. There are several contributing factors to this, like the selection of dust species to fit, the choice on the sizes of the grains, how the opacity curves are calculated (cf. Figs.~1 and 2 in \citealp{Juhasz2010}), and the differences in the underlying disk model. Our silicate fit is not directly comparable to the aforementioned studies, because we dissected the disk into several zones, and obtained a composition for each zone. Still, our values in the disk zone 2 are comparable to the results by \citet{Juhasz2010}, as both feature a large contribution from large olivine and small pyroxene grains, and a preference for more enstatite than forsterite. Although our MATISSE $N$ band spectro-interferometric data have lower S/N than the Spitzer spectrum, we argue that our data have more constraining power in the compositional fits, because we have seven spectra (one single-dish and six correlated spectra) probing the same object. 



\subsection{Comparison with condensation sequences}
\label{sec:disc:ggchem}

\begin{figure*}
  \includegraphics[width=\hsize]{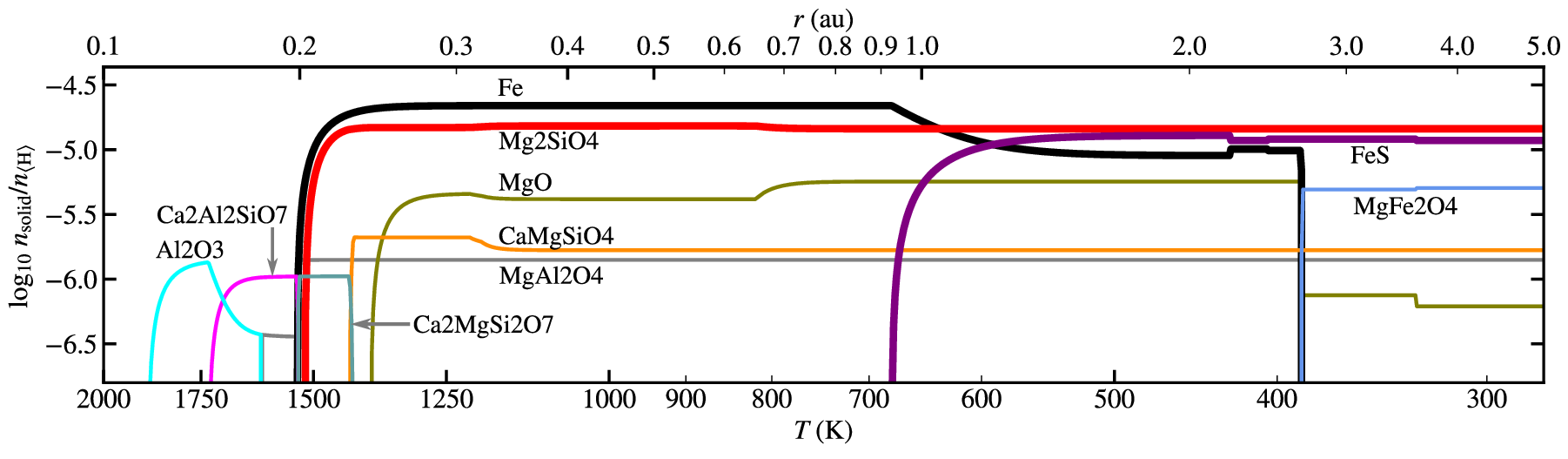}
  \includegraphics[width=\hsize]{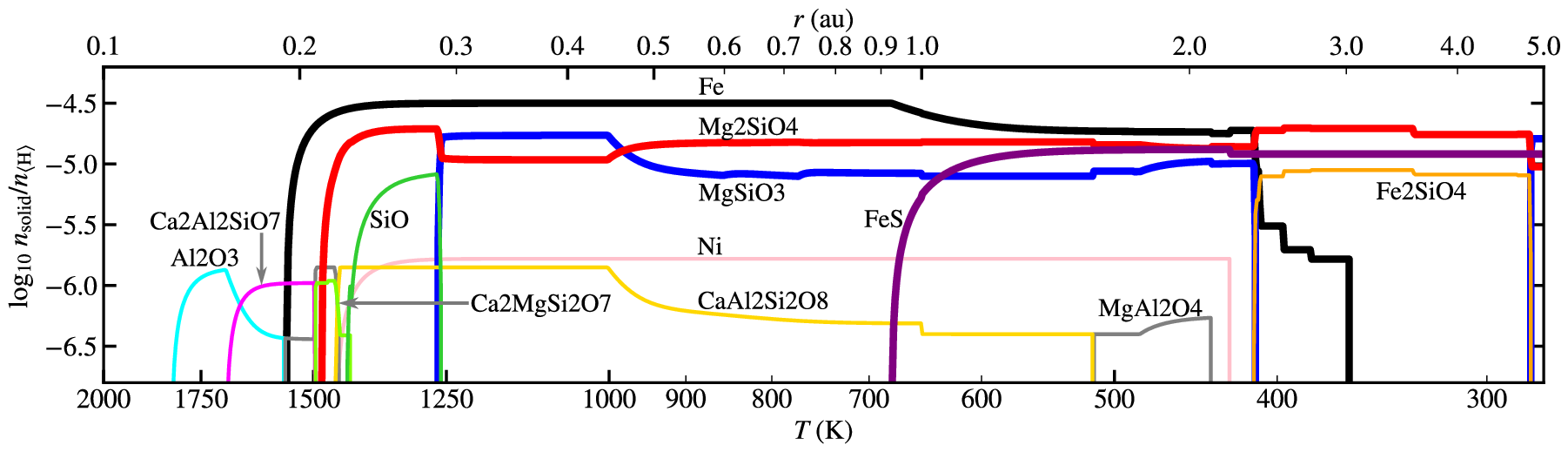}
      \caption{Concentration of various condensates (normalized to hydrogen atom number density) as a function of radius and temperature from our calculation using the thermo-chemical equilibrium code \textsc{GGchem} \citep{Woitke2018}. Top: plot showing the concentration using the photospheric elemental abundances of HD~144432. Bottom: plot showing the concentration using Solar elemental abundances. For clarity, we show a selection of the condensates: we plot only those species which are among the top three most abundant condensates for at least one temperature value. The $T\left(r\right)$ relation follows our best-fit \texttt{TGMdust} model with iron grains. The gas pressure-temperature profile follows a power law with $p=3.16 \times 10^{-2}$~bar at the maximum temperature $T=2000$~K, and $p=1.79 \times 10^{-6}$~bar at the minimum temperature $T=277.5$~K.  } 
         \label{fig:GGchem}
\end{figure*}

\paragraph{{The expected rich mineralogy of disks.}} While only a few dust species have been directly detected in the interstellar and circumstellar medium so far (mainly forsterite, enstatite, silicon carbide, and polycyclic aromatic hydrocarbons), there is evidence that the diversity of minerals in the cosmic dust is far richer. The evidence mainly comes from the analysis of Solar System interplanetary dust (IPD) grains, thought to be of presolar origin \citep[e.g.,][]{MacKinnon1987,Lodders2005,Engrand2023}. For example, carbon and iron are both found in the IPD and in meteorites \citep{Anders1964}. Moreover, the planets in the inner Solar System are mostly composed of refractory elements: the main refractories Mg, Al, Si, Ca, Fe, Ni have a mass fraction of $67\%$ in Earth, and $84\%$ in Mercury \citep{Morgan1980}. Most of these elements had been locked in the dust grains of the presolar nebula, before they got incorporated in planetary bodies. As many of the supposed materials in the cosmic dust have only weak or nonexistent optical-IR spectral features, the rich mineralogy of the interstellar and circumstellar dust mostly remains to be hidden from our observations.

\paragraph{{Modeling condensation sequences with \textsc{GGchem}.}} Chemical modeling of dust condensation offers an alternative way to assess the variety of minerals which could exist under the physical conditions found in circumstellar disks. Here we investigate the expected diversity and distribution of solid state materials in a disk like HD~144432, using the thermo-chemical equilibrium code \textsc{GGchem} \citep{Woitke2018}. \textsc{GGchem} calculates the chemical composition of astrophysical gases, with the option to include the formation of condensates. We aim to address the following questions with \textsc{GGchem}:
\begin{itemize}
    \item The radial distribution of forsterite (Mg$_2$SiO$_4$) and enstatite (MgSiO$_3$)\footnote{\textsc{GGchem} cannot model the crystalline structure of the condensates. Thus, in this section we use the terms forsterite and enstatite to encompass both amorphous and crystalline forms of the materials.}.
    \item The nature of dust species which can exist in the innermost disk regions where $T>1500$~K.
    \item The amount of carbon and iron which condensates to dust grains in the disk.
\end{itemize}
For the input gas composition, we use the stellar abundances of C, O, Fe, Mg, Si reported by \citet{Kama2015} for HD~144432. For the rest of the elements (H, He, N, Na, Al, Ca, Ti, S, Cl, K, Li, F, P, V, Cr, Mn, Ni, Zr, W), we use Solar abundances by \citet{Asplund2009}. \textsc{GGchem} calculates the concentration of the end products as function of the temperature and gas pressure (or hydrogen density). We define a pressure-temperature profile ($p_\text{gas}-T$)  based on the minimum mass solar nebula description by \citet{Hayashi1981}, but for the $T\left(r\right)$ and $\Sigma\left(r\right)$ radial profiles we use the power law exponents found for zone 1 in our \verb|TGMdust| run with iron grains. For the calculation of $p_\text{gas}$, we used Eqs. B.5 to B.9 from \citet{Jorge2022}, assuming that the disk is vertically isothermal. The results are shown in the top panel of Fig.~\ref{fig:GGchem}, indicating that the most abundant condensates between $0.2$ and $1$~au are iron and forsterite. We see no enstatite forming at all, which is also found by \citet{Jorge2022} if the Mg/Si ratio is above the solar value by $0.17$~dex. In the case of HD~144432, the Mg/Si ratio is even higher, $0.26$~dex more than in the Sun. 

We also performed a chemical calculation with Solar abundances taken from \citet{Asplund2009}, shown in the bottom panel of Fig.~\ref{fig:GGchem}. The plot indicates that both forsterite and enstatite form in a solar-like gas mixture, that is consistent with the results of our silicate compositional fits for HD~144432. The figure also shows that the radial distribution of the two main silicate minerals are different: while enstatite can be found between $0.3$ and $2$ au, forsterite has a broader distribution from $0.2$~au to $5$ au and beyond. In our results we do not find evidence for such radial distribution, as our model fit requires significant amounts of enstatite in all three disk zones. Finally, we note that our estimation ($0.08$~dex) for the Mg/Si ratio of the disk (Sect.~\ref{sec:res_sil}) is very close to the solar value ($0.09$~dex).

\paragraph{{The problem of enstatite.}} $N$ band spectroscopic and spectro-interferometric data clearly indicate that there is significant amount of enstatite in the disk around HD~144432, in stark contrast with the prediction of our \textsc{GGchem} calculation with the stellar abundances of HD~144432. We consider two scenarios to resolve this discrepancy: either the equilibrium condensation conditions do not hold in the disk, or the chemical composition of the disk differs from that of the stellar photosphere. The general notion is that the host star abundances are good proxies for the chemical composition of the cloud from which both the star and the disk are formed \citep{Adibekyan2021}. However, \citet{Kama2015} found that stars hosting group I disks, known to have large gaps, are depleted in Fe, Mg, and Si, compared to the Sun and to stars hosting group II (flat, continuous) disks. They suggested that this depletion is caused by giant planets which block the accretion of dust onto the star, while not affecting gas accretion. The same planets can also open gaps in the disk, making it exhibiting a group I type SED. Although HD~144432 has a group II disk, \citet{Kama2015} detected a depletion of refractory elements similar to the stars hosting group I disks. Thus, it seems plausible that the disk of HD~144432 has a solar-like elemental chemical composition, and the depletion of the stellar photosphere in Fe and Si is caused by a giant planet in the inner disk ($<5$~au) halting the radial drift of the dust. The planet, which might be responsible for the inner disk substructure as well, could be located in one of the gap-like regions between two rings (either at $\sim$$1$ or at $\sim$$3$~au). However, there are two problems with this idea: i) small (sub-)$\mu$m-sized dust grains are relatively well coupled to the gas, and ii) we detect enstatite even in the inner zone ($0.2-1$~au), where we do not expect a planet blocking the accretion at the inner edge. 

{Alternatively,} the physico-chemical conditions in disks might be more complex than that we can represent with a simple disk model with equilibrium chemistry. \textsc{GGchem} calculates the gas and dust composition at every distance (temperature) by minimizing the Gibbs free energy. However, the dust will have seen a complex history of growth, settling and mixing, which is not accounted for in an equilibrium calculation. For instance, \citet{Gail2004} concluded that by taking into account radial mixing, the bulk of the grains at temperatures lower than $\sim$$1000$~K should be enstatite. This is because all forsterite grains, which move from the innermost regions outward, pass through a region in which they are converted to enstatite, and at further out these reactions could freeze out, so that conversion back to forsterite would not take place. Furthermore, equilibrium theory cannot well describe the complex nucleation processes of solids, and phase-equilibrium models have no history. As illustrated in Fig.~11 of \cite{Herbort2020}, condensation of $\mu$m-sized grains is intrinsically fast ($<1$~s), but the timescale of annealing is much longer (on the order of $1$~Myr at $T=900$~K). Therefore, if the disk has a constant temperature structure, and the dust is at rest, only the innermost part can provide conditions for phase-equilibrium.  
\citet{Pignatale2016} demonstrated that equilibrium condensation can be a valid assumption for the disk region where $T \gtrsim 600$~K ($r \lesssim 1$~au in our model).

\subsection{Hot dust at the inner rim}
\label{sec:disc_inner_edge}

In Sect.~\ref{sec:res_tau} we noted that our models prefer a temperature at the inner rim that is $100-200$~K above the sublimation temperature ($\sim$$1500$~K) of regular dust grains. From the \textsc{GGchem} calculation we can evaluate what kind of minerals can endure such high temperatures. Calculations with solar and with the HD~144432 abundances show similar results. The first two condensates forming are tungsten at $\sim$$2000$~K, and zirconium dioxide (ZrO$_2$) at $\sim$$1900$~K, but they are not shown in our plots because of their very low concentrations ($\log_{10} \left(n_{\text{solid}}/n_{\left<\text{H}\right>}\right)<-9.4$). At slightly lower temperatures ($\sim$$1850$~K) corundum (Al$_2$O$_3$) forms, followed by perovskite (CaTiO$_3$, not shown because its low abundance) and gehlenite (Ca$_2$Al$_2$SiO$_7$) condensing at $\sim$$1700$~K. This condensation sequence is very similar to what was found by \citet{Woitke2018} at $p_\text{gas} = 1$~bar; the only difference is that in our calculation the condensation occurs at lower temperatures due to the lower pressures in our disk model.


The most abundant condensates (corundum and gehlenite) present at $T>1500$~K have concentrations ($\sim$$-6$) about 30 times less than the major solid state species iron and forsterite. Moreover, the near-IR opacity of corundum ($\sim$$100$~cm$^2$g$^{-1}$) is much lower than that of iron ($\sim$$2000$~cm$^2$g$^{-1}$). Considering the difference in temperature and in the surface area, we estimate that the $3\ \mu$m flux from corundum grains at $T=1800$~K located at $0.15$~au is only $0.3\%$ of the flux emitted by iron grains at $T=1500$~K at $0.2$~au radius. It remains to be determined, whether that apparently weak signal from corundum and other refractory grains could be detected in observational data.
 
\subsection{Iron vs. carbon}

\paragraph{{A favored iron-rich composition.}} Condensation of carbon does not occur in the warm disk regions ($r<5$~au, $T>300$~K) if the gas has a solar C/O ratio ($\text{C}/\text{O}=0.55$) or less. {According to \citet{Kama2015}, HD 144432 has a C/O ratio of $0.24$.} If we artificially increase the C/O ratio to $\sim$$0.9$, we start to see graphite forming near a temperature of $T=1000$~K. If C/O reaches $1$, graphite becomes a major constituent in the dust mix between $850$~K and $1250$~K, corresponding to the radial region between $0.3$ and $0.6$~au. Such high C/O ratio seems to be unlikely in the majority of stars in our Solar neighborhood. 
In the Hypatia Catalog by \citet{Hinkel2014}, which is a compilation of stellar abundance data within $150$~pc of the Sun, only $2\%$ of the stars have $\text{C}/\text{O}>0.9$. Assuming a similar C/O distribution for the nearby star forming regions and YSOs, we suggest that in the warm inner disk regions ($r<5$~au) of planet-forming disks most, if not all, carbon is in the gas phase, while iron and iron sulfide (FeS, troilite) grains are major constituents of the solid mixture, along with forsterite and enstatite. Thus, small (sub-)$\mu$m sized iron grains should be the main source of dust continuum radiation at thermal IR wavelengths in disks. This is in line with the conclusions of \citet{Matter2020} who suggested to replace carbon with iron when modeling inner disk spectra of YSOs. For more discussion on the effects of varying C/O in inner disks, we refer to \citet{Matter2020}. 

Carbon in solid form is expected to be abundant in the cold outer disk, locked up in ices or as pre-stellar grains. Radial drift can cause that material to reach the inner disk. As a result, the inner disk would be populated by a mix of grains which had experienced chemical equilibrium conditions and grains which came from the outer disk, which is a scenario that cannot be ruled out for HD 144432. Indeed, the fact that we detect both crystalline and amorphous silicate grains may already signify that such a mix exists. Still, the inner Solar System planets are highly depleted in carbon, so there should be processes resulting in carbon-depleted inner disk solid reservoirs \citep{Oberg2011}. 
\citet{Klarmann2018} investigated the efficiency of oxidation and photolysis in the removal of carbon grains from the solid phase, and they found that radial transport of dust can easily prevent the depletion of carbon. They conclude that more efficient carbon-depletion mechanisms along with radial drift barriers are needed to prevent the replenishment of carbon in the inner disk. In a carbon-enriched environment, we can expect the presence of various hydrogenated carbon compounds which show spectral features in the mid-IR \citep{Jones2017}. In HD 144432, those carbonaceous features are much weaker than the silicate emission bands \citep{Seok2017}. Further observations by MATISSE using the medium spectral resolution mode ($R \approx 500$) in $L$ band may shed light on the spatial distribution of the carbonaceous dust in the disk of HD 144432. As far as iron is concerned, iron-rich grains should be present both in the population of grains which are produced by recondensation in situ, as well as from the outer disk, drifting inward.

\begin{table}
 \caption[]{\label{tab:comp_comp}{Comparison of elemental abundances between Mercury, Earth, CI chondrites, and our best-fit disk models for HD 144432. References for the abundances: \citet{Morgan1980} (Mercury, Earth), and \citet{Vollstaedt2020} (CI chondrites).}}
 \centering
\begin{tabular}{lccccc}
 \hline \hline
 &
   &
   &
  \multicolumn{2}{c}{HD 144432} &
  \\
 &
   &
   &
 \multicolumn{2}{c}{Disk zone 2} & \\ 
    &
   &
   &
 Fe-rich & C-rich
 & CI
 \\ 
 Element &
  Mercury &
  Earth &
 \multicolumn{2}{c}{model}  & chondrite\\ 
 \hline
Fe & $64.5$ & $32.1$ & $56$ & $0$ & $18.4$\\
O & $14.4$ & $30.1$ & $20$ & $41$ & $46.5$\\
Si & $7.1$ & $15.1$ & $11$ & $22$ & $10.7$\\
Mg & $6.5$ & $13.9$ & $13$ & $27$ & $9.6$\\
C &  trace & $0.04$ & $0$ & $10$ & $3.2$\\
other & $7.5$ & $8.8$ & $0$ & $0$ & $11.6$\\
\hline
\end{tabular}
\end{table}

In our modeling we found evidence for a high abundance of enstatite in the disk of HD 144432. The  combination of enstatite with iron is reminiscent of enstatite chondrites which are believed to be a major building block of the Earth and other inner Solar System bodies \citep[e.g., ][]{Javoy1995,Javoy2010}. 

By deriving elemental abundances from the dust composition that we fit (Tables~\ref{tab:specfit_res} and \ref{tab:fit_res}), we can investigate what kind of rocky planets could form in the inner disk of HD 144432. In Table~\ref{tab:comp_comp} we show our abundance values for HD 144432, compared with elemental abundances of Mercury, Earth, and CI chondrites. We selected the second disk zone ($1.3<r<4.1$~au) for the comparison, as that zone covers large part of the terrestrial planet forming region. 
The table shows that our model with iron-rich dust represents an elemental composition intermediate between that of Mercury and Earth. In contrast, the elemental composition of our model with carbon-rich dust has some similarities with the abundances of CI chondrites, although the correspondence is not particularly good. We note that in our models either we have iron but no carbon, or the other way around. With that we cannot account well for the composition of CI chondrites, as they contain significant amounts from both carbon and iron.

\paragraph{{Carbon content and planet habitability.}} To put the iron vs. carbon dichotomy in a wider context, \citet{Hakim2019_mineralogy} {has} shown that exoplanets forming from carbon-enriched material will have, after differentiation, a graphite outer layer. Such a shell has major effects on the thermal evolution and geochemistry of the planet. Moreover, if the graphite layer is thick enough, it might prevent the silicate-rich mantle material to come to the surface. In this case the planet surface would be devoid of essential life-bearing elements (except of course carbon), making the planet potentially uninhabitable.

\paragraph{{Chemical sorting by grain growth.}} \citet{Johansen2022} explored the grain growth of condensates forming in a gas of solar composition. They found that iron preferentially grows into sparser, bigger pebbles, while enstatite, forming at the same location, has smaller grain sizes. Consequently, the population of large iron grains becomes chemically separated from the silicate grains having Stokes numbers $10$ times lower. They further propose that under such conditions streaming instability leads to the formation of iron-rich planets. They conclude that this formation mechanism can explain why Mercury is so enriched in iron, without the need for experiencing giant, mantle-stripping impacts. This study demonstrated both the crucial role of iron in the formation of terrestrial planets, and the importance of microphysics when dealing with dust evolution. 

To conclude, our observations and modeling of HD~144432 show tentative evidence for the presence of iron in its planet-forming disk. This claim is partly based on the notion that our model with iron-rich dust provides a better fit to the SED, compared to our other model with carbon-rich dust. Furthermore, our chemical calculations in a disk model representing the pressure-temperature conditions in the HD~144432 disk also give support for an iron-rich chemistry, in agreement with numerous literature sources. 

\section{Summary}
\label{sec:summary}

In this paper we studied the planet-forming disk around HD~144432 by modeling an extensive set of photometric and spectro-interferometric data covering the thermal IR wavelengths. Our data set includes new observations with VLTI/GRAVITY ($K$ band) and VLTI/MATISSE ($L$, $M$, and $N$ bands), and archival data from VLTI/PIONIER ($H$ band). We introduced a new disk model, \verb|TGMdust|, to image the interferometric data, and to constrain the disk structure and dust composition. \verb|TGMdust| can represent disk geometries without (single zone) or with substructure. In the latter case the disk is divided into multiple radial zones, each having its own surface density profile and dust composition. Our main results are as follows:
\begin{enumerate}
  \item The disk within $<5$~au is structured. We need at least 3 zones to fit the data. Our best-fit model features an inner zone between $0.15$ and $1.3$~au, a second zone between $1.3$ and $4.1$~au, and an outer zone starting at $4.1$~au. Each zone has a bright inner edge, thus the disk appears to show a three-ringed structure.
  \item From our silicate compositional fits we find that the dust composition of the zones significantly differ from each other, indicating radial mineralogy gradients. Specifically, the mass fraction\footnote{These values are given as fractions of the total silicate dust mass, excluding the dust components responsible for the continuum emission.} of crystalline silicates is about {$61\%$} in the innermost zone ($<1.3$~au), while only $\sim$$20\%$ in the outer two zones. 
  \item By inspecting the fits to the MATISSE $N$ band data, we find that the silicate spectral feature mostly originates from the second disk zone ($1.3-4.1$~au). {We} show that the silicate spectral feature {in the correlated spectra is} mixed with strong spatial modulation signatures, {an effect} caused by the ring-like substructures. We also find that the disappearance of the silicate feature in the longest baseline ($B_p = 129$~m) correlated spectrum is mostly due to spatial modulation effects, not to mineralogy gradients or radiative transfer effects, as was previously thought. As a general note, interpreting correlated spectra as proper spectra should be generally avoided, instead, interferometric modeling or image reconstruction should be applied.
  \item {To identify the dust component responsible for the infrared continuum emission, we explore two cases for the dust composition, one with a silicate+iron, the other with a silicate+carbon mixture. We} can fit the interferometric data with both compositions equally well, however, the iron-rich model provides a better fit to the SED.
  \item We find that the radiation from the inner two disk zones ($r < 4$~au) is mostly optically thin, having $\tau<0.4$ at wavelengths longer than $3\ \mu$m. This finding is consistent with the radiative transfer modeling results of \citet{Chen2016}. 
  \item Assuming that the dark regions in the disk are gaps opened by planets, we estimated the masses of the putative planetary bodies responsible for those structures. We find that both gaps, the one in zone 1 at $\sim$$0.9$~au and the other in zone 2 at $\sim$$3$~au, might be carved by approximately Jupiter mass planets. 
  \item In order to assess the hidden diversity of minerals in a planet-forming disk like HD~144432, we performed chemical equilibrium calculations of dust condensation with \textsc{GGchem}. 
  Using solar elemental abundances in the \textsc{GGchem} calculation results in a dust mixture with major constituents iron, forsterite, and enstatite, in agreement with our modeling with the iron+silicate dust mix. 
  \item \textsc{GGchem} predicts the existence of some refractory minerals which can survive in the $1500-2000$~K temperature range. The most abundant species under those conditions are corundum (Al$_2$O$_3$) forming at $\sim$$1850$~K, and gehlenite (Ca$_2$Al$_2$SiO$_7$) condensing at $\sim$$1700$~K. We estimate that the $3\ \mu$m flux emitted by corundum grains at $T = 1800$~K located at $0.15$~au is only $0.3\%$ of the flux emitted by iron grains at $T = 1500$~K located at $0.2$~au radius. It is still unclear whether that weak emission could be detected with current observational techniques.
  \item Our \textsc{GGchem} calculations show that condensation of carbon does not occur in the warm disk regions ($r<5$~au, $T > 300$~K) if the gas has a C/O ratio $\lesssim0.9$. 
  We propose that in the warm inner regions ($r < 5$~au) of typical planet-forming disks most if not all carbon is in the gas phase, while iron and iron sulfide (FeS, troilite) grains are major constituents of the solid mixture, along with forsterite and enstatite. Thus, small (sub-)$\mu$m sized iron grains should be the main source of dust continuum radiation at thermal IR wavelengths in disks. We repeat the recommendation of \citet{Matter2020} who suggested to replace carbon with iron when modeling inner disk spectra of YSOs.
\end{enumerate}

Our analysis exemplifies the need for detailed studies of the dust in inner disks with multiwavelength high angular resolution techniques. Currently, MATISSE is the only instrument which can resolve disks on au-scales in the $N$ band, probing the warm, forsterite and enstatite rich dust, {while at longer wavelengths} JWST/MIRI may give insights to the cold dust population, {ices, and gas chemistry \citep{Kospal2023,Grant2023,Perotti2023}}. {Additionally,} ALMA could possibly detect larger scale substructures like the $r \approx 4$~au ring of HD~144432 at its highest spatial resolution. {As of mid-2023, there is no high-resolution measurement of our target in the ALMA archive, only a measurement with the 7m array, but that has a low spatial resolution of only $4.5\arcsec$.} In the coming years, more results on the mineralogy of planetary building blocks will be expected from the MATISSE GTO survey of YSOs.

\begin{acknowledgements}
MATISSE was designed, funded and built in close collaboration with ESO, by a consortium composed of institutes in France (J.-L. Lagrange Laboratory -- INSU-CNRS -- C\^ote d’Azur Observatory -- University of C\^ote d'Azur), Germany (MPIA, MPIfR and University of Kiel), the Netherlands (NOVA and University of Leiden), and Austria (University of Vienna). The Konkoly Observatory and Cologne University have also provided some support in the manufacture of the instrument.

{GRAVITY has been developed in a collaboration by the Max Planck Institute for Extraterrestrial Physics, LESIA of Paris Observatory-PSL/CNRS/Sorbonne Université/Université Paris-Cité and IPAG of Université Grenoble Alpes/CNRS, the Max Planck Institute for Astronomy, the University of Cologne, the Centro Multidisciplinar de Astrofisica Lisbon and Porto, and the European Southern Observatory.}

Based on observations collected at the European Southern Observatory under ESO programmes 190.C-0963(D), 190.C-0963(E), 190.C-0963(F), 0100.C-0278(E), 0103.D-0153(C), 0103.C-0347(C), 0103.D-0153(G), 108.225V.003, 108.225V.011, and 108.225V.006.

P. \'Abrah\'am, F. Lykou and J. Varga are funded from the Hungarian NKFIH OTKA project no. K-132406. Á. Kóspál, J. Varga, {P. \'Abrah\'am, F. Lykou, and F. Cruz-Sáenz de Miera have received funding from the European Research Council (ERC) under the European Union's Horizon 2020 research and innovation programme under grant agreement No 716155 (SACCRED).}

The research of J. Varga and M. Hogerheijde is supported by NOVA, the Netherlands Research School for Astronomy.

J. Varga acknowledges support from the Fizeau exchange visitors program. The research leading to these results has received funding from the European Union’s Horizon 2020 research and innovation programme under Grant Agreement 101004719 (ORP).

F. Cruz-Sáenz de Miera received financial support from the European
Research Council (ERC) under the European Union’s Horizon 2020 research and innovation programme (ERC Starting Grant ``Chemtrip", grant agreement
No 949278).

P. Woitke acknowledges funding from the European Union
H2020-MSCA-ITN-2019 under Grant Agreement no. 860470 (CHAMELEON).

N. Morujão acknowledges the financial support provided by FCT/Portugal through grants PTDC/FIS-AST/7002/2020 and UIDB/00099/2020.

This research has made use of the services of the ESO Science Archive Facility.

This research has made use of the VizieR catalog access tool, CDS, Strasbourg, France (DOI : 10.26093/cds/vizier). The original description of the VizieR service was published in 2000, A\&AS 143, 23.

This research has made use of the Jean-Marie Mariotti Center OiDB service available at http://oidb.jmmc.fr .
 
This work has made use of data from the European Space Agency (ESA) mission
{\it Gaia} (\url{https://www.cosmos.esa.int/gaia}), processed by the {\it Gaia}
Data Processing and Analysis Consortium (DPAC,
\url{https://www.cosmos.esa.int/web/gaia/dpac/consortium}). Funding for the DPAC
has been provided by national institutions, in particular the institutions participating in the {\it Gaia} Multilateral Agreement.
This study has made use of the \textsc{GGchem} code, available at \url{https://github.com/pw31/GGchem}.
\end{acknowledgements}

%
%

\bibliographystyle{aa}
\bibliography{ref_MIDI_atlas}

\begin{appendix}

\section{VLTI data plots}
\label{sec:app_dataplots}

\begin{figure}[H]
    \begin{minipage}{0.99\textwidth}  

\includegraphics[width=0.97\hsize]{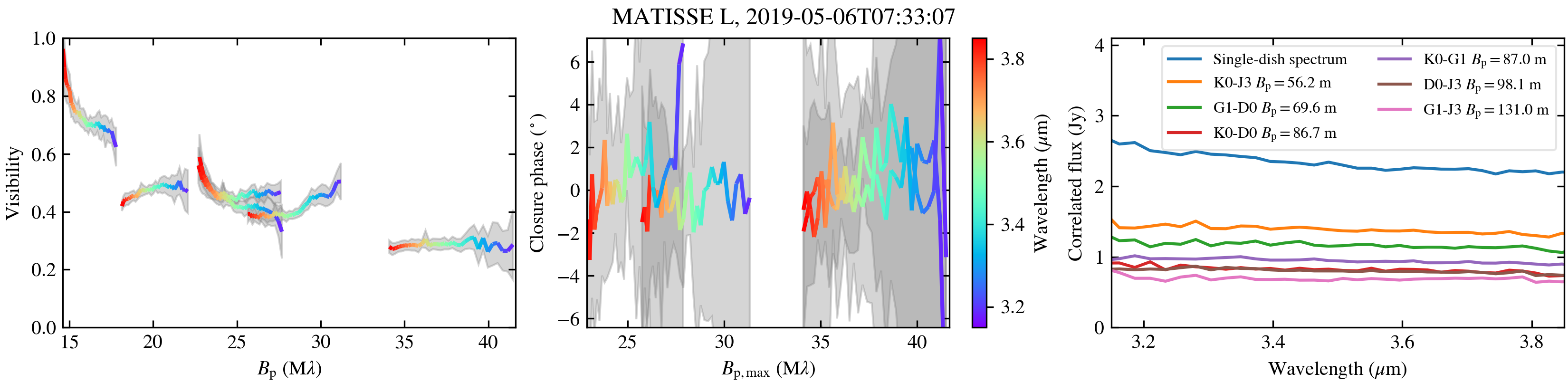}
\includegraphics[width=0.97\hsize]{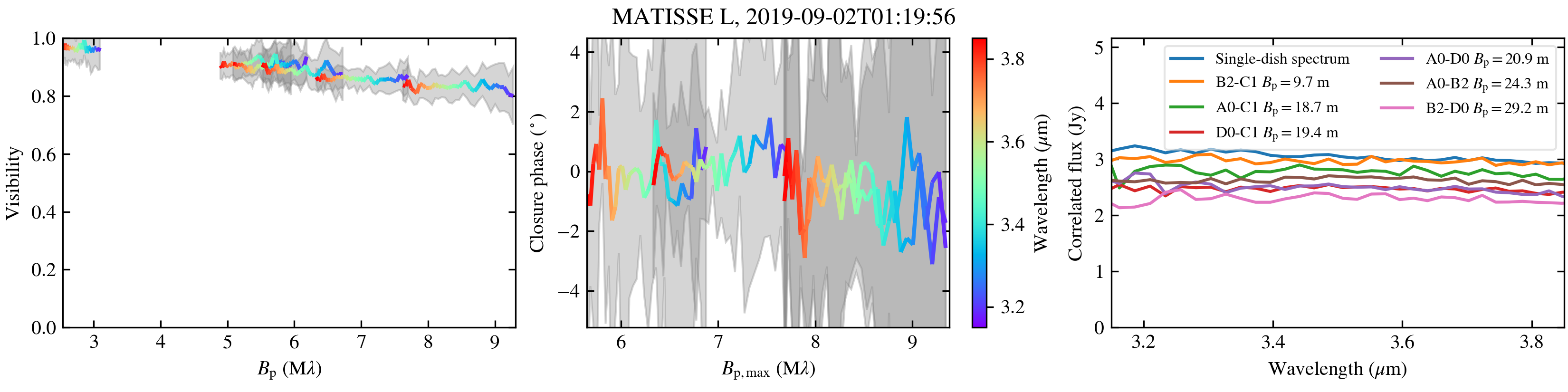}
\includegraphics[width=0.97\hsize]{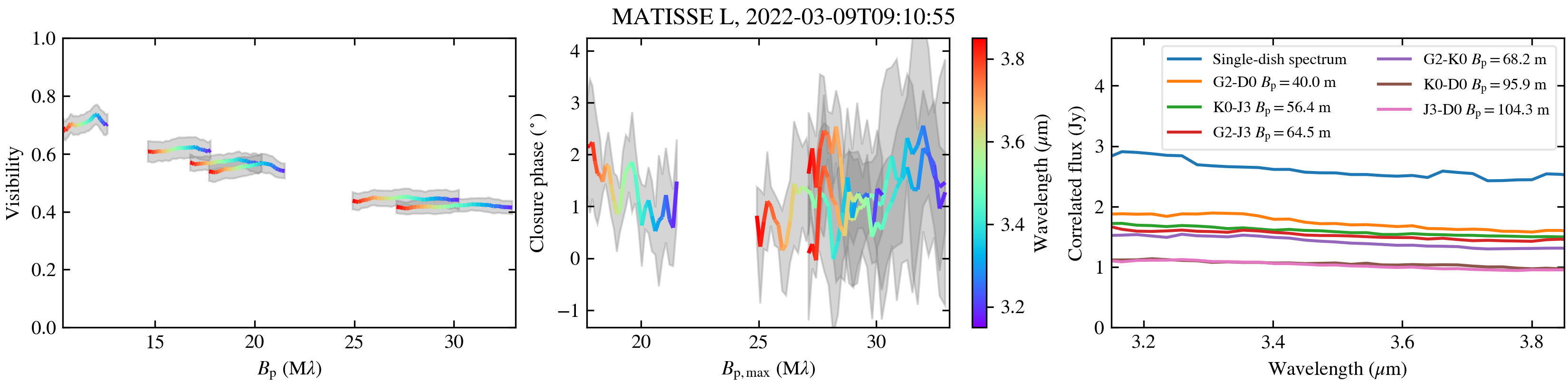}
\includegraphics[width=0.97\hsize]{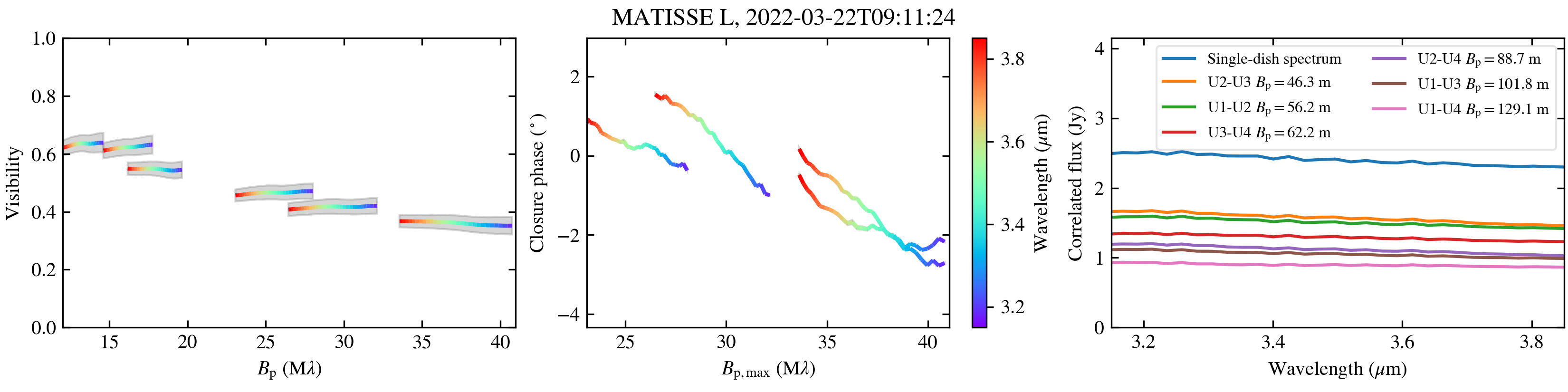}
\includegraphics[width=0.97\hsize]{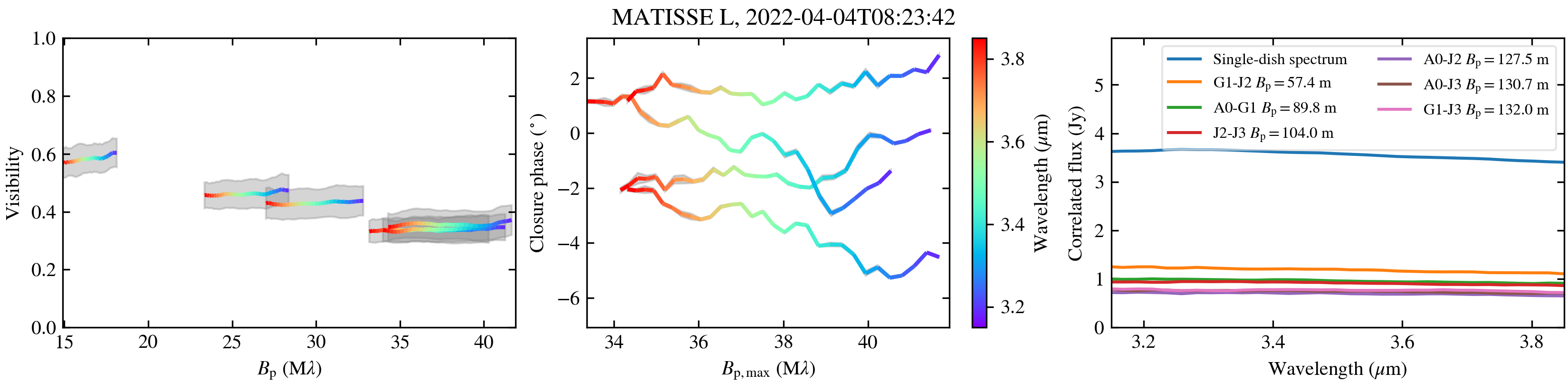}
\caption{{Calibrated MATISSE $L$ band data sets. Left column: visibility as a function of the spatial frequency. Middle column: Closure phase as a function of the spatial frequency corresponding to the longest baselines of the triangles. Right panel: Correlated flux as a function of the wavelength. We also plot the single-dish spectrum, as the zero-baseline correlated spectrum. }}
         \label{fig:data_L}
    \end{minipage}        
\end{figure}

\begin{figure*}
\includegraphics[width=\hsize]{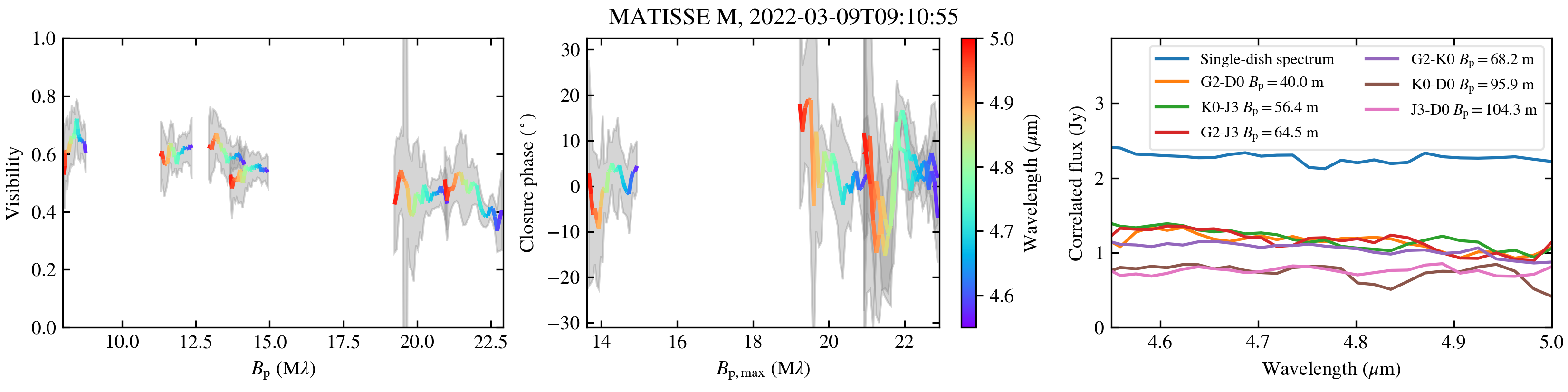}
\includegraphics[width=\hsize]{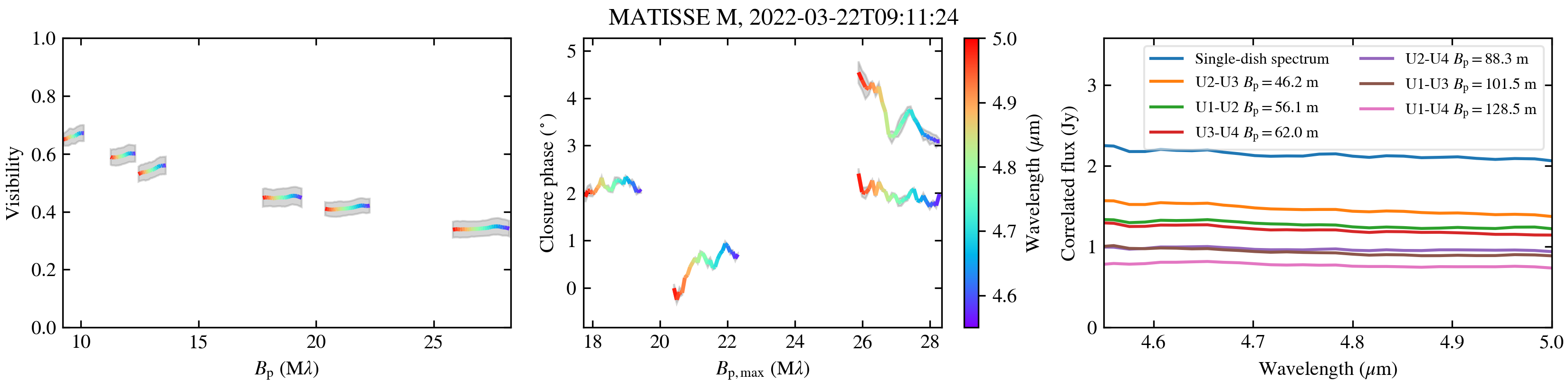}
\includegraphics[width=\hsize]{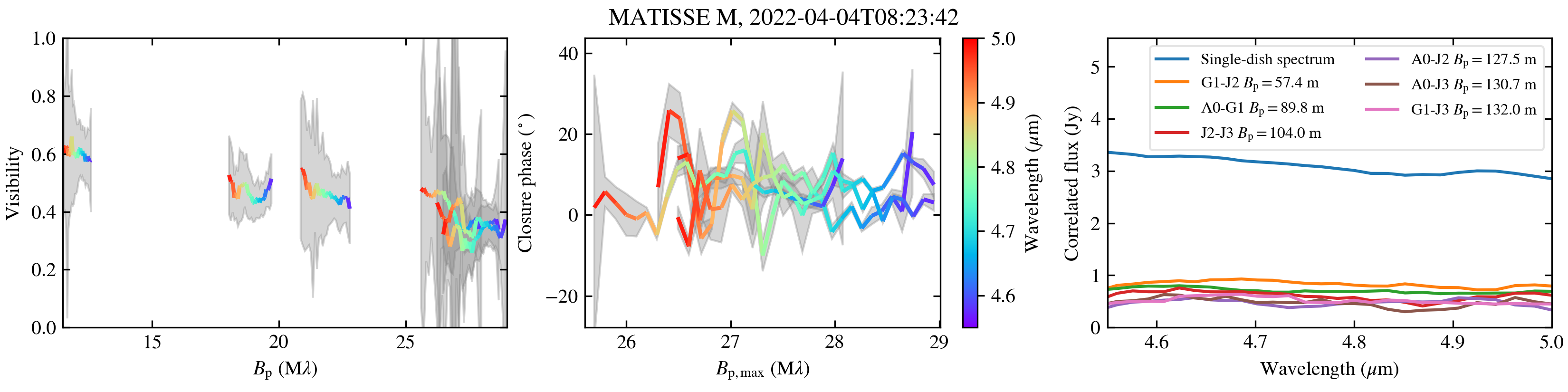}
\caption{{Same as Fig.~\ref{fig:data_L}, but for the MATISSE $M$ band data. }}
         \label{fig:data_M}
\end{figure*}

\begin{figure*}
\includegraphics[width=\hsize]{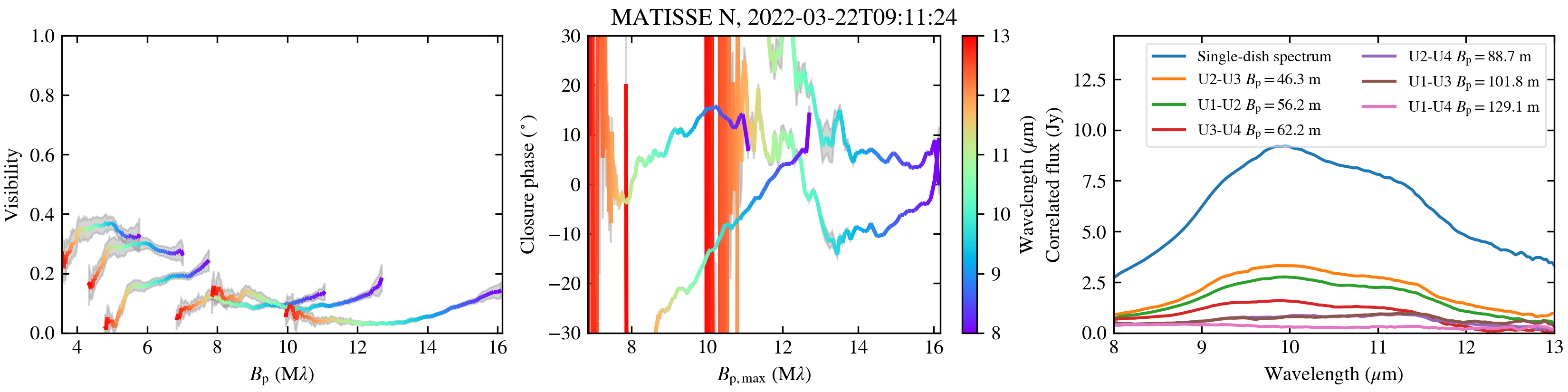}
\caption{{Same as Fig.~\ref{fig:data_L}, but for the MATISSE $N$ band data. We note that, as of DRS version 1.5.8, the sign convention of the $N$ band phases are flipped with respect to the $LM$ band phases \citep[see also][]{Gamez2022}. This causes a $180\degr$ rotation of the $N$ band images. This issue is no concern for us, as we do not use the $N$ band closure phases in our modeling. }}
         \label{fig:data_N}
\end{figure*}

\begin{figure}
\includegraphics[width=\hsize]{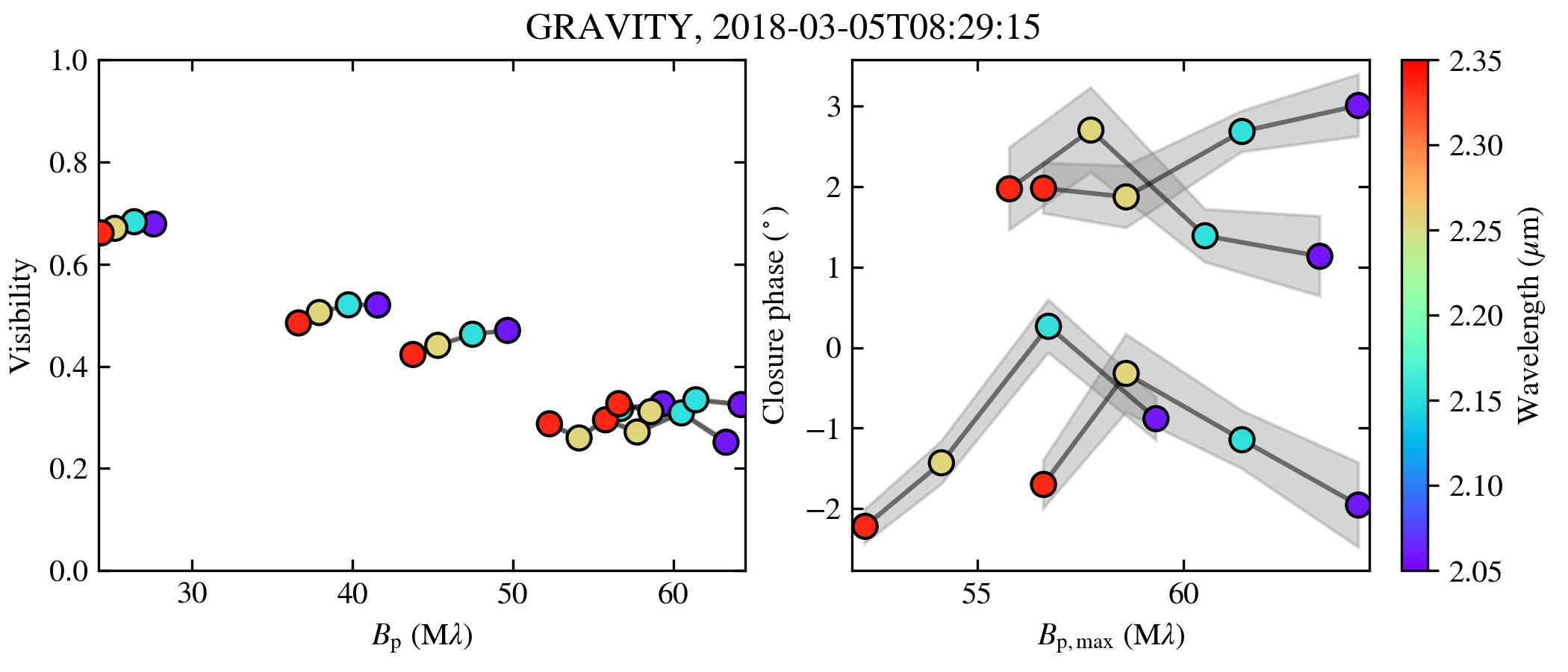}
\includegraphics[width=\hsize]{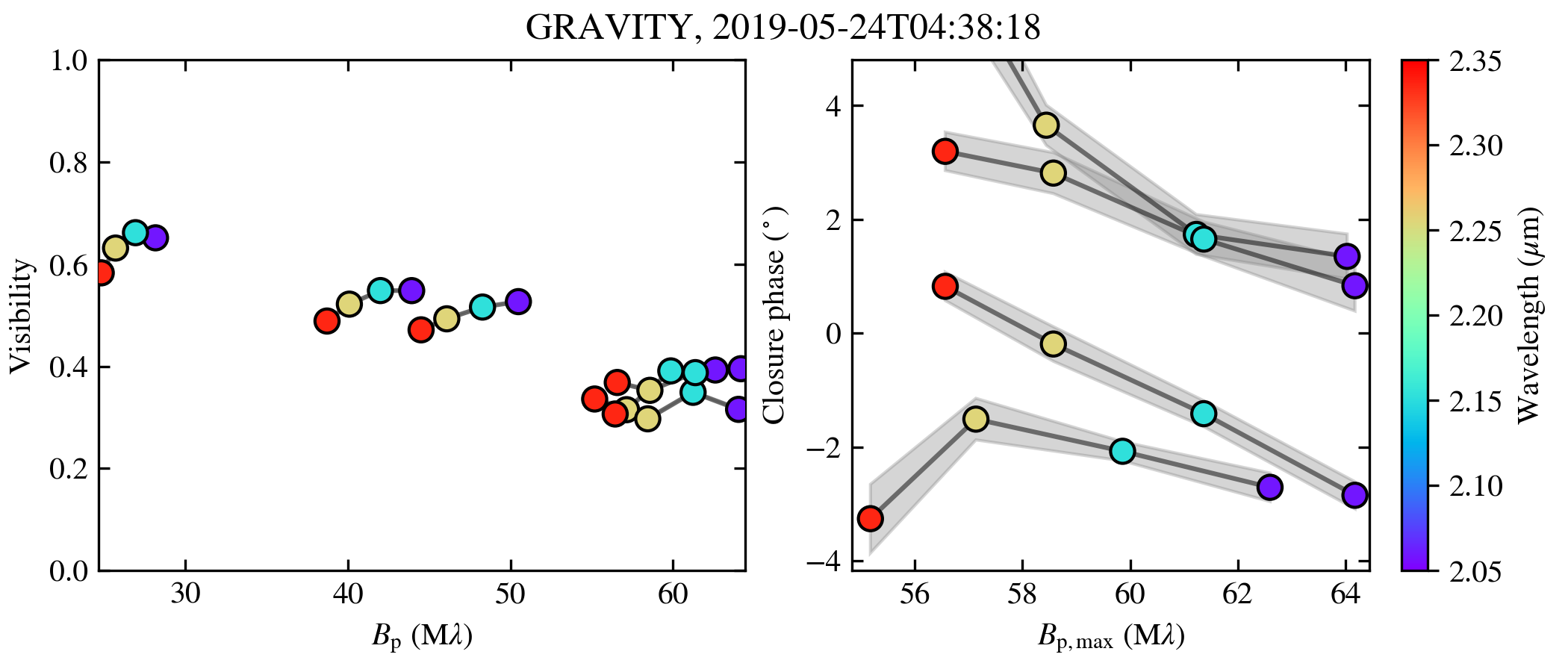}
\caption{{Calibrated GRAVITY data sets. Left column: visibility as a function of the spatial frequency. Right column: Closure phase as a function of the spatial frequency corresponding to the longest baselines of the triangles. }}
         \label{fig:data_K}
\end{figure}

\begin{figure}
\includegraphics[width=\hsize]{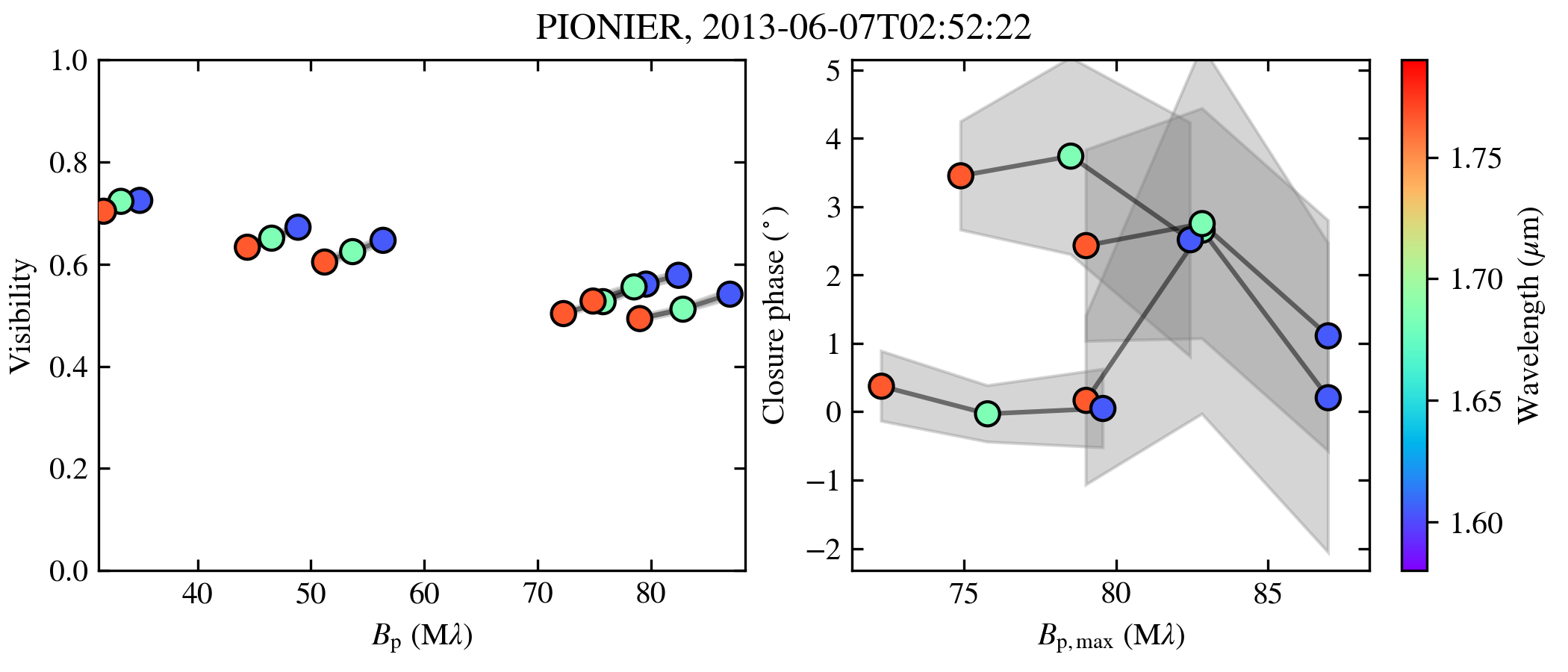}
\includegraphics[width=\hsize]{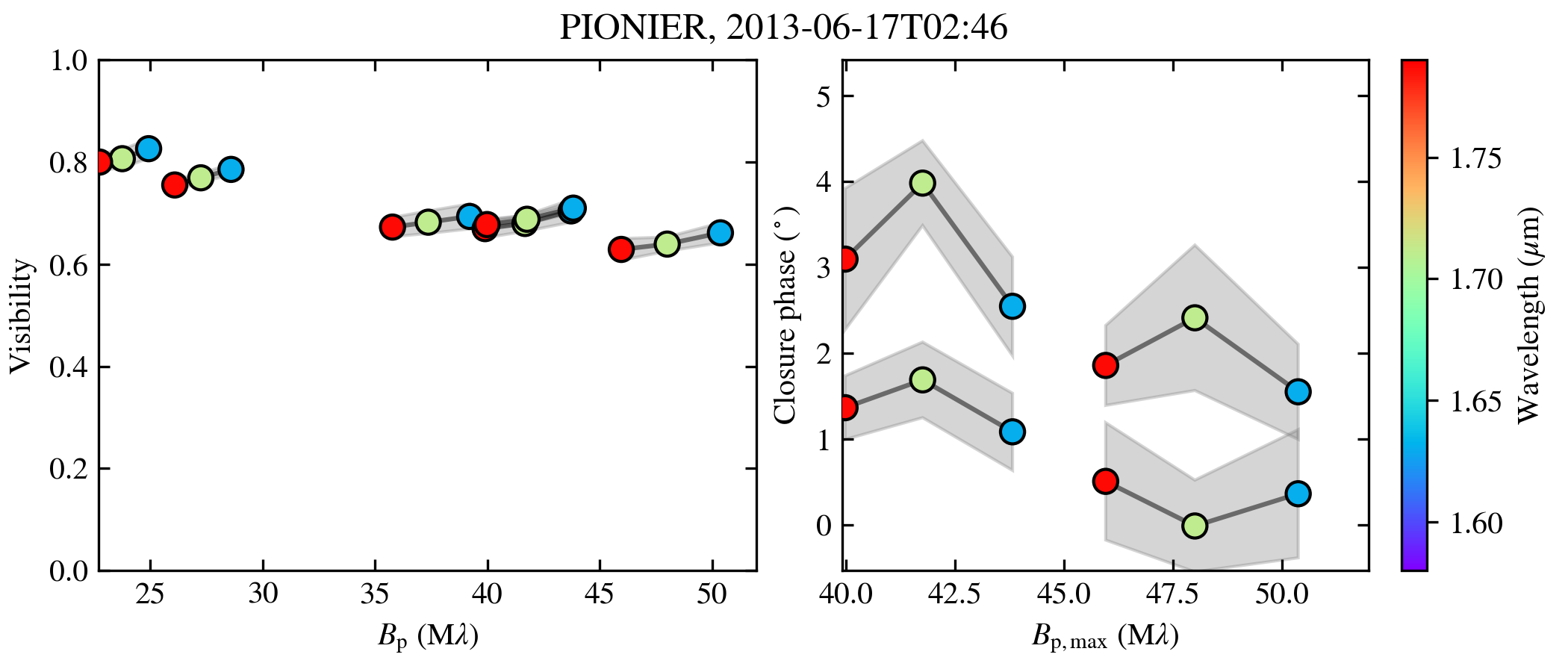}
\includegraphics[width=\hsize]{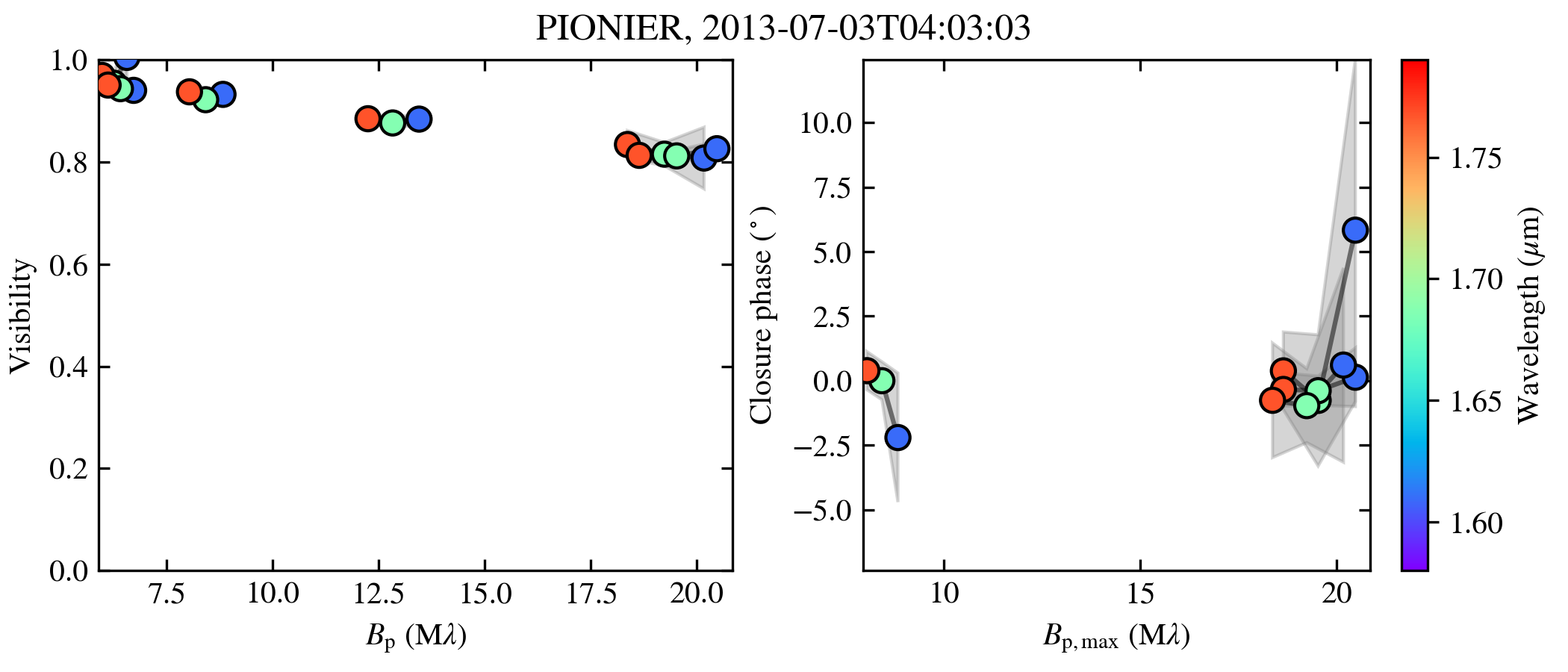}
\caption{{Same as Fig.~\ref{fig:data_K}, but for the PIONIER data. }}
         \label{fig:data_H}
\end{figure}

\FloatBarrier

\section{Additional notes on the MATISSE data processing}
\label{sec:app_dataproc}
While the $L$ band sensitivity of the instrument is more than enough to record good quality data on HD~144432 using the ATs, it is not the case in $N$ band where only the UT data can be used. For obtaining correlated fluxes we apply the so-called coherent processing provided by the DRS pipeline, while the visibilities are processed using the incoherent method. The coherent algorithm has a gain in S/N compared to the incoherent method. Due to this advantage, coherently processed  correlated fluxes are especially useful in the $N$ band. For the \verb|specfit| modeling, we directly used those data. However, the PIONIER and GRAVITY data does not contain correlated flux, only visibilities, thus for a homogeneous modeling of all VLTI data sets we need to use visibilities even in the $N$ band. We opt not to use the incoherently processed $N$ band visibilities, because of their lower quality. Instead we apply a special treatment to the $N$ band correlated fluxes by dividing them with the average single-dish flux, in this way producing coherently processed visibility data. We use these specially calibrated $N$ band visibilities in the \verb|TGMdust| runs.

The $L$ and $M$ band MATISSE data comes in two flavors: chopped and non-chopped. Chopping helps to suppress the sky background, which gets increasingly dominant in the $M$ band. Thus, in $M$ band it is mandatory to use chopped data, while in $L$ band both chopped and non-chopped data may be used. During chopping the frequent switching between the object and sky positions
might cause to lose track of the adaptive optics guiding or the fringe tracking, which results in bad quality data. Thus, when both non-chopped and chopped $L$ band data-sets were available, we carefully inspected both products, and selected the one which we deemed more reliable {(the one with smaller error bars)}. 

\section{Visibility calculation}
\label{sec:app_viscalc}
The visibility function is the Fourier-transform of the object's image on the sky. IR interferometers sample the visibility function at distinct spatial frequencies, set by the baseline length between the telescope pairs. The visibility function is a complex quantity, and in IR interferometry it is usually represented by its absolute value and phase. The absolute value of the visibility can be expressed either as a unitless quantity, or in flux units. In the latter case it is known as the correlated flux. It is the correlated flux which is directly measured by the interferometer. In polar coordinates the Fourier-transform becomes the Hankel-transform. {For an object with circular symmetry,} the correlated flux can be expressed from the radial surface brightness profile {($I_{\nu}\left(r\right)$)} {as follows}:

\begin{equation}
\label{eq:hankel}
F_{\text{corr},\nu}\left(B_{\mathrm{p}}\right) = \int_{R_\text{in}}^{R_\text{out}} 2\pi r  I_{\nu}\left(r\right) J_0\left(2\pi r B_{\mathrm{p}} / \lambda \right) \mathrm{d}r,
\end{equation}
where $B_\text{p}$ is the projected baseline length,  $J_0$ is the Bessel function of the first kind 
{of order zero}, {and $R_\text{in}$, $R_\text{out}$ are the inner and outer radii of the object, respectively}. The $B_\text{p}/\lambda$ is the spatial frequency. The argument of $J_0$ is in radians. In our multi-zone disk model, $F_{\text{corr},\nu,i}$ {is} the correlated flux of the i-th zone, {with the integration limits being the zone boundaries $R_\text{in}=R_\text{i}$ and $R_\text{out}=R_\text{i+1}$}. The formula above assumes a circularly symmetric object image. If the image is elliptic, such as an inclined thin disk, the visibility function also depends on the position angle of the baselines ($\theta_B$), not just on their lengths. In that case we can transform the projected baseline lengths so that we remove the effect of the inclination: 
\begin{equation}
    B_\text{eff} = B_p \sqrt{\cos^2 \left(\Delta\theta\right) + \cos^2 \left(i\right) \sin^2 \left(\Delta \theta\right)},
\end{equation}
where $\Delta \theta = \theta_B-\theta$ is the difference between the position angle of the baseline and the position angle of the major axis of the object image ($\theta$). Using $B_\text{eff}$ removes the $\theta_B$ dependence in the visibility calculation, and we can apply Eq.~\ref{eq:hankel}, as if the object image were circular.

In our model the central star is represented as a point source, thus its correlated flux equals its (single-dish) flux. The correlated fluxes of the disk radial zones ({$F_{\text{corr},\nu,i}$}), and that of the central star {($F_{\nu,\text{star}}$)} are simply added together:
\begin{equation}
F_{\text{corr},\nu}\left(B_{\mathrm{eff}}\right) = \sum_{i=1}^{M} F_{\text{corr},\nu,i}\left(B_{\mathrm{eff}}\right) + F_{\nu,\text{star}}.
\end{equation}
Finally, we calculate the unitless visibility ($V_\nu$), by normalizing the correlated flux with the single-dish flux {($F_{\nu}$)}:
\begin{equation}
V_{\nu}\left(B_{\mathrm{eff}}\right) = F_{\text{corr},\nu}\left(B_{\mathrm{eff}}\right) / F_{\nu}.
\end{equation}

\newpage

\section{Opacity curves}
 
\begin{figure}[H]
    \begin{minipage}{\textwidth}  

 \includegraphics[width=\hsize]{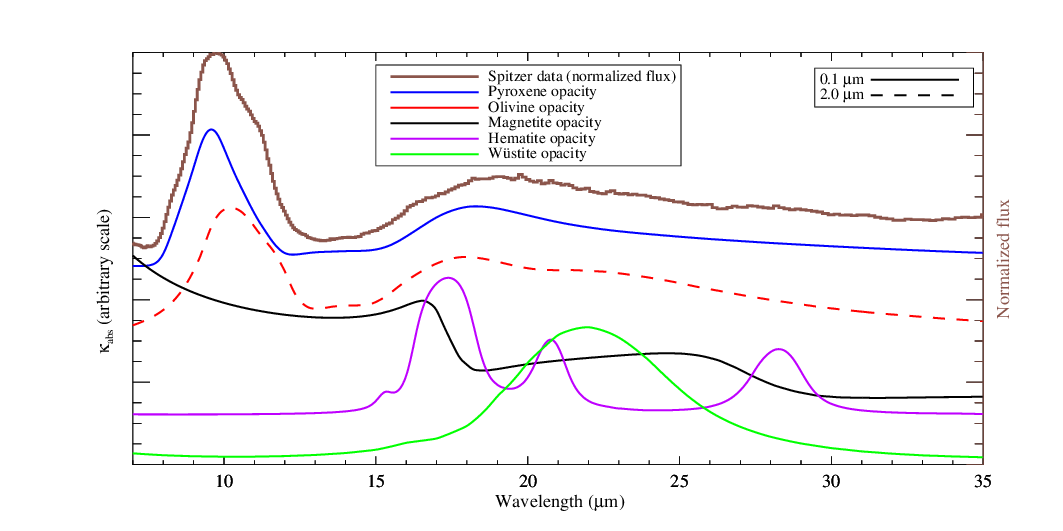}

      \caption{$7-35\ \mu$m opacity curves of pyroxene, olivine, and various iron oxides, compared with the normalized Spitzer spectrum of HD 144432 (brown curve, AOR: 3587072). The Spitzer spectrum was normalized by dividing the data with the source function of our best-fit model. Optical constants of iron oxides were downloaded from the Jena Database of Optical Constants for Cosmic Dust (\url{https://www.astro.uni-jena.de/Laboratory/OCDB/mgfeoxides.html}). The original source for w\"ustite is \citet{Henning1995}. Optical data for hematite and magnetite come from Amaury H.M.J. Triaud (unpublished). The opacity curves were generated with optool, using the distribution of hollow spheres (DHS) scattering theory. The upper boundary of the hollow sphere distribution ($f_\mathrm{max}$) is $0.7$. All curves are arbitrarily scaled and shifted to aid visual comparison.
             }
         \label{fig:app_oxides}
\end{minipage}
   \end{figure}
   
\begin{figure}[H]
   \begin{minipage}{\textwidth}  
 \includegraphics[width=\hsize]{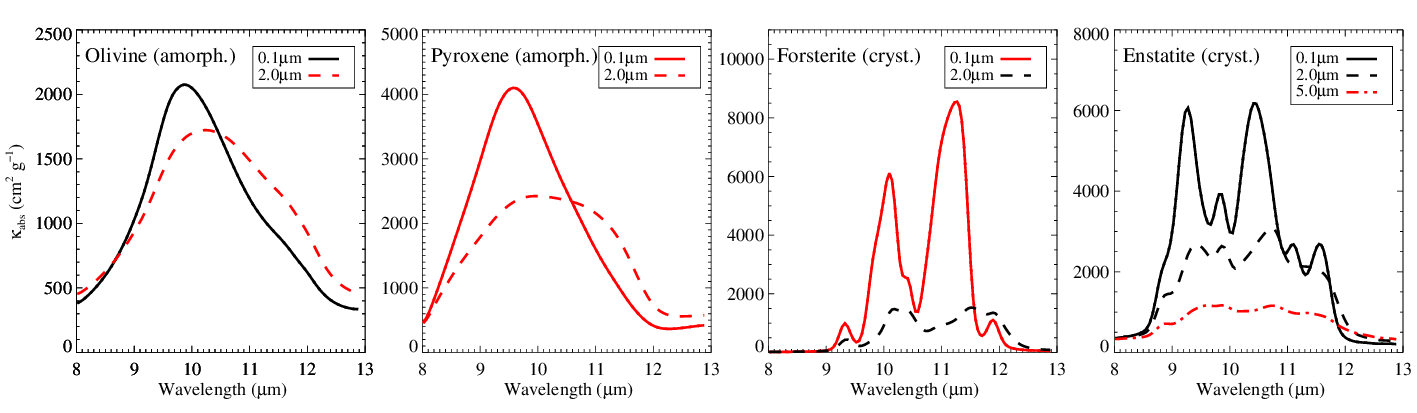}

      \caption{$N$ band opacity curves of the silicate minerals used in this work. Red color indicates the subset used in the final dust modeling. 
             }
         \label{fig:app_opac}
\end{minipage}
   \end{figure}

\twocolumn

 \begin{figure}
 \includegraphics[width=\hsize]{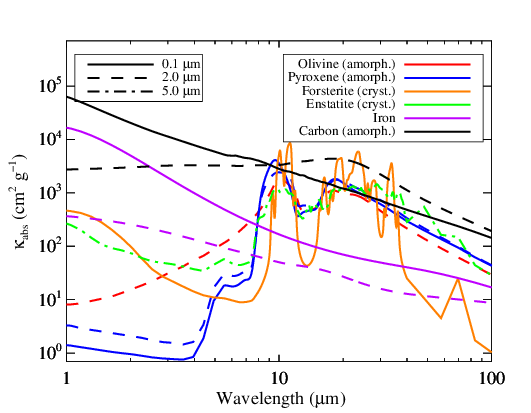}
      \caption{Opacity curves in the full wavelength range of the data. Shown only the subset used in the final dust modeling, including iron and carbon as well. 
             }
         \label{fig:app_opac2}
   \end{figure}

\FloatBarrier

\section{Two-step optimization procedure}
\label{sec:app_optimization}

Our approach for finding the best-fit model is detailed in the following.
First, we constrain the {silicate} composition {and the main structural disk parameters} by fitting only the $N$ band MATISSE data {(six correlated spectra and the single-dish spectrum between $8$ and $13\ \mu$m).} For this purpose we use our own idl software, based on the \verb|specfit| dust compositional fitting tool by \citet{Juhasz2009}. In our code we implemented our multi-zone temperature gradient model, and added the functionality to model interferometric data, while we retained the original fitting procedure of \verb|specfit|. The optimization of the disk structural parameters is done by a genetic optimization algorithm named \verb|pikaia| \citep{Charbonneau1995}. In each optimization step of the genetic algorithm, a spectral decomposition is employed by a bound constrained linear least-squares fitting algorithm \citep{Lawson1974}, in order to evaluate the dust mass fractions. The latter step requires a change in our model when calculating the surface brightness profile in Eq.~\ref{eq:I_nu_profile}:
\begin{equation}
\label{eq:I_nu_profile_alt}
I_{\nu,i} \left( r \right) = \tau_{\nu,i}\left(r\right) B_{\nu}\left( T\left( r \right)\right) = \sum_{j=1}^{N}{c_{i,j} \Sigma_{i}\left( r \right)  \kappa_{\nu,j}} B_{\nu}\left( T\left( r \right)\right).
\end{equation}
{Here we assume the disk to be optically thin ($\tau \ll 1$) at our observing wavelengths. Thus, we can calculate the total emission by simple linear addition of the contributions from the individual dust components. This is needed by the spectral decomposition algorithm in the \texttt{specfit} code.}
The final result {of this modeling} is the set of fit structural parameters and dust mass fractions.

We expect that the N-band data alone cannot differentiate between the dust species with featureless opacity curves (iron and carbon). This is because the opacities of those species look very similar in that spectral region. Thus, we do not include either carbon or iron in our \verb|specfit| runs. Instead we include a hidden dust component with a power law opacity curve to represent the dust responsible for the featureless continuum. The exponent of the power law is a fit parameter, while we fix the opacity\footnote{This represents a typical order-of-magnitude value for submicron-sized grains.} at $8\ \mu$m at $500\ \text{cm}^2/\text{g}$. 
    
In the second modeling step we aim to constrain the disk structure overall. The fitting procedure employed in the first step is very capable at constraining the dust composition, but it can only model optically thin emission. 
The linear spectral decomposition gives erroneous results if in some spectral regions the emission is optically thick. The latter constrain motivated our choice to restrict the fitting with \verb|specfit| to the $N$ band.
To overcome that limitation, we apply our own model fitting tool, \texttt{TGMdust}\footnote{{\texttt{TGMdust} is available for download at \url{https://home.strw.leidenuniv.nl/~varga/pro/tgmdust}.}} {(Temperature Gradient Modeling with dust)}, which is a python implementation of our multi-zone disk model described in Sect.~\ref{sec:model}. For the optimization we use \verb|emcee|, an MCMC ensemble sampler \citep{Foreman-Mackey2013a,Foreman-Mackey2013b}.  
In \verb|TGMdust| we use {our original assumption} for the radiative transfer, described in Eq.~\ref{eq:I_nu_profile}, {capable to represent optically thick emission}. We note that the linear combination assumption employed in the previous step provides nearly equivalent results to the \verb|TGMdust| fits if the dust emission is optically thin. In this step of modeling, we use all data we have, not just the $N$ band. The large number of model parameters (44 in the case of the 3-zone model) might prevent the MCMC sampler from reaching convergence. Thus, with \texttt{TGMdust} we fit a limited set of parameters, and we use the dust mass fractions obtained in the first step by \texttt{specfit} as fixed parameters. 

\FloatBarrier
\twocolumn

\section{Degeneracy in the spectral decomposition}
\label{sec:app_alt_specfit}

\begin{figure}[H]
\begin{minipage}{\textwidth}  
 \includegraphics[width=0.503\textwidth,trim=30 20 40 42,clip]{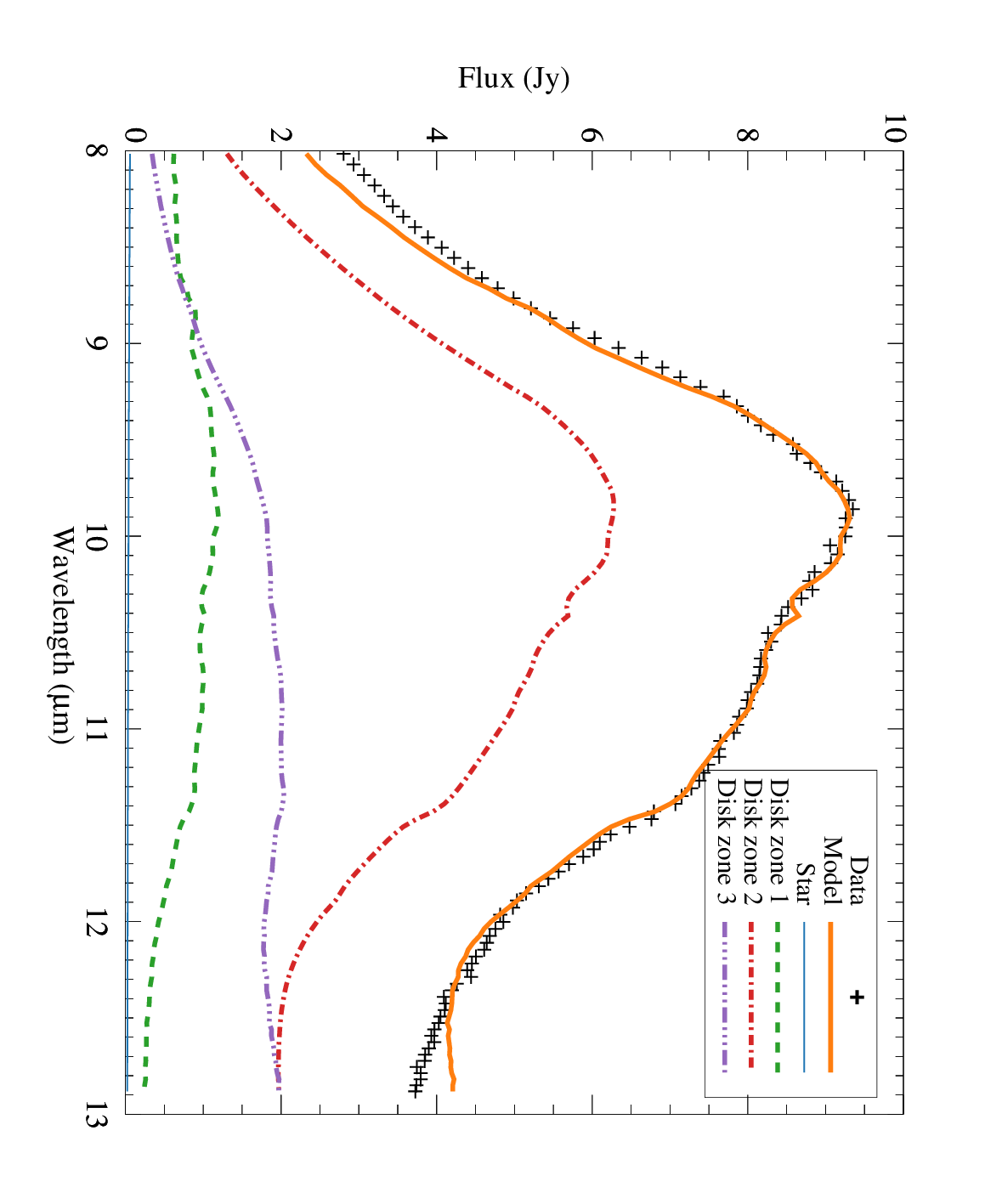}
 \includegraphics[width=0.487\textwidth,trim=18  8 24 20,clip]{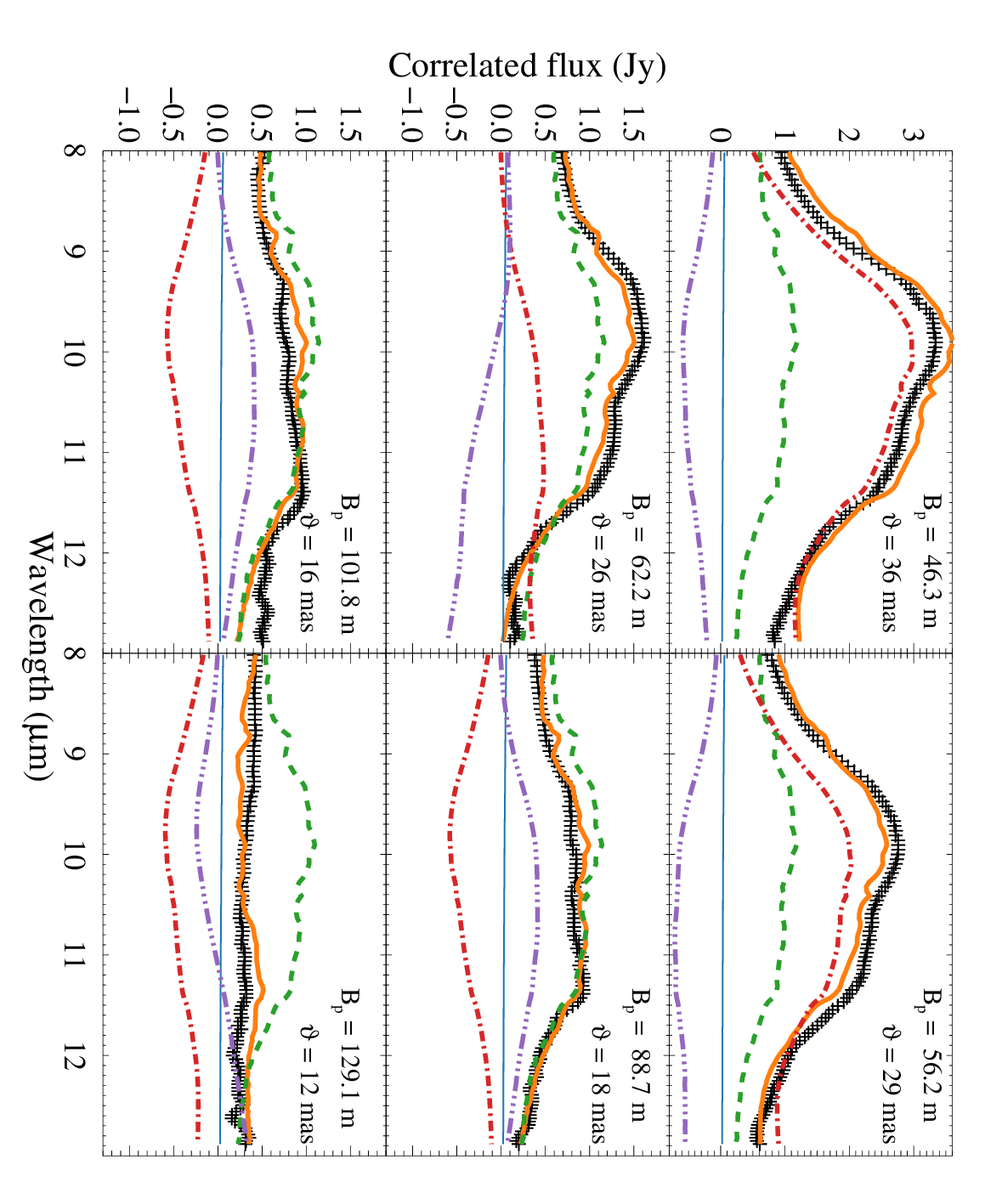}
      \caption{Same as Fig.~\ref{fig:silfit}, but for a \texttt{specfit} solution corresponding to the dust mass fractions listed in Table~\ref{tab:app_alt_specfit_res}.}
         \label{fig:app_alt_silfit}
\end{minipage}
\end{figure}

\begin{figure}
 \includegraphics[width=0.98\columnwidth,trim=20 25 35 22,clip]{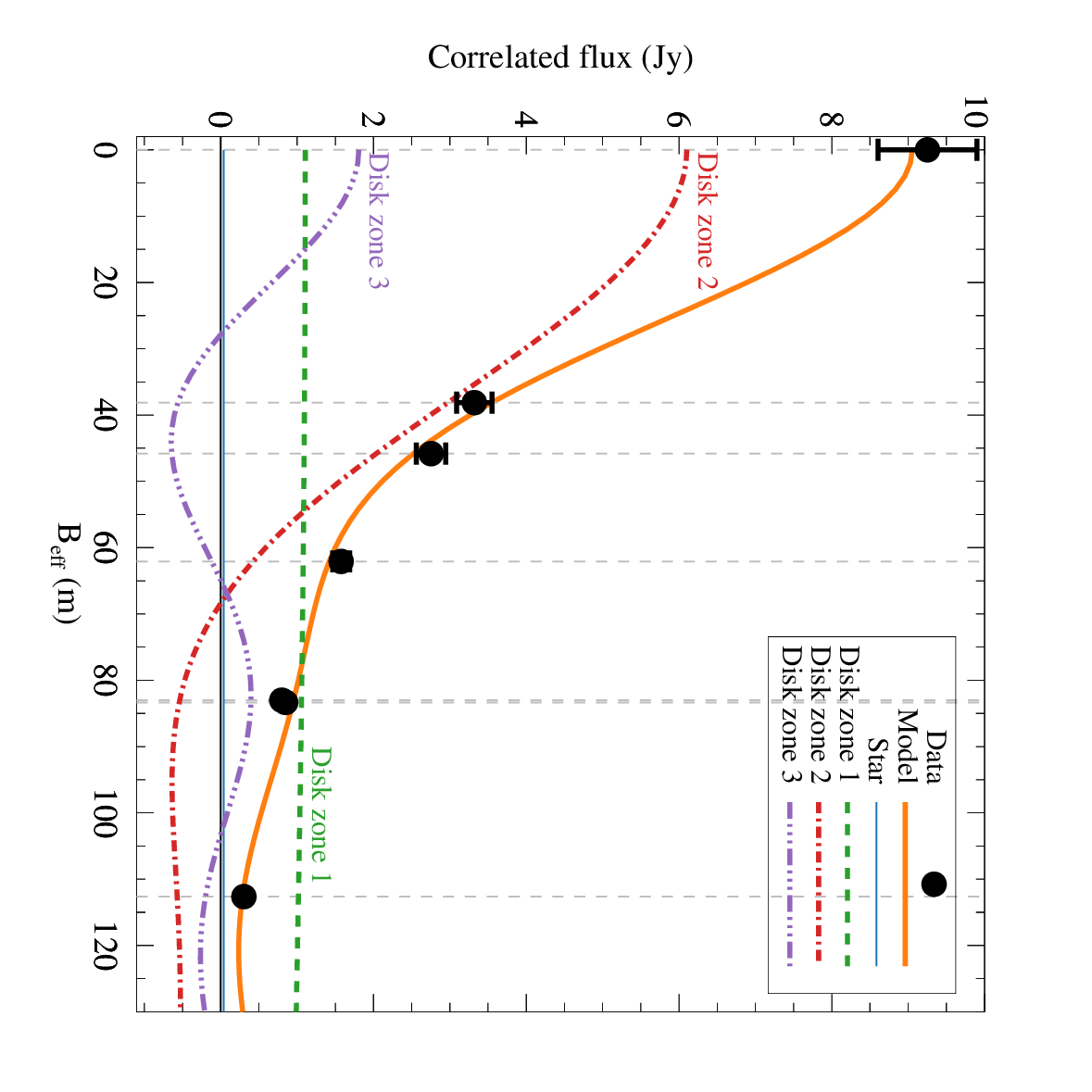}
      \caption{
      Same as Fig.~\ref{fig:silfit_bl}, but for a \texttt{specfit} solution corresponding to the dust mass fractions listed in Table~\ref{tab:app_alt_specfit_res}.} 
         \label{fig:app_alt_silfit_bl}
\end{figure}

As was shown by \citet{vanBoekel2005survey} (see Sect. 5.2 and Fig.~8 therein), the silicate spectral templates used in the spectral decomposition are degenerate to a degree. We investigated this issue in our \texttt{specfit} model run by inspecting the individual samples. We found several good-fitting solutions with a dust composition differing considerably from the best fit. One example is shown in Table~\ref{tab:app_alt_specfit_res}, and in Figures \ref{fig:app_alt_silfit}, \ref{fig:app_alt_silfit_bl}. The $\chi^2$ of the alternative solution is $1.7$ times larger than that of the best fit. The fit is qualitatively similar to that of the best fit, except at the edges of the band ($<8.5\ \mu$m and $>12.5\ \mu$m) where there is a larger deviation from the data. The zone boundaries ($0.2$, $0.9$, and $4.0$~au) are close to the values of the best fit, while the mass fraction values indicate degeneracies between large pyroxene and enstatite in zone 2, and between large olivine, small pyroxene, and enstatite in zone 3. These findings imply that despite the degeneracies in the silicate spectral templates, the disk substructrures (zone boundaries) are robustly constrained by our modeling.

\begin{table}[H]
 \caption[]{\label{tab:app_alt_specfit_res} An example for an alternative good-fitting solution for the silicate spectral decomposition. The abundances are 
 given as fractions of the total dust mass in percent, excluding the {hidden} dust component responsible for the continuum emission.} 
    \begin{tabular}{p{2.2cm}ccc} 
    \hline \hline 
    Species & 
    Zone 1 & 
    Zone 2 & 
    Zone 3 \\ 
    & $0.2-0.9$~au & $0.9-4.0$~au & $>4.0$~au \\
    \hline
$\text{Olivine}\ 2.0\ \mu\text{m}$ &  $14.9$  & $38.7$  & $0.0$  \\[1mm]
$\text{Pyroxene}\ 0.1\ \mu\text{m}$ &  $0.0$  & $43.4$  & $50.9$  \\[1mm]
$\text{Pyroxene}\ 2.0\ \mu\text{m}$ &  $9.3$  & $15.8$  & $0.0$  \\[1mm]
$\text{Forsterite}\ 0.1\ \mu\text{m}$ &  $2.9$  & $1.7$  & $1.2$  \\[1mm]
$\text{Enstatite}\ 5.0\ \mu\text{m}$ &  $72.9$  & $0.4$  & $47.9$  \\[1mm] 
    \hline
    \hline 
{Crystallinity fraction} & {$75.8$} & {$2.1$} & {$49.1$} \\[1mm]
    \end{tabular} \\
\end{table}

\FloatBarrier
\twocolumn

\section{Supplementary model plots}
\label{sec:app_model_plots}

\subsection{\texttt{TGMdust} model run with iron grains}

\begin{figure}[H]
   \begin{minipage}{\textwidth}  
 \includegraphics[width=\hsize]{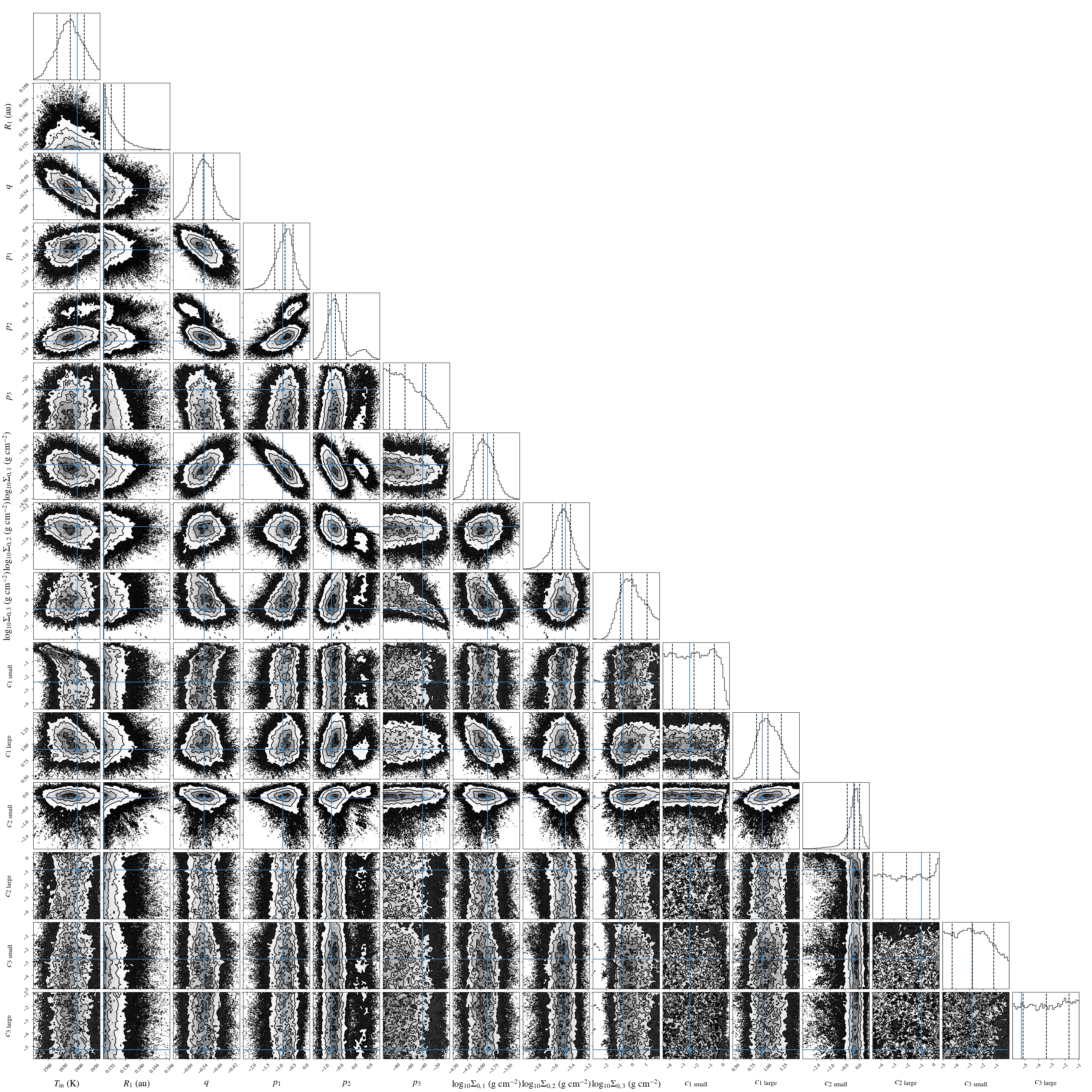}
      \caption{Corner plot showing the posterior distributions of the MCMC samples from the run with iron grains. We note that in case of the surface densities and mass fractions we fit the logarithm of the quantities.  
             }
         \label{fig:cornerplot_Fe}
  \end{minipage}
   \end{figure}

\FloatBarrier
\twocolumn

\subsection{\texttt{TGMdust} model run with carbon grains}
\label{sec:app_plots_C}

\begin{figure}[H]
   \begin{minipage}{\textwidth} 
 \includegraphics[width=\hsize]{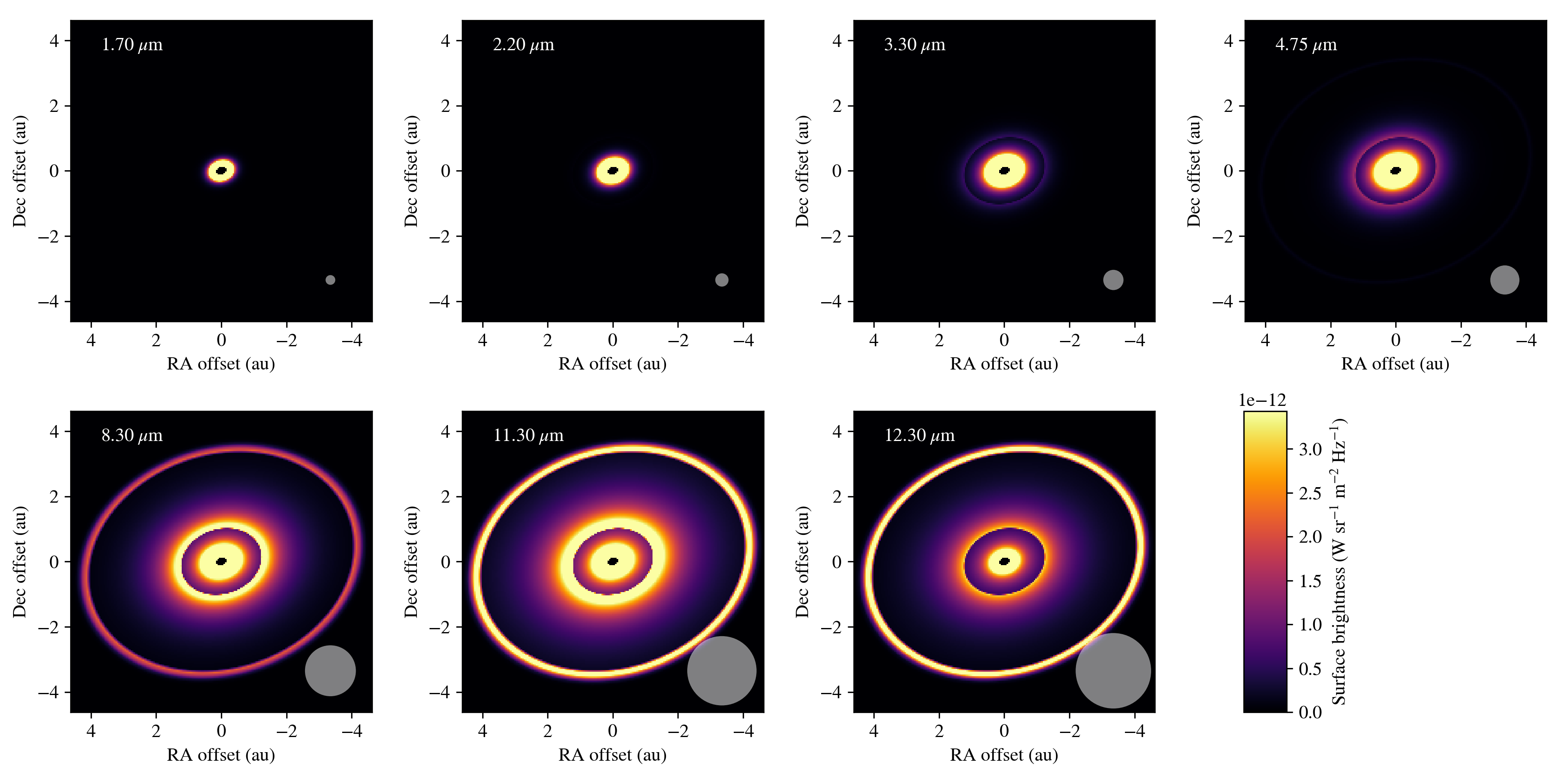}
 \caption{Best-fit model images at various data wavelengths with the \texttt{TGMdust} run with carbon grains. The brightness scaling is linear and homogeneous across all wavelengths. {For a better perception} of the fainter structural features, the images become saturated at $3.4\times10^{-12}$ W~m$^{-2}$~Hz$^{-1}$~sr$^{-1}$. The gray circles in the bottom left corner indicate the approximate beam size {(estimated as $0.77 \lambda / (B_\text{p,max})$, where $B_\text{p,max}$ is the maximum baseline length).}}
         \label{fig:modelimg_C}
  \end{minipage}
   \end{figure}

\begin{minipage}{0.49\textwidth} 

\begin{figure}[H]
 \includegraphics[width=\hsize]{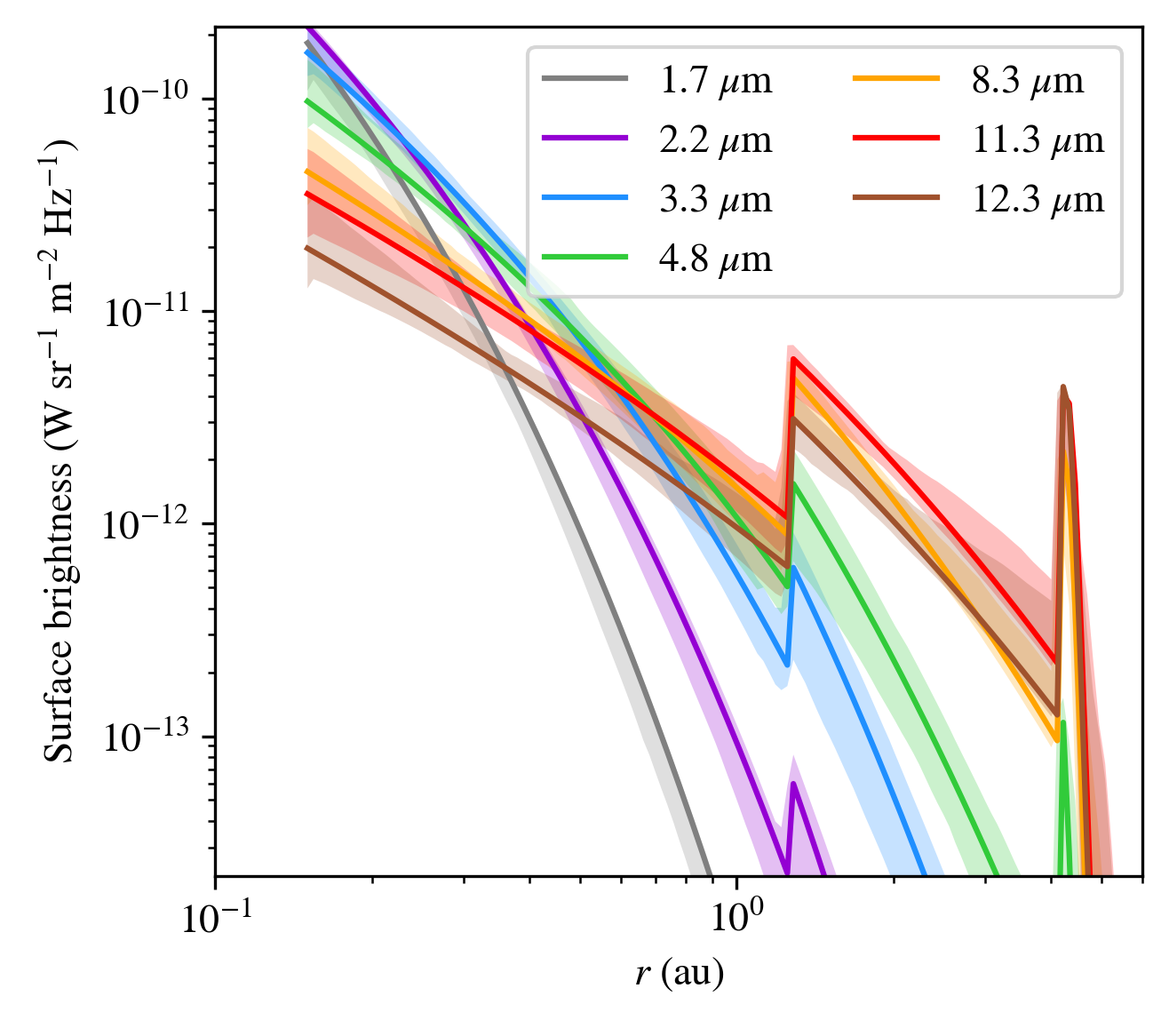}
      \caption{Radial surface brightness profiles of the disk corresponding to the best-fit \texttt{TGMdust} model with carbon grains. The shaded area around each curve is the $16-84$ percentile range from the last $5000$ samples of the chain. 
      } 
         \label{fig:I_nu_prof_C}
   \end{figure}
\end{minipage}

\noindent
\begin{minipage}{0.49\textwidth} 
\vspace{14.7cm}
\begin{figure}[H] 
 \includegraphics[width=0.495\hsize] {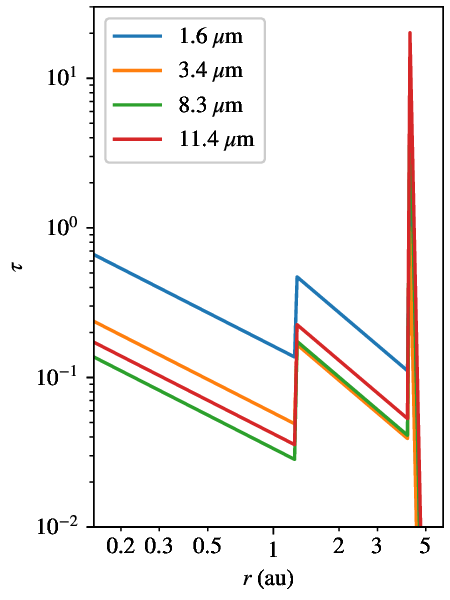}
  \includegraphics[width=0.495\hsize]{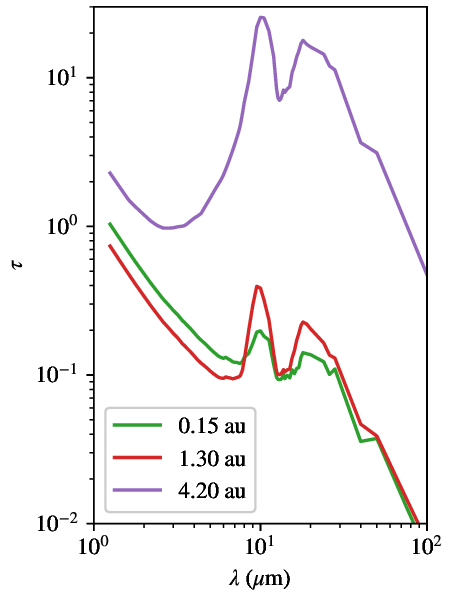}
      \caption{Vertical optical depth of the best-fit disk model with carbon grains. Left panel: Radial optical depth profiles at selected wavelengths. Right panel: Optical depth spectra at the inner edges {of each zone}.  } 
         \label{fig:tau_C}
   \end{figure}
\end{minipage}

 \begin{figure*}
 \includegraphics[width=\hsize]{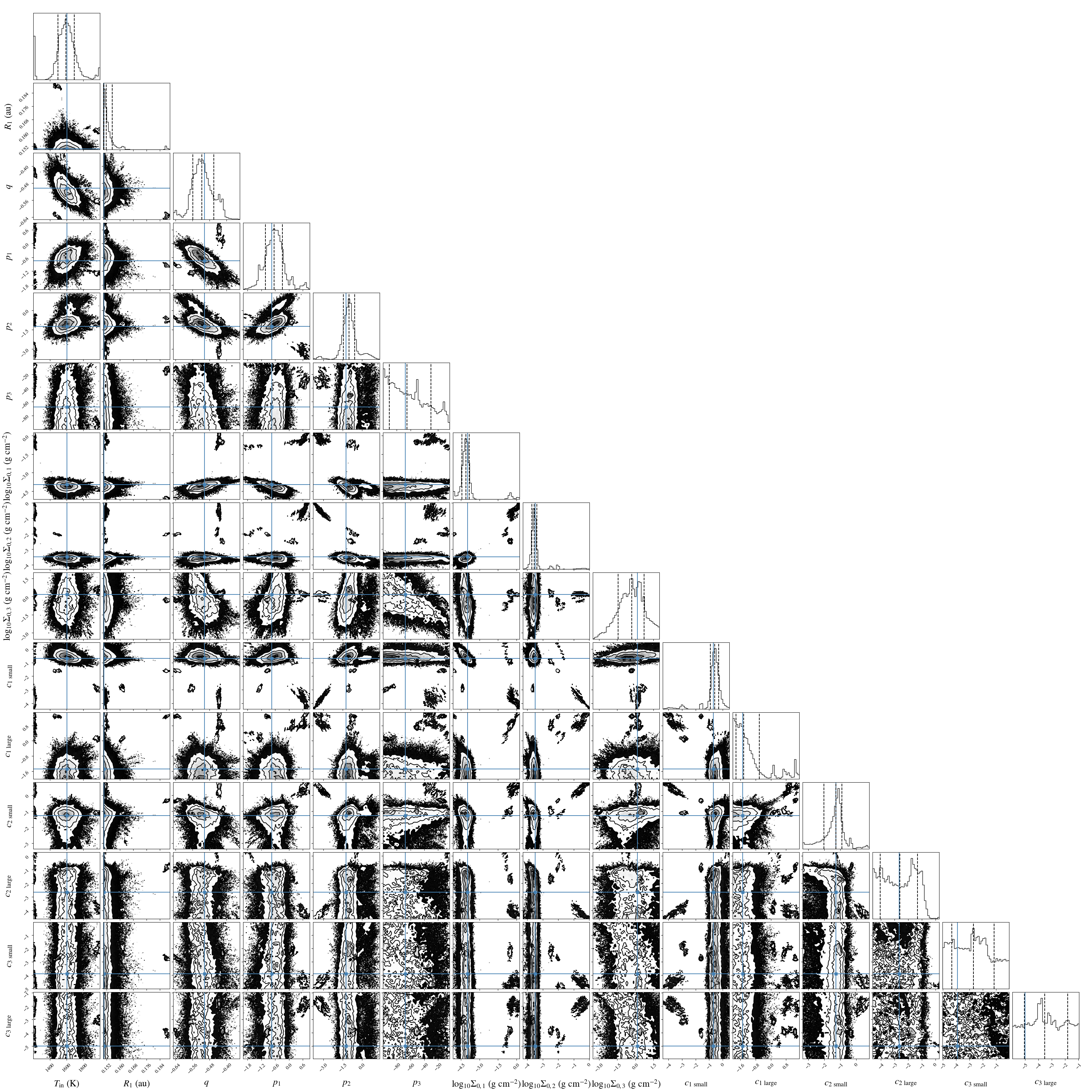}
      \caption{ Corner plot showing the posterior distributions of the MCMC samples from the run with carbon grains. We note that in case of the surface densities and mass fractions we fit the logarithm of the quantities.  
             }
         \label{fig:cornerplot_C}
   \end{figure*}

\end{appendix}
\end{document}